\documentclass{aa}  
\newcommand{\resgammafit}{$\gamma = -15.091_{-0.021}^{+0.019}\;\mathrm{km/s}$} 
\newcommand{\resqfit}{$q = 0.7663_{-0.0018}^{+0.0020}$} 
 
\newcommand{\resgammawilson}{$\gamma = -15.092 \pm 0.029\;\mathrm{km/s}$} 
\newcommand{\resqwilson}{$q = 0.767 \pm 0.002$} 
 
\usepackage{graphicx}
\usepackage{txfonts}
\usepackage[table,xcdraw]{xcolor}
\usepackage{hyperref}
\usepackage{multirow}
\usepackage{tablefootnote}
\usepackage{threeparttable}
\usepackage{longtable}
\usepackage{booktabs}
\usepackage{float}
\usepackage{orcidlink}
\begin{document} 
% -------------------------------------------
% -------------------------------------------
% -------------------------------------------
\title{Spectroscopic Characterization of LOFAR Radio-emitting M\,dwarfs\thanks{Based on observations obtained with the Hobby-Eberly Telescope (HET), which is a joint project of the University of Texas at Austin, the Pennsylvania State University, Ludwig-Maximillians-Universitaet Muenchen, and Georg-August Universitaet Goettingen. The HET is named in honor of its principal benefactors, William P. Hobby and Robert E. Eberly.}}

% -------------------------------------------
% -------------------------------------------
% -------------------------------------------
% list of authors 
\author{
E. Koo\orcidlink{0009-0007-0740-0954}\inst{1}\and
G. Stefansson\orcidlink{0000-0001-7409-5688}\inst{1}\and
R. D. Kavanagh\orcidlink{0000-0002-1486-7188}\inst{2,1}\and
M. Delamer\orcidlink{0000-0003-1439-2781}\inst{3,4}\and
S. Mahadevan\orcidlink{0000-0001-9596-7983}\inst{3,4,5}\and
J. R. Callingham\orcidlink{0000-0002-7167-1819}\inst{2,1}\and
H. Vedantham\orcidlink{0000-0002-0872-181X}\inst{2,6}\and
P. Robertson\orcidlink{0000-0003-0149-9678}\inst{7}\and
D. Bruijne\orcidlink{0009-0006-0218-6269}\inst{1}\and
C. F. Bender\orcidlink{0000-0003-4384-7220}\inst{8}\and
C. I. Cañas\orcidlink{0000-0003-4835-0619}\inst{9}\and
S. Diddams\orcidlink{0000-0002-2144-0764}\inst{10,11}\and
J. I. Espinoza-Retamal\orcidlink{0000-0001-9480-8526}\inst{12,13,1}\and
R. B. Fernandes\orcidlink{0000-0002-3853-7327}\inst{3,4}\thanks{President's Postdoctoral Fellow}\and
S. Halverson\orcidlink{0000-0003-1312-9391}\inst{14}\and
S. Kanodia\orcidlink{0000-0001-8401-4300}\inst{15}\and
D. Krolikowski\orcidlink{0000-0001-9626-0613}\inst{8}\and
A. S. J. Lin\orcidlink{0000-0002-9082-6337}\inst{16}\and
B. J. S. Pope\orcidlink{0000-0003-2595-9114}\inst{17}\and
A. Roy\orcidlink{0000-0001-8127-5775}\inst{18}\and
C. Schwab\orcidlink{0000-0002-4046-987X}\inst{17}\and
R. Terrien\orcidlink{0000-0002-4788-8858}\inst{19}\and
J. T. Wright\orcidlink{0000-0001-6160-5888}\inst{3,4,20}
}

% -------------------------------------------
% -------------------------------------------
% -------------------------------------------
% Affiliations here
\institute{
Anton Pannekoek Institute for Astronomy, University of Amsterdam, Science Park 904, 1098 XH Amsterdam, The Netherlands
\and
ASTRON, Netherlands Institute for Radio Astronomy, Oude Hoogeveensedijk 4, Dwingeloo 7991 PD, The Netherlands
\and
Department of Astronomy \& Astrophysics, 525 Davey Laboratory, Penn State, University Park, PA, 16802, USA
\and
Center for Exoplanets and Habitable Worlds, 525 Davey Laboratory, Penn State, University Park, PA, 16802, USA
\and
Astrobiology Research Center, 525 Davey Laboratory, Penn State, University Park, PA, 16802, USA
\and
Kapteyn Astronomical Institute, University of Groningen, PO Box 800, 9700 AV Groningen, The Netherlands
\and
Department of Physics \& Astronomy, The University of California, Irvine, Irvine, CA 92697, USA
\and
Steward Observatory, University of Arizona, 933 N.\ Cherry Ave, Tucson, AZ 85721, USA
\and
NASA Goddard Space Flight Center, Greenbelt, MD 20771, USA
\and
Electrical, Computer \& Energy Engineering, 440 UCB, University of Colorado, Boulder, CO 80309, USA
\and
Department of Physics, 390 UCB, University of Colorado, Boulder, CO 80309, USA
\and
Instituto de Astrof\'isica, Pontificia Universidad Cat\'olica de Chile, Av. Vicu\~na Mackenna 4860, 782-0436 Macul, Santiago, Chile
\and
Millennium Institute for Astrophysics, Santiago, Chile
\and
Jet Propulsion Laboratory, California Institute of Technology, 4800 Oak Grove Drive, Pasadena, California 91109
\and
Earth and Planets Laboratory, Carnegie Science, 5241 Broad Branch Road, NW, Washington, DC 20015, USA
\and
Department of Astronomy, California Institute of Technology, 1200 E California Blvd, Pasadena, CA 91125, USA
\and
School of Mathematical \& Physical Sciences, 12 Wally’s Walk, Macquarie University, NSW 2113, Australia
\and
Astrophysics \& Space Institute, Schmidt Sciences, New York, NY 10011, USA
\and
Carleton College, One North College St., Northfield, MN 55057, USA
\and
Penn State Extraterrestrial Intelligence Center, 525 Davey Laboratory, Penn State, University Park, PA, 16802, USA
}

\date{}
\abstract{Recent observations with the LOw Frequency ARray (LOFAR) have revealed 19 nearby M\,dwarfs showing bright circularly polarised radio emission. One of the possible sources of such emission is through magnetic star-planet interactions (MSPI) with unseen close-in planets. We present initial results from a spectroscopic survey with the Habitable-zone Planet Finder (HPF) and NEID spectrographs designed to characterize this sample and further investigate the origin of the radio emission. We provide four new insights into the sample. I) We uniformly characterize the stellar properties, constraining their effective temperatures ($T_{\mathrm{eff}}$), surface gravities ($\log g$), metallicities ([Fe/H]), projected rotational velocities ($v \sin i_\star$), rotation periods ($P_\text{rot}$), stellar radii ($R_\star$), and stellar inclinations ($i_\star$) where possible. Further, from a homogenous analysis of the HPF spectra, we infer their chromospheric activity and spectroscopic multiplicity states. From this, we identify GJ 625, GJ 1151, and LHS 2395 as single, quiescent stars amenable to precise RV follow-up, making them strong MSPI candidates. II) We show that the distribution of stellar inclinations are compatible with an isotropic distribution, providing no evidence for a preference to pole-on configurations. III) We refine the radial velocity solution for GJ~625~b, the only currently known close-in planet in the sample, reducing the uncertainty in its orbital period by a factor of three, to facilitate future phase-dependent radio analysis. IV) Finally, we identify GJ 3861 as a spectroscopic binary with an orbital period of $P=14.841181_{-0.00010}^{+0.00011}$ d and with a mass ratio of $q = 0.7663^{+0.0020}_{-0.0018}$, making it the only confirmed binary with a relatively short orbit in the sample, where we surmise the radio emission is likely related to magnetospheric interactions between the two stars. These results advance our understanding of radio-emitting M\,dwarfs and establish an observational foundation for identifying MSPI.}


\maketitle

% -------------------------------------------
% -------------------------------------------
% -------------------------------------------

\section{Introduction}
The interaction between a planet and its host star’s magnetosphere offers the possibility to study the magnetic environments in which planets orbit \citep{Zarka2018, Callingham2024_nov}. These `Magnetic Star-Planet Interactions' (MSPIs) also provide a unique opportunity to probe the magnetic field of the planet itself \citep{Cauley2019, Kavanagh2022}, which is known to play a key role in regulating atmospheric erosion \citep{Owen2014, Garcia-Sage2017, Presa2024}, and thereby exoplanet habitability \citep{Shields2016}.

MSPI is expected to occur analogous to the well-studied interaction between Jupiter and its moon Io \citep{Burke1955, Bigg1964}. If a planet were to orbit within the host star's Alfv\'en surface \citep[typical orbital periods of $\lesssim$10 days for M\,dwarfs;][]{kavanagh21}, electrons can be accelerated along the stellar magnetic field lines \citep{saur13, kavanagh21}. As these electrons travel along converging field lines toward the stellar surface, some enter a so-called `loss cone' and precipitate into the star’s atmosphere, while others are reflected due to magnetic mirroring. This anisotropic pitch-angle distribution ultimately generates bright, circularly-polarized radio emission via the electron cyclotron maser instability (ECMI) mechanism \citep{Dulk1985, Treumann2006}. The emission is strongly beamed along an `emission cone', making detections highly dependent on the system's geometry, including the relative orientations of the magnetic, rotational, and orbital axes \citep{Hess2011,kavanagh2023}. The energetics of the interaction depends, among other things, on the magnetic field in the vicinity of the planet \citep[e.g.,][]{saur13}. Therefore, the detection and characterization of MSPI-driven radio emission could not only provide a new way to discover exoplanets, but could also enable us to characterize their magnetic environments \citep{Callingham2024_nov}.

The expected occurrence of MSPI remains largely based on theoretical predictions. MSPI has been observed to manifest as orbital phase-modulated variations in chromospheric activity indicators or flare rate, modulated at the orbital period of the planet \citep[][]{Shkolnik2004, Ilin2025}. However, only tentative observations of MSPI-driven radio emission have been reported for a select few stellar systems, including HAT-P-11 \citep{LecavelierdesEtangs2013}, $\tau$~Boötis \citep{Turner2021}, GJ 1151 \citep{Vedantham2020}, Proxima Centauri \citep{Perez-Torres2021}, and YZ~Ceti \citep{Pineda2023, Trigilio2023}. Although tantalizing, none have decisively been confirmed to be MSPI. Definite confirmation requires detecting periodic radio emission matching either the planet’s orbital period or the synodic period between the stellar rotation and planetary orbit \citep{Callingham2024_nov}. It has been challenging to observe more candidate systems because the low radio frequencies associated with MSPI fall below the sensitivity and bandwidth of most radio facilities. This is now changing with the improved sensitivity and low-frequency range (10-240 MHz) of the LOw Frequency ARray \citep[LOFAR;][]{Haarlem2013}, the upgraded Giant Metrewave Radio Telescope \citep[uGMRT;][]{Gupta2017}, and with the advent of upcoming low-frequency radio facilities like the Square Kilometer Array (SKA) \citep{Johnston2008}.

LOFAR surveys the Northern sky as part of the LOFAR Two-Metre Sky Survey (LoTSS) \citep{Shimwell2019}. Based on the survey’s circularly polarised data, two distinct samples of stellar systems have been identified to exhibit bright low-frequency radio emission \citep{Callingham2021}. The first group comprises a sample of extremely active stars, where the emission is likely either generated from activity-induced plasma emission mechanisms \citep{Dulk1985, Callingham2021}, and/or magnetospheric interactions in close-in RS CVn-like binaries or rapidly rotating stars \citep{Toet2021,Vedantham2022}. The second group comprises 19 nearby M\,dwarfs, some of which are inactive and therefore promising MSPI candidates \citep{Callingham2021}. However, the radio emission in M\,dwarfs can generally occur through various other scenarios, such as through particle acceleration in flares or through co-rotation breakdown. Thus, the underlying radio emission mechanism in this sample remains to be further confirmed. Photometric and spectroscopic observations are essential to distinguish between these emission mechanisms, as they provide insights into stellar activity, multiplicity, and the presence of planetary companions consistent with MSPI. Photometric TESS follow-up has previously been conducted by \citet{Pope2021}.

Here we present first results from a spectroscopic survey using the Habitable-zone Planet Finder \citep[HPF;][]{Mahadevan2012, Mahadevan2014} and NEID \citep{Schwab2016} spectrographs to target the LOFAR radio-emitting M\,dwarf sample of \cite{Callingham2021}. HPF is a high resolution, fiber-fed \citep{Kanodia018}, temperature-stabilized \citep{Stefansson2016} spectrograph on the Hobby-Eberly Telescope \citep[HET][]{Ramsey1998}. NEID is an ultra-precise stabilized \citep{Stefansson2016, Robertson2019}, fiber-fed \citep{Kanodia2023} spectrograph on the WIYN 3.5 m Telescope\footnote{The WIYN Observatory is a joint facility of the University of Wisconsin–Madison, Indiana University, NSF NOIRLab, the Pennsylvania State University, and Princeton University.}. With these spectrographs, the survey aims to help determine the origin of the radio emission, with the ultimate goal of detecting, or ruling out, close-in planets consistent with the MSPI scenario. We discuss the observations and data reduction in Section \ref{sec:observations_and_data_reduction}. While constraints on potential planets will be addressed in future work, this paper characterizes the sample as a whole and identifies promising MSPI targets that could be the focus of follow-up radio observations. We investigate the following four key aspects of the sample:

(I) By obtaining HPF spectra of the full sample, we perform a homogeneous characterization of stellar spectroscopic parameters (effective temperature ($T_{\mathrm{eff}}$), surface gravity ($\log g$), metallicity ($[\mathrm{Fe}/\mathrm{H}]$), rotational velocity ($v \sin i$), rotation period ($P_{\text{rot}}$), stellar radius ($R_\star$), stellar inclination ($i_\star$)), and their activity and multiplicity states. This analysis allows us to identify inactive stars amenable to precise RV follow-up. We consider these systems to be prime MSPI candidates, as they offer the best opportunity to detect and characterize sub-Alfvénic (i.e. close-in) planets, for systems where alternative radio emission mechanisms than MSPI are unlikely. This analysis is presented in Section \ref{sec:stellar_properties}.

(II) We furthermore report the sample's stellar rotation period ($P_{\text{rot}}$) and stellar radius ($R_\star$). By combining these values with the spectroscopically characterized rotational velocity ($v \sin i$) we infer the sample's distribution of stellar inclination angles ($i_\star$). This allows us to test whether the sample shows a preference for any clustering in inclination, as the MSPI scenario suggests. For example, recently \citep{kavanagh2023} highlighted that MSPI could be preferentially biased towards pole-on stellar inclinations, a hypothesis we examine in this work.

(III) For the only known relatively close-in (P $\lesssim15$ d) planet in our sample---GJ~625~b, a super-Earth with a minimum mass of $2.82 \pm 0.51 M_\oplus$ in a 14.6 day orbit around a nearby star at 6.5~pc \citep{SurezMascareo2017}---we combine new NEID RV measurements with literature RVs to refine the system's orbital and physical parameters (Section \ref{sec:GJ_625_results}). The improved orbital parameter constraints facilitate investigations into potential orbital phase dependencies in current and future radio observations of the system. We further discuss the system as a promising MSPI candidate in Section~\ref{sec:discussion_GJ625}.

(IV) We identify GJ 3861---an M2.5-type star at a distance of 18.47 pc \citep{Callingham2021}--- as a spectroscopic binary. We present a full characterization of its orbital solution and mass ratio in Section ~\ref{sec:GJ_3861_results}.

We discuss the above results and their implications in Section \ref{sec:discussion}, and conclude with a summary of our findings in Section \ref{sec:conclusions}.

% --------------------------------------------------
% --------------------------------------------------
% --------------------------------------------------
\section{Observations and Data Reduction}
\label{sec:observations_and_data_reduction}

% --------------------------------------------------
\subsection{The Habitable-zone Planet Finder}
As part of the survey, we carried out spectroscopic observations of the 19 LOFAR radio M\,dwarf targets with the Habitable-zone Planet Finder \citep[HPF;][]{Mahadevan2012, Mahadevan2014}. HPF is a high resolution (R$\sim$ 55,000), fiber-fed \citep{Kanodia018}, temperature-stabilized \citep{Stefansson2016} spectrograph, operating in the near-infrared (NIR) from 8079–12786 $\AA$. It is located on the 10m Hobby-Eberly Telescope \citep[HET;][]{Ramsey1998,Hill2021} at the McDonald Observatory, Texas.

We acquired the observations through the HET observing queue \citep[][]{Shetrone2007}. Each visit typically consisted of two consecutive observations, with each exposure generally 969 seconds in duration (with variations up to a minute in duration). These exposure times were chosen to achieve a high signal-to-noise ratio (SNR), while optimizing the likelihood of successful scheduling within the HET queue. For the brightest targets we employ alternative observing strategies, designed for a total exposure time of $\sim$1800 s per visit, while avoiding saturation: GJ 450 was observed with six exposures of 330 seconds, GJ~625 was observed using three exposures of 650 seconds, AD Leo was observed with 12 exposures of either 160 or 170 seconds, GJ 860 B was observed with five exposures of 394 seconds, and GJ 9552 was observed with three exposures of 650 seconds. We visited each target a variety of times, obtaining a median of nine visits per target. The targets GJ~625 and GJ~9552 were visited twice, while all other targets were observed at least three times. We obtained numerous observations both before and after 2022 May, when engineering work on the instrument introduced a vacuum break, resulting in an RV offset between the `pre' and `post' observing eras. Consequently, pre- and post-break observations must be treated as originating from distinct instruments. Note that we visited GJ~625 twice, where one visit was obtained pre-break and one visit was obtained post-break, thus obtaining two visits with two different offsets. These HPF observations are therefore not taken into account into GJ~625's orbit analysis (see Section \ref{sec:GJ_625_results}).

We used bracketed Laser-Frequency Comb (LFC) observations to derive the wavelength solution following \citet{Stefansson2020_feb}, which avoids any possibility of contamination in science spectra by calibration light. The 1D spectra were extracted using the HPF pipeline which uses the \texttt{HxRGproc} code \citep[][]{Ninan2019}. Using the 1D wavelength corrected spectra, we extracted radial velocities using spectral-matching code \texttt{HPF-SERVAL} \citep{Stefansson2020_feb,Stefansson2023}, which builds on the SpEctrum Radial Velocity AnaLyser code \citep[SERVAL;][]{Zechmeister2018}. The median SNR (in HPF order index 5) and median RV precision of all HPF observations are reported in Table~\ref{tab:designations}.
 
% --------------------------------------------------
\subsection{NEID}
To better characterize the only known close-in planet in the sample---GJ 625~b---in addition to the HPF spectra, we obtained five high precision RVs with the NEID spectrograph \citep{Schwab2016}. NEID is an ultra-precise stabilized \citep{Stefansson2016, Robertson2019}, fiber-fed \citep{Kanodia2023}, optical spectrograph on the WIYN 3.5 m Telescope at Kitt Peak Observatory in Arizona. It has a resolving power of R $\sim$110,000, and covers an extended wavelength range from 380 nm to 930 nm. The red-optical wavelength coverage is ideal for observing brighter earlier-type M\,dwarf targets in our sample, resulting in a high RV precision.

We used 900 second exposures (varying up to a few seconds in duration), with the exception of one observation, which lasted 632 seconds. The median SNR obtained at NEID order index 102 (covering wavelengths from 8551 to 8715$\AA$) was 91. Due to the faintness of the target, we did not use simultaneous etalon calibration during observation. The 1D spectra were reduced via the NEID data reduction pipeline (v1.3.0). We extracted RVs using the \texttt{NEID-SERVAL} spectral matching code \citep{Stefansson2022_jun}, which builds on the original \texttt{SERVAL} code \citep{Zechmeister2018}, but is customized to the NEID spectrograph. As all of the GJ~625 observations were all obtained before the Contreras fire in Summer 2022, we homogenously extracted all of the data using a data driven template in SERVAL, built from all of the NEID observations. In doing so, we obtain a median RV precision of 0.50 m s$^{-1}$. The available GJ~625 RVs and their uncertainties are reported in Table \ref{tab:gj625rvs} (Appendix \ref{sec:rvs}).

% --------------------------------------------------
\subsection{Other Spectroscopic Observations}
For the orbital analysis of GJ~625 b, in addition to the HPF and NEID data discussed above, we also use available RV observations from the High Accuracy Radial velocity Planet Searcher for the Northern hemisphere (HARPS-N) \citep{Cosentino2012}, the Calar Alto high-Resolution search for M\,dwarfs with Exoearths with Near-infrared and optical Echelle Spectrographs (CARMENES) \citep{Quirrenbach2020}, and the High Resolution Echelle Spectrometer (HIRES) \citep{Vogt1994}. We elected to use the SERVAL code \citep{Zechmeister2018} to extract high precision radial velocities of the HARPS-N data, which better leverages the high information content in the M\,dwarf spectra than the cross-correlation technique used by the HARPS-N pipeline. The median extracted RVs of GJ~625 are improved in precision from 1.01 to 0.94 m s$^{-1}$, and we obtain in total 200 RVs. We furthermore obtain 33 CARMENES RVs directly from the CARMENES DR1 data products \citep{Ribas2023}, which include readily available SERVAL RV extractions, corrected for instrumental drift, secular acceleration, and nightly zero points (designated as `AVC' in the CARMENES DR1 data products). Finally, we obtained 53 HIRES RVs from the The California Legacy Survey \citep{rosenthal2021}, of which 16 were obtained before the HIRES CCD upgrade in 2004, and 37 were obtained after.

% --------------------------------------------------
% --------------------------------------------------
% --------------------------------------------------
\section{Stellar Properties}
\label{sec:stellar_properties}

% --------------------------------------------------
\subsection{Stellar Properties from Spectral Matching}
\label{sec:stellar_properties_general}
To homogenously derive the stellar spectroscopic parameters of the whole LOFAR radio M\,dwarf sample, we use the \texttt{HPF-SpecMatch} code \citep{Stefansson2020_feb} to constrain the stellar effective temperature (T$_{\text{eff}}$), surface gravity (log $g$), and metallicity ([Fe/H]). As described in \citep{Stefansson2020_feb}, \texttt{HPF-SpecMatch} infers stellar parameters of a target star spectrum by comparing it with a library of 166 high SNR as-observed HPF spectra for stars with well-characterized properties, ranging T$_{\text{eff}}$ values of 2700 to 5990 K, log $g$ values of 4.29 to 5.26, and [Fe/H] values of -0.49 to 0.53. The five best matching library spectra are weighted and combined to obtain the stellar parameters of the observed star. We follow \cite{Stefansson2020_feb} and adopt the values extracted from HPF order index 5 which covers 8534–8645 $\AA$ and is the order least affected by tellurics. The uncertainties in the stellar parameters are determined using a cross-validation procedure following \cite{Stefansson2020_feb} and \cite{jones2024}, focused on the M dwarfs in the library (i.e. targets with $T_\text{eff}<4500$ K). We obtain the uncertainties $\sigma_{\mathrm{Teff}} =$ 59 K, $\sigma_{\mathrm{\log g}} =$ 0.04, and $\sigma_{\mathrm{Fe/H}} =$ 0.16, where we note that constraining the metallicity of M\,dwarfs remains challenging, as reflected in the relatively large uncertainty in $\mathrm{[Fe/H]}$. An overview of the stellar properties we derive for the whole sample is shown in Table \ref{tab:general_properties}. We note that we do not list any stellar parameters for systems that we identify as spectroscopic multiples (see Section \ref{sec:bf}) as the \texttt{HPF-SpecMatch} algorithm assumes the target star is a single star in the $\chi^2$ spectral matching process. 

% --------------------------------------------------
\subsection{Spectral Types}
\label{sec:stellar_properties_spectraltypes}
We derive the spectral types of our sample using the empirical relationships established in \citet{Kiman2019}, which relate Gaia magnitudes and distances (G–G$_\text{RP}$, G$_\text{BP}$–G$_\text{RP}$, G$_\text{BP}$–G, $M_\text{G}$, $M_\text{G$_\text{RP}$}$, and $M_\text{G$_\text{BP}$}$) to spectral type. Each of the six photometric relations yields slightly different spectral type estimates, and we therefore adopt the standard deviation across these estimates as the uncertainty in the derived spectral type. The G~240-45 system lacks a reported Gaia color (likely due to it being an unresolved binary), rendering five out of the six relations unusable. For this object, we instead adopt the spectral type reported in \citet{Callingham2021}. Our adopted spectral types are listed in Table \ref{tab:general_properties}. Our spectral types are in good agreement with the spectral types adopted in \citet{Callingham2021}, except for two objects: 2MASS J09481615+5114518, and LHS 2395 are both later in spectral type than adopted in \cite{Callingham2021}, with spectral types of $\mathrm{M6.1\pm0.6}$, and $\mathrm{M7.0\pm0.7}$, instead of M4.5, and M5.5, respectively. The spectral types we adopt are in good agreement with the effective temperatures we derive from the \texttt{HPF-SpecMatch} analysis of $3063\pm59\;\mathrm{K}$ and $2869\pm59\;\mathrm{K}$, respectively.

% --------------------------------------------------
\subsection{Rotational Velocities}
\label{sec:stellar_properties_vsini}
To measure the projected rotational velocities $v \sin i_\star$ of the LOFAR radio M\,dwarfs, we used three different approaches. First, \texttt{HPF-SpecMatch} measures a $v \sin i_\star$ during the spectral-matching process. Due to the resolution of HPF ($R \sim 55,000$), we are unable to reliably measure $v \sin i_\star$ values smaller than $\sim$2~km~s$^{-1}$ \citep{Stefansson2020_feb}, which we adopt as the lower limit. However, as \texttt{HPF-SpecMatch} uses the full extent of the HPF orders, the measured rotational broadening is an average across many spectral lines, some of which can be significantly affected by magnetic Zeeman broadening in young and highly active stars \citep{Shulyak2010}. Thus only for the inactive stars, where the Zeeman broadening is expected to be negligible, do we adopt the $v \sin i_\star$ values from \texttt{HPF-SpecMatch}. 

For the highly active rapidly rotating stars, we employ a more refined Cross Correlation Function (CCF) technique. This technique follows \cite{Reiners2007}, where we focus this process on the two magnetically insensitive ($g$ = 0) FeH lines at air wavelengths 9941.9$\AA$ and 9950.400$\AA$, which are neither affected by Zeeman broadening, nor affected by any nearby telluric lines. This technique requires a comparison with an artificially broadened slowly rotating reference template star. In short, we generate a calibration curve that maps the FWHM of the CCF of the reference star for known input $v \sin i_\star$ values. To estimate the $v \sin i_\star$ of the target star, we then measure the FWHM of the CCF of the target star, and use the calibration curve to estimate the corresponding $v \sin i_\star$. Uncertainties are estimated using the derivative of the slope of the calibration curve, while following \cite{stefansson2020k225}, and setting a minimum uncertainty of the $v \sin i_\star$ estimate of $0.7$~km/s. For template stars, we used known inactive slowly rotating stars that were similar in spectral type to the target star. Here we use the same reference stars for the broadening function analysis and summarized in Figure \ref{fig:bfs} in the Appendix.

Third, for one of the targets---2MASS J14333139+3417472---the broadening was so large that both of the above methods struggled to provide an accurate accounting of the broadening, struggling to converge the fit to the lines instead of the broader continuum. For this star, we instead used the value from the broadening function (BF) analysis, which is further discussed in Section \ref{sec:bf}. 

The final values of all $v \sin i_\star$ measurements are presented in Table \ref{tab:general_properties}, with the corresponding methods denoted as follows: $\alpha$ for results from \texttt{HPF-SpecMatch}, $\beta$ for the CCF method, and $\gamma$ for values derived from the broadening function.

% --------------------------------------------------
\subsection{Stellar Inclinations}
\label{sec:stellar_properties_inclinations}
To measure the stellar inclinations, $i_\star$, we follow \cite{masuda2020} and \cite{stefansson2020k225}, and combine the $v \sin i_\star$ measurements from the spectral broadening analysis with estimates of the stellar radii, $R_\star$, and stellar rotation periods. To estimate the $R_\star$, we apply the $K_s-R_\star$ empirical relation established by \citet{Mann2015} for M\,dwarfs, which yields a typical uncertainty in $R_\star$ of $\sim$$3\%$. We queried the $K_s$-band magnitudes from the TESS input catalog \citep{Stassun2018}, and used distances from \citep{Bailer-Jones2021}. One target, GJ~860~B, did not have a listed $K_s$-magnitude, so we used the Gaia color--magnitude relation from \citet{GaiaDR3_documentation_ch5} to derive an absolute $K_s$ band magnitude of $6.87 \pm 0.04$.

Stellar rotation periods have been measured for 13 of the 19 M\,dwarfs in the sample by \cite{Pope2021}, \citet{Newton2017}, and \citet{Morin2008}. For the remaining 6 targets, we attempted to measure stellar rotation periods using publicly available data from the Zwicky Transient Facility (ZTF) \citep{Masci2019}, which conducts a 2-night cadence survey of the entire northern sky in g \& r bands, and for which the latest data-release (DR23) spans March 2018 to October 2024. Here we use the $z_g$ filter, as the $z_r$ filter lacked sufficient amount of data for a periodogram analysis. Figure \ref{fig:ztf} shows the ZTF photometry for 4 of those 6 targets (2MASS J09481615+5114518, LSPM J1024+3902E, 2MASS J10534129+5253040 and GJ 3861) along with generalized Lomb-Scargle periodograms of the data. Clear rotation signals (less than 0.1\% false alarm probability) are detected for 2MASS J09481615+5114518, LSPM J1024+3902E, and GJ 3861. However, no significant rotation period is detected for 2MASS J10534129+5253040. The remaining two targets, LHS 2395, and G 240-45 did not have sufficient ZTF data to perform a reliable periodogram analysis. The final values of $R_\star$, $P_{\text{rot}}$, and $i_\star$ are listed in Table \ref{tab:general_properties}.

% ---------------------------------------------------------
\begin{table*}[]
\centering
\caption{Stellar parameters derived for the V-LoTSS M\,dwarf sample, with targets grouped according to the designations in Table \ref{tab:designations} and then ordered according to spectral type. Rotation periods are adopted from \cite{Pope2021} unless otherwise noted. All other values are from this work unless otherwise noted. The cross-validation uncertainties for the stellar effective temperature, $\log g$, and the metallicity are $\sigma_{T_{\mathrm{eff}}} =$ 59 K, $\sigma_{\log g}$ = 0.04, and $\sigma_{\mathrm{Fe/H}} =$ 0.16. The method used to derive $v \sin i_\star$ is annotated, where $\alpha$ is a result from \texttt{HPF-SpecMatch}, $\beta$ is obtained through the CCF method and $\gamma$ is obtained from the broadening function. References are annotated with letters a-d.}
\begin{tabular}{llccccccc}
\hline
\hline
Target & Sp Type & $T_{\mathrm{eff}}$ (K) & $\log g$ & $\mathrm{[Fe/H]}$ & $v \sin i_\star$ (km/s) & $P_{\mathrm{rot}}$ (d) & $R_\star (R_\sun)$ & $i_\star$ \\ \hline
GJ~625 & M2.5 $\pm$ 0.4 & 3475 & 4.90 & -0.49 & <2$^\alpha$ & $80 \pm 8^{a}$ & $0.338\pm0.010$ &  \\
GJ 1151 & M5.2 $\pm$ 0.5 & 3140 & 5.07 & 0.01 & <2$^\alpha$ & $125 \pm 13 ^{a}$ & $0.202\pm0.006$ &  \\
LHS 2395 & M7.0 $\pm$ 0.7 & 2869 & 5.14 & -0.12 & <2$^\alpha$ & - & $0.140\pm0.004$ &  \\
\midrule
GJ 450 & M1.7 $\pm$ 0.1 & 3581 & 4.77 & -0.14 & <2$^{\alpha}$ & $23 \pm 2.3^{a}$ & $0.461\pm0.013$ &  \\
AD Leo & M2.8 $\pm$ 0.5 & 3242 & 4.89 & -0.15 & $2.7 \pm 0.7$  $^{\beta}$ & $2.2399 \pm 0.0006^{c}$ & $0.427\pm0.012$ & $18_{-4}^{+4}$ \\
GJ 3647 & M3.5 $\pm$ 0.4 & 3187 & 4.92 & -0.19 & $4.1 \pm 1.0$ $^{\beta}$ & $7.77 \pm 0.8^{a}$ & $0.326\pm0.009$ & $74_{-18}^{+11}$ \\
NLTT 20858 & M5.5 $\pm$ 0.6 & 3080 & 5.05 & -0.22 & $6.9 \pm 0.8$$^{\beta}$& $1.7 \pm 0.17^{a}$ & $0.182\pm0.005$ & $79_{-11}^{+8}$ \\
NLTT 24199 & M5.6 $\pm$ 0.6 & 3033 & 5.05 & 0.14 & <2$^{\alpha}$ & $61 \pm 6^{a}$ & $0.225\pm0.006$ &  \\
2M J143$^{1,2}$ & M5.9 $\pm$ 1.1 & 2884 $\pm$ 157 & 4.934 $\pm$ 0.007 & - & $71 \pm 7$$^{\gamma}$ & $0.14 \pm 0.014^{a}$ & $0.289\pm0.008$ & $45_{-6}^{+9}$ \\
2M J094$^{1}$ & M6.1 $\pm$ 0.6 & 3063 & 5.05 & -0.17 & $2.4 \pm 0.7$$^{\beta}$ & $31 \pm 3^{b}$ & $0.195\pm0.006$ & $67_{-24}^{+16}$ \\
\midrule
GJ 860 B & M3.7 $\pm$ 2.1 & 3289 & 5.04 & -0.41 & <2$^{\alpha}$ & $0.41 \pm 0.04^{a}$ & $0.326\pm0.009$ & $3.6_{-1.1} ^{+1.3}$ \\
LSPM J102$^{1}$ & M4.3 $\pm$ 0.4 & 3184 & 4.97 & 0.01 & $4.3 \pm 1.0$$^{\beta}$ & $2.1 \pm 0.2^{b}$ & $0.276\pm0.008$ & $45_{-10}^{+14}$ \\
GJ 412 B & M7.2 $\pm$ 0.7 & 2867 & 5.16 & 0.07 & $7.9 \pm 1.8$$^{\beta}$ & $0.78 \pm 0.08^{a}$ & $0.128\pm0.004$ & $69_{-18}^{+15}$ \\
\midrule
GJ 9552 & M0.8 $\pm$ 1.0 & - & - & - & - & $1.98 \pm 0.2^{a}$ & - & - \\
GJ 3861 & M2.3 $\pm$ 0.4 & - & - & - & - & $14.84 \pm 0.15 ^{b}$ & - & - \\
GJ 3729 & M3.2 $\pm$ 0.7 & - & - & - & - & $13.58 \pm 1.4^{d}$ & - & - \\
GJ 3789 & M3.2 $\pm$ 1.2 & - & - & - & - & $0.11 \pm 0.01^{a}$ & - & - \\
2M J105$^{1,3}$ & M3.2 $\pm$ 0.8 & 3195 & 4.90 & -0.12 & $5.5 \pm 0.8$$^{\beta}$ & - & - &  \\
G 240-45$^{3}$ & M4.0$^{e}$ & 3177 & 5.06 & -0.19 & <2$^{\alpha}$ & - & - & \\

 \hline
 \hline
\end{tabular}
\begin{tablenotes}
\footnotesize
\item $^{1}$Targets shortened for space: 2MASS~J09481615+5114518 (2M J094), 2MASS~J14333139+3417472 (2M J143), LSPM~J1024+3902E~(LSPM J102), and 2MASS~J10534129+5253040 (2M J105).
\item $^{2}$The highly broadened lines of target 2MASS~J14333139+3417472 makes it difficult for \texttt{HPF-SpecMatch} to find matching spectra, and thus produces highly uncertain results. Therefore, we here show the stellar properties listed in the Tess Input Catalog.
\item $^{3}$G 240-45 and 2MASS~J10534129+5253040 are likely binaries, but do not show obvious double lines and so we report the \texttt{HPF-SpecMatch} results here. We suggest caution in interpreting these results.
\end{tablenotes}
\tablebib{$^{a}$ \cite{Pope2021}, $^{b}$ This work, $^{c}$ \cite{Morin2008}, $^{d}$\cite{Newton2017}, $^{e}$\cite{Callingham2021}.}
\label{tab:general_properties}
\end{table*}

% --------------------------------------------------
\subsection{Multiplicity and Broadening Functions}
\label{sec:bf}
To assess if the stars are single or spectroscopic multiples, we homogenously examined spectral-line broadening functions \citep{Rucinski1992} of the HPF spectra of all of the 19 M\,dwarfs in the sample. The broadening function is the convolution kernel that, when applied to a template spectrum, best reproduces the observed target spectrum. The broadening functions thus represent the average photospheric absorption line profile in the spectrum. For a single star, the broadening function has a single peak. For a double star, there will generally be two peaks, unless the two radial velocities are too close to be resolved or if the secondary component is faint. Furthermore, the broadening functions allow us to obtain the rotational velocities $v \sin i_\star$ \citep[e.g.,][]{Tofflemire2019}, and as noted in Section \ref{sec:stellar_properties_vsini}, 2MASS
J14333139+3417472 showed very large broadening, making it difficult to measure the $v \sin i_\star$ through either the \texttt{HPF-SpecMatch} or CCF-based techniques, so we adopt the $v \sin i_\star$ value from the broadening function analysis in this case instead. 

% BFs more details
To calculate the HPF broadening functions, we followed the methodology outlined \cite{stefansson2025}, which we summarize shortly here. Following \cite{stefansson2025}, we use the \texttt{saphires}\footnote{\url{https://github.com/tofflemire/saphires}} code \citep{Tofflemire2019}, which we have adapted for HPF data. The code uses singular-value decomposition to perform the convolution inversion following \cite{Rucinski1999}. As reference template spectra we used slowly rotating inactive spectra, which were good matches to the target star from the \texttt{HPF-SpecMatch} analysis. By using observed data instead of synthetic spectra as templates, we bypassed the need for a model of instrumental broadening. We show the broadening functions for stars we identify as spectroscopic multiples in Figure \ref{fig:bfsdouble}, as well as the broadening functions for several close visual or spectroscopic binaries (i.e. both components falling within the HPF fiber). The broadening functions for the single-lined stars are shown in Figure \ref{fig:bfs} in Appendix \ref{sec:bfssingle}, along with the template spectra used.

To obtain a data-driven estimate of the RV uncertainties in the broadening functions, we calculated broadening functions for four of the HPF order indices least contaminated by tellurics: order indices 5, 6, 16 and 17, covering wavelength ranges from 866--889\,nm and 1027--1058\,nm. This allowed us to obtain independent estimates of the RVs, and we assign the RV uncertainties as the standard deviation of the four independent RV estimates from the different HPF orders.

Overall, we find that the sample contains six close visual or spectroscopic multiples (see Figure \ref{fig:bfs}), which we discuss in further detail here.

First, GJ 3789 is a known close-in binary system \citep[$\rho = 0.174\arcsec$;][]{Beuzit2004}. However, due to its rapid rotation ($P_{\text{rot}}=0.11\pm0.01$ days; see Table \ref{tab:general_properties}) and its long orbital period \citep[$P_{\text{orb}} \approx 7$ yr;][]{Beuzit2004}, the resulting Keplerian velocities are small (a few km/s), and the spectral lines are highly broadened with a $v \sin i = 51.5\pm4$km/s \citep{Delfosse1998}. Given that the $v \sin i_\star$ values are much larger than the Keplerian velocities, the HPF broadening function for GJ 3789 does not exhibit clear two distinct components (see Figure \ref{fig:bfsdouble}). 

Second, G 240-45 has a companion at a separation \citep[$\rho = 0.89 \pm 0.07\arcsec$;][]{lamman2020} too close for the HPF fiber to resolve, suggesting it should also appear as a spectroscopic binary. However, this target also does not show a clear double peak in its broadening function similar to GJ 3789, presumably because of small velocity separations due to a long-period orbit. 

Third, Figure~\ref{fig:bfsdouble} shows that GJ~3729 exhibits three distinct peaks in its broadening functions, revealing that it is a spectroscopic triple system. As we currently have only 15 HPF observations, we lack the data necessary to robustly characterize its orbital configuration, which we leave for future work.

Fourth, GJ 9552 is a known spectroscopic binary, with an estimated separation of $\rho \sim 0$\arcsec$.148$\citep{Tamazian2008}. This is clearly seen also in its broadening functions, which show two distinct, nearly equal-strength peaks. Previous estimates of the system’s orbital period vary, but generally suggest it is on the order of several years \citep{Tamazian2008, Shkolnik2010, Sperauskas2019}. Note, however, that the primary peak may itself consist of two components, raising the possibility that GJ 9552 is a hierarchical triple system. We acquired only two HPF observations of this target, and thus more observations are needed to confirm the existence of such a third component, as well as obtaining an accurate full orbital solution.

Fifth, for the GJ 3861 system, the distinct flux peaks between binary components allowed us to arrive at a unique orbital period and constrain the full binary orbital solution across 10 HPF observations. We present this analysis in further detail in Section~\ref{sec:GJ_3861_results}.

Sixth, we note that 2MASS J10534129+5253040 currently does not have a known stellar companion in the literature. However, we note that Gaia DR3 reports a high Gaia RUWE value of 2.4, while targets with RUWE values >1.25 are considered to likely not be a single star \citep{Penoyre2022}. The RUWE value thus suggests the presence of a binary companion with a detectable astrometric orbit over the 34-month Gaia DR3 observing baseline \citep{Gaia2020_DR3}. It is also over-luminous in the color-magnitude diagram (see Figure \ref{fig:HR_GB}A) compared to the main sequence, further pointing to a likely binary companion. At the system distance of 45~pc, companions orbiting at $<$45 AU are well within the HPF fiber diameter of $\sim1.7\arcsec$, which is very likely the case for this system. Despite this, the HPF broadening functions shown in Figure~\ref{sec:bfssingle} do not exhibit clear evidence of double peaks. We suspect the lack of a detection of double lines in the HPF broadening functions could be due to i) low Keplerian velocities due to a long orbital period, or ii) a close-to face-on orbit. Nevertheless, for the reasons given above, we assign it as a  `close visual or spectroscopic binary' in Table \ref{tab:designations}.

We furthermore note that we cannot definitively determine whether the broad line profile of the target 2MASS J14333139+3417472 arises from the blending of lines from two rapidly rotating M\,dwarfs, rather than from a single object. However, given that there is no known close companion and the Gaia RUWE value is not significantly elevated (1.08), we find no compelling evidence to suggest that this is a binary system. We therefore assume the target to be a single star.

Finally, we note that some of the targets in the sample are known wide-separated binaries: GJ 412 B (WX UMa) \citep{Gould2004}, GJ 860 B \citep{Hartkopf1997}, and LSPM J1024+3902E \citep[with companion LSPM J1024+3902W;][]{ElBadry2021}. Overall, the sample altogether consists of 53\% single stars, 16\% well-separated binaries, and 31\% HPF spectroscopic multiples (see also Table \ref{tab:designations} and Fig. \ref{fig:HR_GB}).

%----------------------------------------------------
\subsection{Chromospheric Activity Analysis}
\label{sec:stellar_properties_cairt}
The targets' chromospheric activity has previously been probed through the $L_{\mathrm{H}\alpha}/L_{\mathrm{bol}}$ activity indicator \citep{Callingham2021}, and the stellar flare rates seen by TESS \citep{Pope2021}, and X-ray emission. These indicators help identify whether a star’s radio emission is more likely driven by flares--which is expected in chromospherically active stars--or via MSPI, which is more expected for quiescent stars. However, these activity indicators conflict for the target GJ~450, with the $L_{\mathrm{H}\alpha}/L_{\mathrm{bol}}<0.09\times10^{-4}$ indicator favoring a quiescent state \citep{Newton2017}, whereas the high TESS flare rates indicates an active state. Additionally, the target LSPM J1024+3902E neither has a published $L_{\mathrm{H}\alpha}/L_{\mathrm{bol}}$ value, nor a TESS flare rate.

The HPF near-infrared spectra provide access to another activity indicator, namely the Ca II infrared triplet (Ca II IRT) \citep{Robertson2016,stefansson2020k225}. Analogous to the Ca II H \& K doublet in the optical, the Ca II IRT line exhibits excess emission in active chromospheres relative to quiescent stars of similar $T_{\text{eff}}$, $\log g$, and [Fe/H] \citep{Martin2017}. As a result, the Ca II IRT has become a commonly used activity diagnostic across NIR spectroscopic programs, where traditional optical indicators such as H$_\alpha$ and the Ca II H \& K lines are inaccessible \citep[e.g.,][]{Schofer2019, Lanzafame2023}.

To assess the level of chromospheric activity in our targets, we examined the three Ca II IRT triples lines across the HPF observations of single-lined systems. We elected to exclude the spectroscopic multiples that show double peaks from this analysis, due to ambiguities in determining if the spectra contain emission from multiple stars. To identify `high' or `low' Ca II IRT emission in a model-independent manner, we follow past work \citep[e.g.,][]{kanodia2022} employing a reference star subtraction technique. This method involves subtracting the spectrum of an inactive reference star from the target star's spectrum, where the reference star is matched in spectral type (or effective temperature) and artificially broadened to mimic the target's rotational broadening. Because the reference star exhibits no Ca II IRT emission, any residual peak corresponds to intrinsic emission from the target's chromosphere. We then calculate the equivalent width (EW) of the Ca II IRT feature and its associated uncertainty. For the calculation of the 8664$\AA$ Ca II IRT line, we follow \cite{stefansson2020k225} and define the EW region as the region as the vacuum wavelengths from 8664.086–8664.953$\AA$ (highlighted in red in Figure \ref{fig:cairt}). For the EW calculation, we follow \cite{stefansson2020k225} and use the region from 8670.300–8673.190$\AA$ as a normalization region, so the median residual of that region is unity. Hereby we do not perform any telluric correction, as these wavelength ranges are not affected by telluric contamination.

The results of the Ca II IRT line analysis at 8664$\AA$ is shown in Figure \ref{fig:cairt}, which also lists the reference stars used. The results for the two other Ca II IRT triplet lines are the same, so we do not to show them. The figure compares the 95\% highest SNR target spectra in blue, compared to the slowly rotating inactive reference star templates in black, where the template is artificially broadened to match the target $v \sin i_\star$ values. To determine whether the targets show significant evidence of excess emission, we for each target compute the median signal-to-noise ratio of the EW of the residual emission ($\mathrm{SNR_{Ca\ II\ IRT, EW}} = \text{EW}/\sigma_\text{EW}$). We classify targets with $\text{EW}/\sigma_\text{EW} < 10$ as showing `low emission', and those with $\text{EW}/\sigma_\text{EW} > 10$ as showing `high emission', where the dividing line was derived by visually inspecting which stars showed clear emission in the Ca II IRT residuals. We further note that two targets yield ambiguous results. For GJ~412~B, the reference spectrum does not provide a satisfactory match to the target, which is why we do not attempt to interpret the $\text{EW}/\sigma_\text{EW}$ value in this case. We suspect that this is due to the fact that GJ 412 B (also known as WX UMa) is known to be highly magnetically active \citep{Morin2008} and shows clear Zeeman broadening effects. We note that this does not impact the overall assessment of the activity of the star, listing it as `Active' in Table~\ref{tab:designations}. Additionally, 2MASS~J14333139+3417472 shows significant broadening, making it difficult to detect excess emission with confidence.

Overall, we see that stars previously labeled as active---based on TESS flare rates \citep{Pope2021} and $L_{H_\alpha}/L_{\text{bol}}$ values \citep{Callingham2021}---also tend to show Ca IRT emission in the HPF data. The notable exception is GJ~860~B, which lacks detectable Ca IRT emission despite exhibiting significant flare activity (0.27 $\pm$ 0.07 d$^{-1}$), X-ray flux (see Figure \ref{fig:HR_GB}), and $L_{\mathrm{H}\alpha}/L_{\mathrm{bol}}$ values ($L_{\mathrm{H}\alpha}/L_{\mathrm{bol}} = 0.64\times10^{-4}$). Conversely, GJ~450 clearly shows Ca IRT emission, reinforcing its classification as active in agreement with TESS data, though contradicting the low $L_{\mathrm{H}\alpha}/L_{\mathrm{bol}}<0.09\times10^{-4}$ value. The Ca IRT analysis also newly identifies LSPMJ1024+3902E as chromospherically active, in alignment with its rapid rotation of $2.1 \pm 0.2$ days as highlighted in Table \ref{tab:designations}.

Note that GJ 1151 has shown varying levels of activity across TESS sectors, as indicated by an increase in flare rate as determined more recently by \cite{fitzmaurice2024}. This suggests that we could be observing a system with a strong activity cycle, consistent with an observed polarity reversal, which occurred between 2019 and 2022 \citep{lehmann24}. However, as seen in figure \ref{fig:cairt}, our Ca IRT observations do not show any significant variations in activity.

Based on all available activity indicators, we provide updated activity designation of each target in Table \ref{tab:designations}, expanding from \citet{Callingham2021} to take into account also the target flare rates, rotation periods and Ca IRT in addition to the $L_{\mathrm{H}\alpha}/L_{\mathrm{bol}}$. We divide the sample into four stellar groups: `Inactive single', `Active single', `Active, wide-separated binary', and `Close visual or spectroscopic multiple'. For the activity states, we adopt a conservative classification scheme: a target is labeled as `active' if any available activity indicator shows signs of activity. Based on this definition, the most quiescent targets in the sample are GJ 625, GJ 1151, and LHS~2395.

% --------------------------------------
\subsection{Promising MSPI candidates in the sample}
\begin{figure*}
\centering
\includegraphics[width=0.9\linewidth]{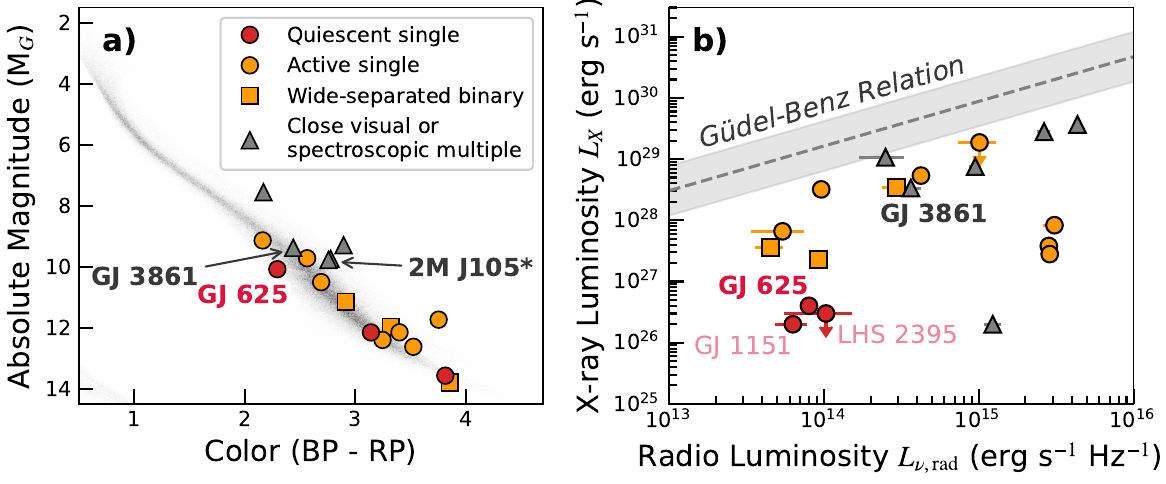}
\vspace{-0.4cm}
\caption{The 19 radio M\,dwarfs studied in this work, along with the designations adopted from this work (Table \ref{tab:designations}). Targets are designated according to activity and multiplicity, and overplotted on \textit{a) }a color-magnitude diagram, where we use the Gaia Catalogue of Nearby Stars \citep{Gaia20201} to indicate the main sequence in grey, and \textit{b)} the empirical G\"{u}del-Benz relation \citep[as described in][]{Williams2014} with the 1$\sigma$ spread indicated in the surrounding gray shaded region. The target G~240-45 is hereby not shown on the color-magnitude diagram, as its color is not listed in Gaia. The names of the suggested best MSPI candidates are all listed, and the systems GJ~625 and GJ~3861---which are further discussed in Sections \ref{sec:GJ_625_results} and \ref{sec:GJ_3861_results}---are highlighted in bold.}
\label{fig:HR_GB}
\end{figure*}

We categorize our targets based on their activity levels and multiplicity, as outlined in Table \ref{tab:designations} (see also Sections \ref{sec:bf} and \ref{sec:stellar_properties_cairt}). Using these classifications, we plot the targets on a color-magnitude diagram and in relation to the G\"{u}del-Benz relation (Figure \ref{fig:HR_GB}). The G\"{u}del-Benz relation describes an empirical correlation between X-ray and radio luminosities observed in many active stars, typically associated with magnetic reconnection events \citep{Benz1994}.

Targets most consistent with the MSPI emission scenario are expected to be under-luminous in X-rays relative to the G\"{u}del-Benz relation, exhibit low chromospheric activity, and possess long rotation periods ($P_\text{rot} > 2$ d) \citep{Callingham2021}. Based on these criteria GJ~625, GJ~1151, and LHS~2395 stand out as slowly rotating and inactive stars, and thus as particularly promising MSPI candidates where the low-frequency radio detections are less likely to be due to particle acceleration in flares or through co-rotation breakdown. These targets furthermore show a high median RV precision (see Table \ref{tab:designations}), making them amenable to finding planets with the RV technique.

In the LOFAR radio M\,dwarf sample, only two systems have known planetary companions: GJ~1151~c, and GJ~625~b. GJ~1151 was the first LOFAR uncovered radio target \citep{Vedantham2020} and has been extensively observed with RVs by \cite{Pope2021}, \cite{mahadevan2021}, \cite{perger2021} and \cite{Blanco-Pozo2023}, showing that it hosts a likely sub-Neptune (minimum mass of $m \sin i = 10.62^{+1.31}_{-1.47} M_\oplus$) on a relatively distant orbit with an orbital period $P = $389.7$^{+5.4}_{-6.5}$ d \citep{Blanco-Pozo2023}. The planet’s wide orbit likely places it beyond the Alfvénic surface---a requirement for MSPI to occur \citep{kavanagh21}. The available RV observations place stringent upper limits on the presence of close-in planets, ruling out the presence of $\gtrsim 1 M_\oplus$ planets in close-in 1-10~day orbits. As such, it remains unclear what could have caused the LOFAR detected radio emission in the GJ 1151 system.

As highlighted in \cite{Blanco-Pozo2023}, the system also shows signs of a far-out, massive companion---possibly a massive planetary companion or a brown-dwarf. Brown dwarfs are known to emit at low frequency \citep{Vedantham2020_nov} radio wavelengths, so one possible scenario is that if a brown dwarf is present in the system that that could be the source of the radio emission. Recent high contrast imaging observations with JWST/NIRCAM in Cycle 3 observations have been conducted to further characterize the nature of this companion \citep{Stefansson2024_jwst}, which will be discussed in future work. Alternatively, if MSPI is operating, a close in small planet below the detection threshold of the RV measurements could also produce the observed radio signature. 

In contrast, GJ~625 hosts a close-in planet with an orbital period of $P_\text{orb} = 14.6$ days \citep{SurezMascareo2017}, making it a plausible candidate for MSPI. In the following section, we present an updated orbital fit for this planet, and in Section \ref{sec:discussion_GJ625}, we further explore the potential for MSPI in this system.

\begin{table*}[]
\centering
\caption{The median SNR (in the 5th order index) and median radial velocity error ($\tilde{\sigma}_\text{HPF}$) of the HPF observations. Furthermore, we report the stellar rotation periods (see references in Table \ref{tab:general_properties}), TESS flare rates \citep{Pope2021}, $L_{H_\alpha}$/$L_{bol}$ values \citep{Callingham2021}, whether the Ca II infrared triplet (Ca IRT) is observed to be in high emission (this work), and their GAIA DR3 RUWE values. We categorize the targets as active/inactive and single/multiple.}
\begin{tabular}{llcllllll}
\hline \hline
Target & SNR & \multicolumn{1}{l}{\begin{tabular}[c]{@{}l@{}}$\widetilde{\sigma}_\text{HPF}$\\ (m s$^{-1}$)\end{tabular}} & $P_{\mathrm{rot}}$ (d) & \begin{tabular}[c]{@{}l@{}}Flare Rate \\ (d $^{-1}$)\end{tabular} & \begin{tabular}[c]{@{}l@{}}$L_{H_\alpha}$/$L_{bol}$ \\ ($\times10^{-4}$)\end{tabular} & \begin{tabular}[c]{@{}l@{}}Ca IRT\\ Emission\end{tabular} & RUWE &Designation \\ \hline \hline
GJ~625 & 126 & 3.3 & 80 ± 8 & 0.015(\textless{}0.036) & 0.06 & Low & 1.0 & \multirow{3}{*}{Inactive single} \\
GJ 1151 & 116 & 2.6 & 125 ± 13 & \textless{}0.0059 & 0.07 & Low & 1.0 &  \\
LHS 2395 & 44 & 4.1 & - & \textless{}0.024 & - & Low & 1.4 &  \\ \hline
GJ 450 & 186 & 1.9 & 23 ± 2.3 & 0.20(\textless{}0.29) & <0.09 & High &1.0 & \multirow{7}{*}{Active single} \\
AD Leo & 217 & 1.3 & 2.2399 ± 0.0006 & - & 1.72 & High & 1.2 &  \\
GJ 3647 & 116 & 5.4 & 7.77 ± 0.8 & 1.1 ± 0.2 & 1.82 & High & 1.1 &  \\
NLTT 20858 & 15 & 32 & 1.7 ± 0.17 & 0.29 ± 0.1 & - & High & 1.2 &  \\
NLTT 24199 & 48 & 5.5 & 61 ± 6 & 0.35 ± 0.1 & 0.84 & High & 1.4 &  \\
2M J143$^{1}$ & 20 & 240 & 0.14 ± 0.014 & 0.38 ± 0.1 & 5.95 & - & 1.1 &  \\
2M J094$^{1}$ & 18 & 10 & 31 ± 3 & 0.063(\textless{}0.14) & - & High & 1.2 &  \\ \hline
GJ 860 B & 302 & 1.0 & 0.41 ± 0.04 & 0.27 ± 0.07 & 0.64 & Low & - &\multirow{3}{*}{\begin{tabular}[c]{@{}l@{}}Active, \\ wide-separated binary\end{tabular}} \\
LSPM J102$^{1}$ & 64 & 7.6 & 2.1 ± 0.2 & - & - & High & 1.2 &  \\
GJ 412 B & 81 & 5.2 & 0.78 ± 0.08 & 0.23 ± 0.1 & 1.21 & - & 1.0 & \\ \hline
GJ 9552 & 311 & - & 1.98 ± 0.2 & 1.7 ± 0.1 & 1.38 & - & 39 &\multirow{6}{*}{\begin{tabular}[c]{@{}l@{}}Close visual or \\spectroscopic multiple\end{tabular}} \\
GJ 3861 & 160 & 3.6$^{2}$ & 14.84 ± 0.15 & 0.42 ± 0.09 & 0.67 & - & 1.7 &  \\
GJ 3729 & 105 & - & 13.58 ± 1.4 & - & 1.64 & - & 3.3 &  \\
GJ 3789 & 169 & - & 0.11 ± 0.01 & 0.75 ± 0.2 & 3.03 & - & - & \\
2M J105$^{1,3}$ & 95 & 9.5 & - & - & 3.74 & High & 2.4 &  \\
G 240-45 & 45 & 7.8 & - & 0.0069 (\textless{}0.015) & - & Low & 1.3 & \\ \hline
\hline
\end{tabular}
\begin{tablenotes}
\footnotesize
\item $^{1}$Targets shortened for space: 2MASS~J09481615+5114518 (2M J094), 2MASS~J14333139+3417472 (2M J143), LSPM~J1024+3902E~(LSPM J102), and 2MASS~J10534129+5253040 (2M J105).
\item $^2$During the orbit analysis of GJ~3861 we obtain RV uncertainties of the primary and secondary component. Here, we report the median uncertainty obtained for the primary component. Note that this uncertainty is not the RV uncertainty obtained through spectral-matching with SERVAL, as are the other values in this column.
\item $^3$The target 2MASS J10534129+5253040 is not listed in the Washington Double Star Catalog, but is listed in GAIA DR3 with a high ruwe value of 2.4 and it is overluminous compared to the main sequence. Although no double line feature was seen in the HPF spectroscopic data, we therefore still include it as a close visual multiple.

\end{tablenotes}
\label{tab:designations}
\end{table*}

%--------------------------------------------------
\begin{figure*}
\centering
\includegraphics[width=0.9\linewidth]{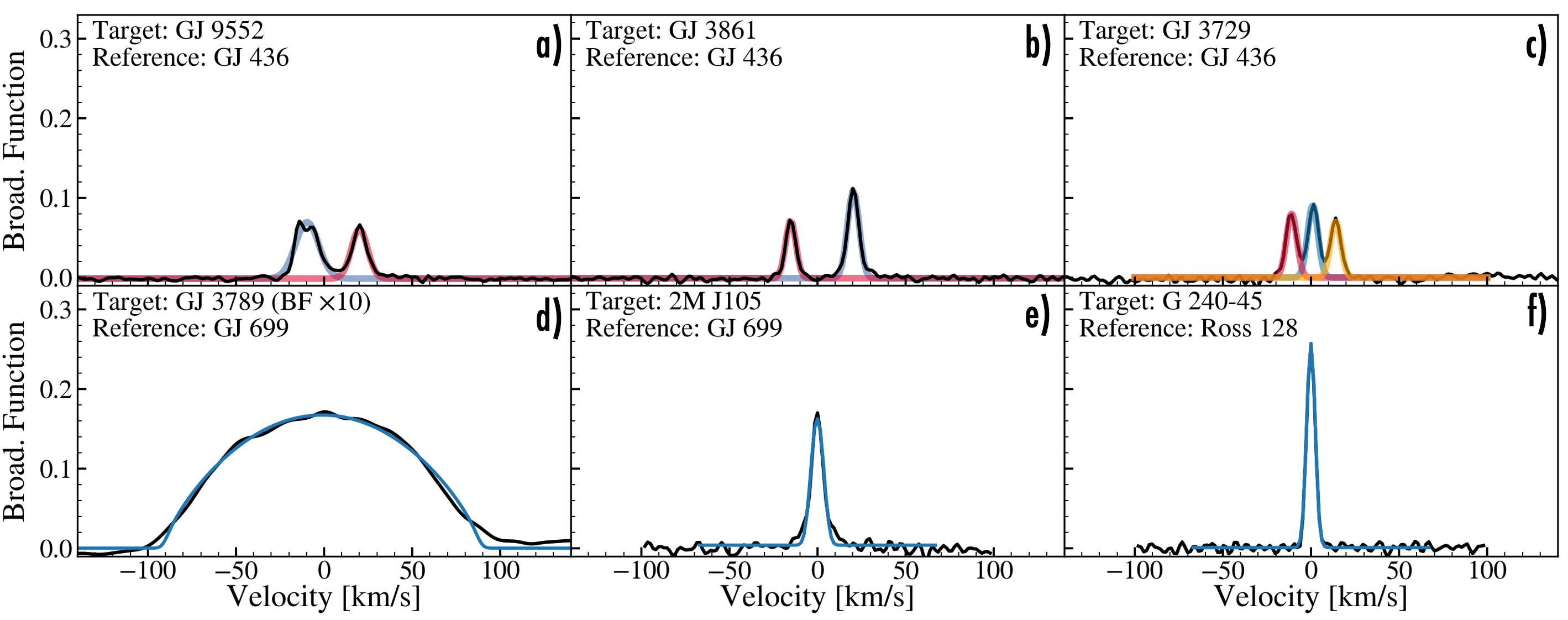}
\vspace{-0.3cm}
\caption{The broadening functions of the targets that showed significant evidence being spectroscopic multiples, i.e., showed evidence of two lines (GJ 3861, GJ 9552), or three lines (GJ 3729). We additionally show GJ 3789, and G 240-45 which do not show a double line profile in the HPF broadening functions, but are known binary systems with close ($<$1.7\arcsec) companions \citep{Beuzit2004, ElBadry2021}. Finally, we also show the broadening function of 2MASS J10534129+5253040, which has a high GAIA DR3 RUWE value of 2.4, which we thus assume to be a close visual binary. The blue curves show a best-fit Gaussian to the primary (i.e. brightest) peak, the red curve shows the secondary peak (if any), and the yellow peak highlights a fit to a tertiary peak (if any). The GJ 3789 broadening function is multiplied by a factor of 10 in the y-axis for enhanced readability.}
\label{fig:bfsdouble}
\end{figure*}

% % --------------------------------------------------

% --------------------------------------------------
% --------------------------------------------------
% --------------------------------------------------
\section{Refining the planet parameters of GJ~625 b}

% -----------------------------------------------------------
\label{sec:GJ_625_results}
GJ~625 is an M2.5V-type star (see Table \ref{tab:general_properties}) located at a distance of 6.5\,pc from Earth \citep{Floriano2015}. As outlined above, it is a promising candidate for MSPI: it is under-luminous in X-rays relative to the G\"{u}del-Benz relation, its activity indicators suggest minimal chromospheric activity, and it is slowly rotating. The radio emission is therefore less likely to be caused by particle acceleration in flares or through co-rotation breakdown, and it is amenable to precise RV observations. This target furthermore stands out in the LOFAR M\,dwarf radio sample, as it is the only target with a confirmed planet in a close-in orbit. The close-in orbit is key for driving MSPI, as such interactions only occur when the planet resides within the Alfv\'en surface.

GJ~625's planetary companion was originally detected by \cite{SurezMascareo2017} using 151 HARPS-N observations over the span of 3.5 years, revealing a likely super-Earth with a minimum mass of $M_p \sin i = 2.82 \pm 0.51 M_\oplus$ at the inner edge of the habitable zone with an orbital period of $P_{\text{orb}} = 14.628_{-0.013}^{+0.012}$ days. To refine the planet’s orbital parameters, we obtained additional radial velocity (RV) measurements using the NEID spectrograph and incorporated both the original 151 HARPS-N RVs from \citet{SurezMascareo2017} and 49 additional HARPS-N data retrieved from the TNG archive\footnote{\url{http://archives.ia2.inaf.it/tng/}}. We also include publicly available RV measurements from the CARMENES \citep{Ribas2023} and HIRES \citep{rosenthal2021} spectrographs, which respectively adds 33 and 53 new RV points. In total, this yields 140 new RV data points that have not been previously analysed. Details of the RV extraction process are provided in Section \ref{sec:observations_and_data_reduction}.

\begin{figure}
\centering
\includegraphics[width=\linewidth]{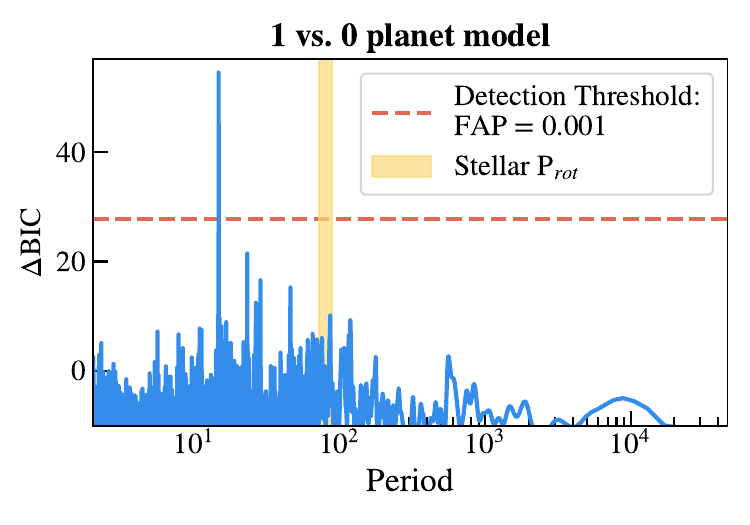}
\vspace{-0.75cm}
\caption{$\Delta$BIC periodograms of GJ~625, for the 1-planet vs. 0-planet model The orange horizontal line shows our adopted False Alarm Probability (FAP) detection threshold of 0.1\%, showing that the 14.6\,d signal is significantly detected at $\Delta$BIC $\sim55$. The stellar rotation period ($P_{\text{rot}} = 80\pm8$ d) is indicated in yellow.}
\label{fig:gj625_per}
\end{figure}

To search for and identify significant periodic signals in the expanded dataset, we used the \texttt{RVSearch} package \citep{rosenthal2021}, which can generate periodograms accounting for eccentric Keplerian orbits and zero-point offsets between various datasets. The resulting periodograms of the data are shown in Figure \ref{fig:gj625_per}, where the Bayesian Information Criterion (BIC) is used for model comparison. The BIC is calculated from the maximum likelihood ($\hat{L}$) of the model as $\mathrm{BIC} = k \ln(n) - 2 \ln(\hat{L})$, where $k$ is the number of free parameters and $n$ is the number of data points. We adopt a detection threshold according to an empirical False Alarm Probability (FAP) threshold of 0.1\%. Figure~\ref{fig:gj625_per} shows a comparison between a one-planet and a zero-planet model. A significant signal with $\Delta\mathrm{BIC} \sim 55$ is clearly seen at 14.6 days, which does not correspond to the stellar rotation period reported in Table \ref{tab:general_properties}. A comparison between the one-planet and two-planet models revealed no sign of other significant signals.

To refine the orbital parameters of the 14.6-day signal, we performed a Keplerian fit to the radial velocity (RV) data using the \texttt{radvel} package, an MCMC-based RV orbit fitting code \citep{Fulton2018}. Initial parameter estimates were taken from the results of \texttt{RVsearch}. For the \texttt{radvel} MCMC analysis we used 50 walkers, running for a total of 880,000 steps. We assessed convergence through visual inspection of the chains, where the first 200,000 steps were discarded as burn-in. The chains were deemed well-mixed, supported by a Gelman–Rubin statistic within 1\% of unity, and $>$40 independent samples.

\begin{figure*}
\centering
\includegraphics[width=\linewidth]{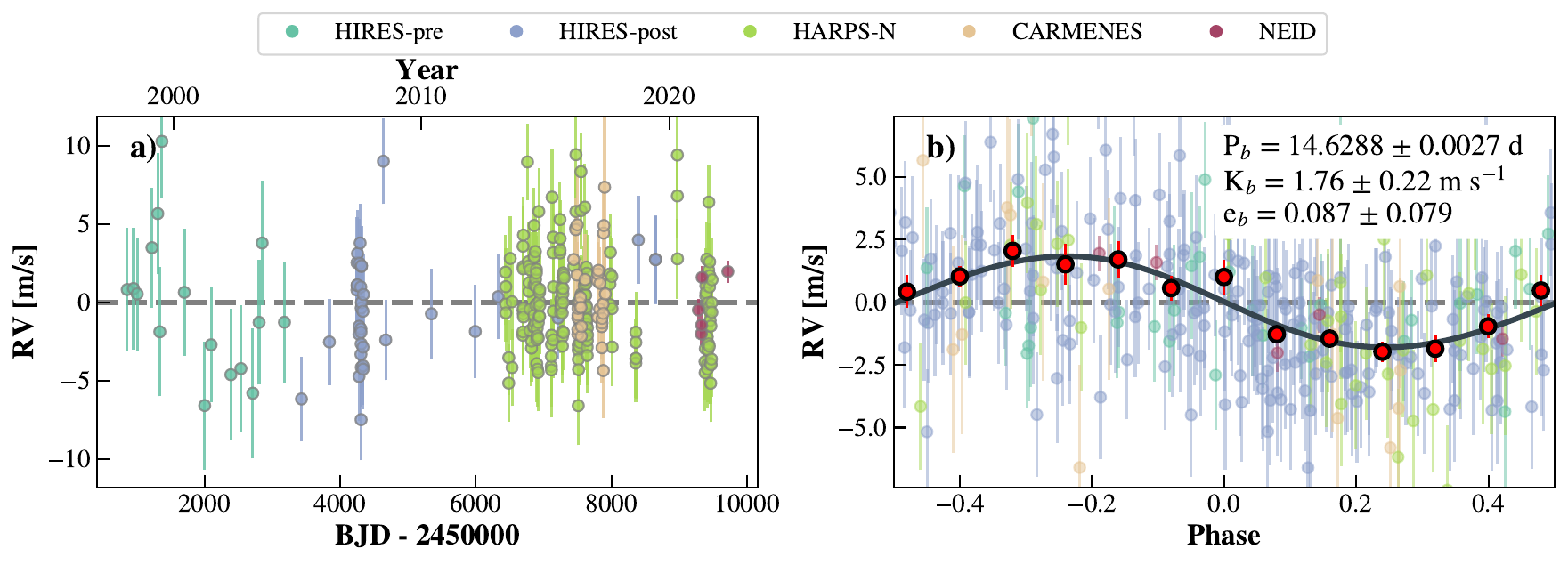}
\vspace{-0.5cm}
\caption{RV observations of GJ 625. \textit{A)} RV time series of all available data of GJ 625, color-coded by spectrograph. \textit{B)} Phase folded RV data of GJ 625 phase-folded on the best-fit period of our model of 14.6288days. The color coding is the same as in the plot on the left. Binned RV points are overplotted in red. The best-fit Keplerian model is shown in black.}
\label{fig:gj625_rvsearch_orbit}
\end{figure*}

We show the orbital fit for the one-planet model in Figure~\ref{fig:gj625_rvsearch_orbit}, with the updated planetary parameters listed in Table~\ref{tab:gj625posteriors}. While the revised parameters remain consistent with those reported by \citet{SurezMascareo2017}, the uncertainty in the orbital period has been reduced by a factor of three---from $P_{\text{orb}} = 14.628_{-0.013}^{+0.012}$ to $P_{\text{orb}} = 14.6288 \pm 0.0027$. This improvement in precision is particularly valuable for enabling future phase-resolved radio observations of upcoming radio data (see also Section~\ref{sec:discussion_GJ625}).

% --------------------------------------------------------

% --------------------------------------------------------

\begin{table}[]
\centering
\caption{Best-fit posterior parameters for GJ~625~b.}
\begin{tabular}{llc}
\hline\hline
Parameter & Description & Posterior Value \\ \hline
$P$ & Orbital period (days) & 14.6288 $\pm$ 0.0027 \\
$t_p$ & Time of periastron (BJD) & 2457303.1 $\pm$ 4.5 \\
$t_c$ & Time of conjunction (BJD) & 2457302.9 $\pm$ 0.42 \\
$e$ & Eccentricity & 0.087 $\pm$ 0.079 \\
$K$ & RV semi-amplitude (m s$^{-1}$) & 1.76 $\pm$ 0.22 \\
$\omega$ & \begin{tabular}[c]{@{}l@{}}Longitude of \\ periastron (rad)\end{tabular} & 0.5 $\pm$ 2.1 \\
%\\
$\gamma_{\mathrm{HIRES\text{-}post}}$ & Absolute RV offset (m s$^{-1}$) & 0.32 $\pm$ 0.48 \\
$\sigma_{\mathrm{HIRES\text{-}post}}$ & Jitter (m s$^{-1}$) & 2.37 $\pm$ 0.41 \\
%\\
$\gamma_{\mathrm{HIRES\text{-}pre}}$ & Absolute RV offset (m s$^{-1}$) & $-0.003 \pm 0.96$ \\
$\sigma_{\mathrm{HIRES\text{-}pre}}$ & Jitter (m s$^{-1}$) & 4.00 $\pm$ 1.00 \\
%\\
$\gamma_{\mathrm{HARPS\text{-}N}}$ & Absolute RV offset (m s$^{-1}$) & 0.78 $\pm$ 0.18 \\
$\sigma_{\mathrm{HARPS\text{-}N}}$ & Jitter (m s$^{-1}$) & 2.30 $\pm$ 0.14 \\
%\\
$\gamma_{\mathrm{CARMENES}}$ & Absolute RV offset (m s$^{-1}$) & $-0.10 \pm 0.42$ \\
$\sigma_{\mathrm{CARMENES}}$ & Jitter (m s$^{-1}$) & 1.77 $\pm$ 0.42 \\
%\\
$\gamma_{\mathrm{NEID}}$ & Absolute RV offset (m s$^{-1}$) & 0.70 $\pm$ 0.54 \\
$\sigma_{\mathrm{NEID}}$ & Jitter (m s$^{-1}$) & 0.98 $\pm$ 0.66 \\
\hline
\end{tabular}
\label{tab:gj625posteriors}
\end{table}

% ------------------------------------------------------
% ------------------------------------------------------
% ------------------------------------------------------
\section{Spectroscopic multiple: Characterization of GJ 3861}
\label{sec:GJ_3861_results}
% ------------------------------------------------------
As highlighted in the broadening function of GJ 3861 (Figure \ref{fig:bfsdouble}), the HPF spectra reveal a clear set of two lines. This motivated us to obtain spectra to characterize the system in further detail. In total, we obtained 10 observations, with a median S/N of 255 at 1$\mu$m. We extracted RVs for each BF for both the primary and secondary peaks, and as the primary and secondary peaks have clearly distinct flux ratios and heights, there is no confusion on which is the primary or the secondary peak in any of our observations. The resulting RVs for the primary and secondary peaks are listed in Table \ref{tab:gj3861rvs} in the Appendix.

\begin{figure}
\centering
\includegraphics[width=0.98\columnwidth]{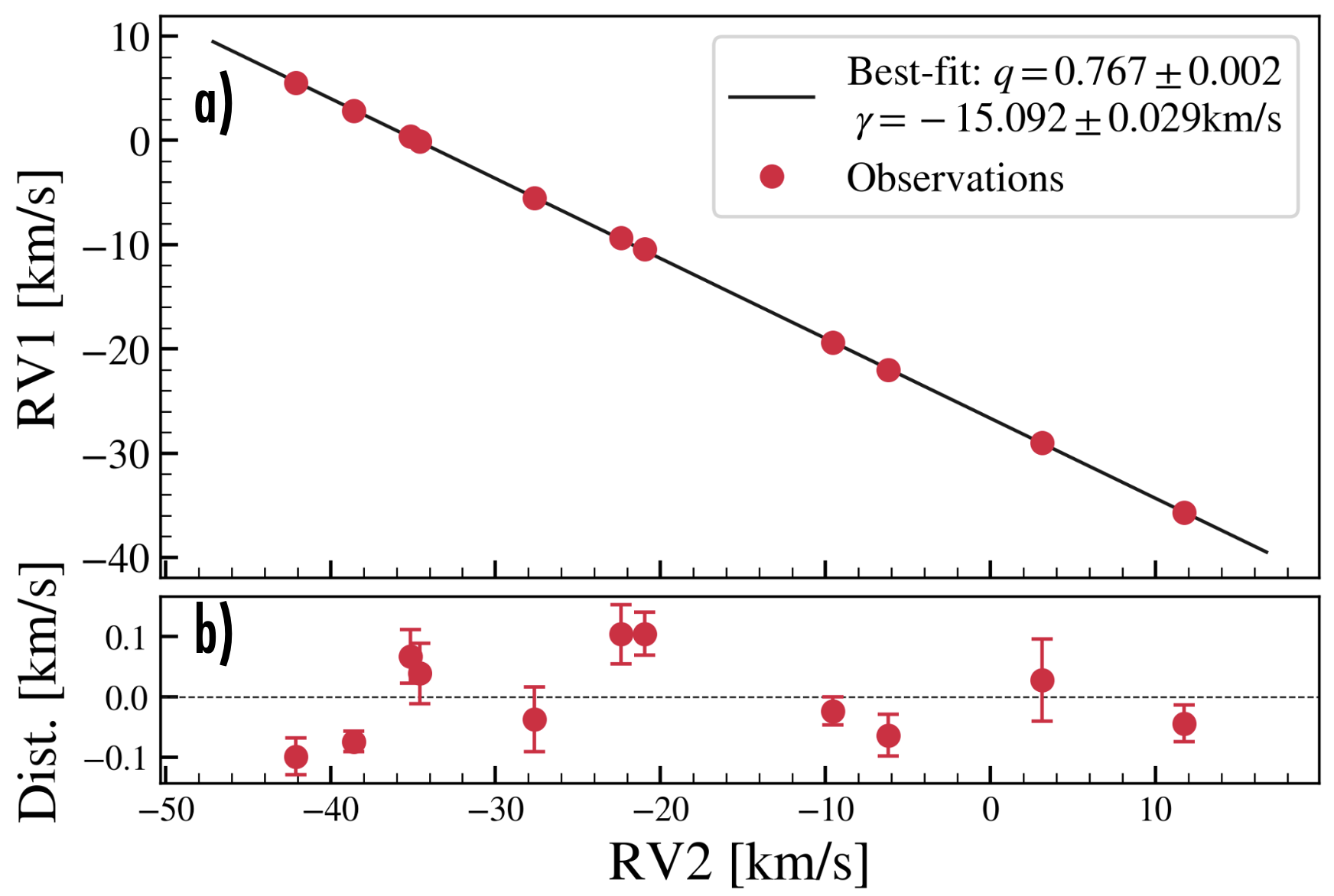}
\vspace{-0.2cm}
\caption{a) Wilson plot of GJ 3861's RVs, showing the primary RV ($RV_1$) as a function of the secondary RV ($RV_2$). The best-fit line (red line) suggests a mass ratio of \resqwilson, and a velocity offset of \resgammawilson. b) Perpendicular distance residuals to the best-fit line shown in a).}
\label{fig:gj3861wilson}
\end{figure}

% Mass ratio from Wilson plot
We used the method presented in \cite{wilson1941} to measure the mass ratio ($q=M_2/M_1$) of GJ 3861 via orthogonal distance regression of the primary velocity as a function of the secondary velocity. We implemented the orthogonal distance regression using the \texttt{scipy.odr} package. Figure \ref{fig:gj3861wilson} shows the `Wilson plot' of the GJ 3861 system, yielding a mass ratio of \resqwilson, and a velocity offset of \resgammawilson. 

\begin{figure*}
\centering
\includegraphics[width=0.9\textwidth]{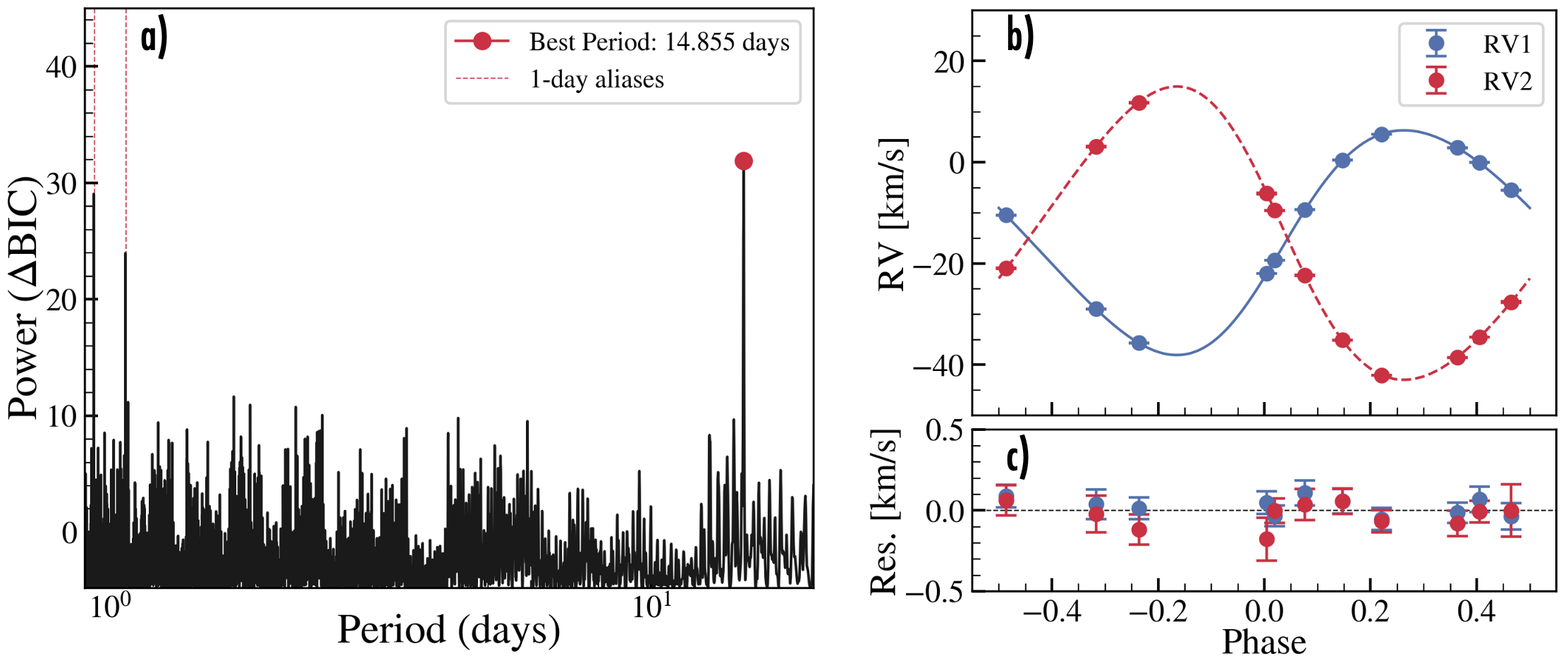}
\vspace{-0.2cm}
\caption{a) Periodogram of the primary star RVs of GJ 3861. The highest peak corresponding to a period of 14.855 days is highlighted with the red dot. The 1-day aliases of the best-fit period are highlighted with the red dashed lines. b) HPF RVs of the GJ 3861 binary system. Red and blue points show the RVs of the primary and secondary, respectively. The best-fit model is shown with the red and blue curves for the primary and secondary, respectively. c) Residuals from the best-fit. RV error bars are shown with the best-fit jitter values applied in quadrature.}
\label{fig:gj3861fit}
\end{figure*}

% RV modeling: Keplerian model
To model the RVs using a two-component Keplerian model, we broadly follow the methodology outlined in \cite{fernandez2017} and \cite{skinner2018}. First, we evaluated and tested different orbital periods by generating Keplerian periodograms with \texttt{RVSearch}. Figure \ref{fig:gj3861fit}a shows the periodogram of the primary star in GJ 3861 system (the periodogram of the secondary star looks similar), where 14.48 days shows the highest peak. We interpret this as the orbital period of the system. We further follow to calculate a `velocity coverage' parameter $v_{\mathrm{cov}}$ from \cite{fernandez2017} and \cite{skinner2018}, which is given by:
\begin{equation}
v_{\mathrm{cov}} = \frac{N}{N-1} (1 - \frac{1}{(v_{\mathrm{max}}-v_{\mathrm{min}})^2}\sum_{i=1}^N (v_{i+1} - v_{i})^2),
\label{eq:vcov}
\end{equation}
where $N$ is the number of observations, and $v_{\mathrm{max}}$ and $v_{\mathrm{min}}$ are the maximum and minimum RV values. As noted by \cite{fernandez2017}, the summation in Equation \ref{eq:vcov} assumes the RVs are sorted and that the mean RV has been subtracted. This parameter quantifies how much of the full RV extent the observations have covered, where \cite{skinner2018} and \cite{fernandez2017} found that systems with at least 8 observations and $v_{cov} > 80\%$ were sufficient to provide good enough sampling of a binary orbit to be able to accurately constrain the binary orbital properties. Using Equation \ref{eq:vcov}, we estimate $v_{\mathrm{cov}} = 95\%$, for GJ 3861, suggesting that we have sufficient velocity coverage to accurately constrain the binary orbital properties.

% --------------------------------------------------------

% Keplerian fit 
Next, we performed a Keplerian fit to the primary and secondary RVs, where the radial velocities of the primary ($v_1$) and the secondary ($v_2$) are:
\begin{equation}
v_{1} = K \left( \cos(\nu + \omega) + e \cos(\omega) \right) + \gamma,
\end{equation}
\begin{equation}
v_{2} = -\frac{K}{q} \left( \cos(\nu + \omega) + e \cos(\omega) \right) + \gamma,
\end{equation}
where $\nu$ is the true anomaly, which is a function of the eccentricity $e$ , the orbital period $P$, and time of periastron $t_p$, $\omega$ is the longitude of periastron, $K$ is the RV semi-amplitude of the primary star, and $\gamma$ is the RV offset of the system. In total, the set of orbital parameters we fitted for are $K$, $e$, $\omega$, $t_p$, $P$ and $\gamma$. 

% MCMC sampling and convergence
Placing an informative prior on the period (normal prior with a median of 14.86154 days and a standard deviation of 0.1 day) from the periodogram analysis, we performed a further optimization in finding a best-fit using the differential-evolution algorithm implemented in \texttt{PyDE} \cite{pyde}. From this best-fit solution we initialized 100 MCMC walkers around the best-fit solution using the \texttt{emcee} \citep{dfm2013} code, where we ran the chains for 50,000 steps, removing the first 2,000 steps as burn-in. We assessed convergence by visually inspecting the chains, and verified that the Gelman-Rubin statistic is within 1\% of unity, and that we have sampled at least 50 independent samples using the autocorrelation timescale feature in \texttt{emcee}. We therefore consider the chains well-mixed. The code we used to implement the Keplerian fit is available on GitHub\footnote{\url{https://github.com/gummiks/rvbinfit/}}.

% Showing resulting fit
Figure \ref{fig:gj3861fit}b shows the RVs of GJ 3861 system along with the best-fit model, and Table \ref{tab:gj3861posteriors} summarizes the priors and posteriors from the fit. From this, we see that the fit suggests a \resqfit, and \resgammafit, in good agreement with the values from the Wilson plot analysis in Figure \ref{fig:gj3861wilson} which suggested \resqwilson, and \resgammawilson.

% Trying additional periods
Additionally, we tested models corresponding to the 1-day aliases of the 14.48 day period, corresponding to 1.07 days and 0.94 days (see Figure \ref{fig:gj3861fit}a). The 1.07 day and 0.94 day fits have much larger residuals of a few km/s and further suggest a mass ratio of $q=0.6645 \pm 0.0007$, and $q = 0.9136 \pm 0.001$, respectively, both of which are inconsistent with the mass ratio derived from the Wilson plot of \resqwilson. Therefore, we conclude that 14.48 days is the orbital period of the system, as listed in Table \ref{tab:gj3861posteriors}.

% Circular vs eccentric
We further note that we tried both circular and eccentric fits for the GJ 3861 RV data. The circular fit resulted in significantly higher residuals at the $\sim$2.5km/s level---much higher than the uncertainties in the measurements---while the eccentric fit yielded residuals consistent with the scatter from the uncertainties resulting in a $\Delta \mathrm{BIC} = 169$ in favor of the eccentric model. We therefore adopt the eccentric model parameters, which are listed in Table \ref{tab:gj3861posteriors}.

% Robo AO
Finally, we note that the RoboAO survey \citep{lamman2020} observed GJ 3861, which did not detect any signs of companions with $\Delta \mathrm{mag} < 4$ from $0.45" - 1.6"$, and any companions with $\Delta \mathrm{mag} < 5$ from $1.2" - 1.6"$. A non-detection of other companions in RoboAO is compatible with the short-period orbit we determine, which would suggest an orbit with a projected orbital separation of fractions of an arcsecond, well below the sensitivity of RoboAO.

The relatively short orbital period suggests that the system may resemble RS Canum Venaticorum (RS CVn) type binaries, which we discuss further in Section~\ref{sec:gj3861_discussion}.

% --------------------------------------------------------
\begin{table}[h]
\caption{GJ 3861 best-fit posteriors.}
\begin{tabular}{cll}
\hline \hline
Parameter      &  Description                     & Posterior Value \\ \hline
$P$            &  Orbital Period (days)           & $14.841181_{-0.00010}^{+0.00011}$ \\
$t_p$          &  Time of periastron              & $2460324.817_{-0.031}^{+0.034}$ \\ 
$e$            &  Eccentricity                    & $0.1196_{-0.0018}^{+0.0017}$ \\
$\omega$       &  Longitude of periastron (deg)   & $252.01_{-0.82}^{+0.90}$ \\
$K_1$          &  RV Semi-amplitude (km/s)        & $22.202_{-0.041}^{+0.040}$ \\
$\gamma$       &  Absolute RV offset              & $-15.091_{-0.021}^{+0.019}$ \\
$\sigma_1$     &  Jitter for RV1                  & $0.061_{-0.027}^{+0.039}$ \\
$\sigma_2$     &  Jitter for RV2                  & $0.066_{-0.039}^{+0.047}$ \\
$q$            &  Mass Ratio ($M_2/M_1$)          & $0.7663_{-0.0018}^{+0.0020}$ \\ 
\hline
\end{tabular}
\label{tab:gj3861posteriors}
\end{table}

% ------------------------------------------------------
% ------------------------------------------------------
% ------------------------------------------------------
\section{Discussion}
\label{sec:discussion}
% ------------------------------------------------------
\subsection{Stellar Inclination Distribution}
\begin{figure}
\centering
\includegraphics[width=\columnwidth]{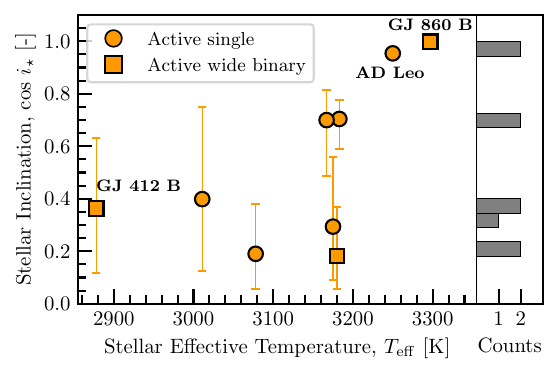}
\vspace{-0.5cm}
\caption{Stellar inclinations ($\cos i_\star$) of the subsample where we have $v \sin i_\star$, $R_\star$, and $P_{rot}$ measurements. Left shows the $\cos i_\star$ over the stellar effective temperature. Right shows a histogram of the $\cos i_\star$ values. The distribution of stellar inclinations is indistinguishable from an isotropic distribution.}
\label{fig:inclinations}
\end{figure}

In Section \ref{sec:stellar_properties}, we used the values found for $R_\star$, $P_{\text{rot}}$ and $v \sin i_\star$ to derive the stellar inclinations ($i_\star$) of our targets. We show the corresponding $\cos i_\star$ values in Figure \ref{fig:inclinations}. The stellar inclination is a key characteristic in understanding the sample of radio-emitting M\,dwarfs, as \cite{kavanagh2023} have shown that the geometry of the stellar magnetic field and planetary orbit determine the likelihood of detecting radio signatures of MSPI. Their results suggest that if all magnetic obliquities are equally likely, blind radio surveys like V-LoTSS are more likely to observe face-on systems, where the planet orbits along the plane of the sky and the stellar rotation is observed pole-on.

Figure \ref{fig:inclinations}, shows no obvious bias to pole-on inclinations. However, as highlighted in Figure \ref{fig:inclinations}, the sample with measured stellar inclinations is solely composed of active, relatively rapidly rotating targets. Therefore, the radio emission from these targets are more likely to be associated with chromospheric activity processes or from rapid rotation, rather than needing to invoke a planetary companion to drive the observed emission. Rapid rotation can also induce ECMI, which results in strongly beamed radio emission, and thus observations of this are also expected to depend on the stellar inclination \citep[e.g.][]{Bloot2024}. In contrast, radio emission driven by particle acceleration flares tends to be isotropic, and thus the impact of system geometry may be less pronounced.

To further quantify whether the observed inclinations show any form of clustering---i.e., inconsistent with an isotropic distribution (uniform in cos $i_\star$)---we performed a modified Anderson–Darling (AD) test. The AD statistic \citep{Anderson1952} is similar to the widely-used Kolmogorov-Smirnov statistic \citep{Feigelson2012}, but more robustly accounts for sensitivity to distribution edge effects. We broadly follow the approach of \citet{Ilin2022} and \citet{Ilin2024}, who applied the AD test to assess clustering in stellar spin orientations. We refer the reader to the details presented in \citet{Ilin2022} and \citet{Ilin2024}, but summarize our methodology briefly here.

We assess consistency with an isotropic distribution by comparing the AD statistic computed from our observed data ($A^2_{\mathrm{obs}}$) to a reference distribution of $A^2$ values expected under the null hypothesis of isotropically distributed stellar inclinations. This reference distribution is generated by calculating $A^2$ for 100,000 synthetic datasets, each consisting of eight $\cos i_\star$ values randomly drawn from a uniform distribution over $[0,1]$. Unlike \citet{Ilin2022} and \citet{Ilin2024}, where measurement uncertainties were negligible, we account for the significant uncertainties in our sample. Instead of computing a single $A^2_{\mathrm{obs}}$, we generate 100,000 $A^2_{\mathrm{obs}}$ values, by sampling from the full posterior distributions of each target's $\cos i_\star$ values. We find that 96\% of our posterior-based $A^2_{\mathrm{obs}}$ values exceed the critical $p$-value ($p>0.05$), indicating that the null hypothesis of an isotropic distribution cannot be rejected. 

Interestingly, two of the stars in the sample, AD Leo and GJ 860 B, are observed to have close to pole-on configurations, with inclinations of 17.5$^{+4.0}_{-4.4}$ and 3.5$^{+1.1}_{-1.3}$, respectively. Further, both stars are bright nearby stars, that show high median RV precisions, making them amenable to ground-based RV observations. However, as the stars are known to be highly active, even in the case of detection of close-in planets, this could complicate attributing any possible radio emission as being due to MSPI. We note that AD Leo has had claims of a planet detected at the rotation period of the star at 2.24 days \citep[][]{Tuomi2018}, but this claim has been disputed, as that signal shows clear signatures of being correlated with stellar activity indicators, with the most likely explanation being that the RV variations are due to stellar activity \citep{Carleo2020, Robertson2020}.

% With only eight systems having measured stellar inclinations so far, the current sample remains relatively small. Future detections from LOFAR and upcoming facilities like the SKA will help expand the dataset. This will enable a more robust investigation of potential geometric effects within the radio-detected population.

% ------------------------------------------------------

% ------------------------------------------------------
\subsection{Prospects for magnetic star-planet interactions on GJ~625}
\label{sec:discussion_GJ625}
In Section~\ref{sec:GJ_625_results}, we refined the orbital parameters of GJ~625~b, deriving a period of $P = 14.6288 \pm 0.0027$ days and a time of conjunction $T_c = $2457302.9 $\pm$ 0.4. Based on this updated ephemeris, we estimate that the LOFAR radio detection reported by \citet{Callingham2021} (2014-05-26) occurred at an orbital phase of $-0.10\pm0.06$. Since that initial detection, GJ~625 has been the subject of continued LOFAR monitoring, being coincidentally located in a LoTSS deep field \citep[ELAIS-N1; ][]{Oliver2000, Sabater2021}. In total, this target has been observed in at least 21 independent LOFAR pointings over two years. The refined orbital solutions presented here facilitate a detailed phase-resolved analysis of these LOFAR observations, which will be discussed in future work.

Such an analysis is motivated by theoretical expectations that MSPI-induced radio emission is modulated by orbital phase, as the emission beam sweeps past the observer when the system's geometry aligns favorably \citep{kavanagh2023}. In the Jupiter–Io system, for instance, radio emission is strongest near orbital quadrature (phases $\pm 0.25$). In the YZ Ceti system, bursts were detected closer to phases $\pm 0.15$ \citep{Pineda2023}. 

Before such a phase analysis can be meaningfully interpreted, we must assess whether conditions in the GJ~625 system are favorable for MSPI to occur. The prospects for MSPI depend primarily on the magnetic field strength at the stellar surface, and the mass-loss rate of the stellar wind. These quantities regulate the size of the Alfv\'en surface \citep{kavanagh21}, which the planet must orbit inside for MSPI to occur.

While the magnetic fields of M~dwarfs manifest at both small and large spatial scales \citep{kochukhov21}, the global stellar wind conditions are predominantly determined by the large-scale magnetic features on the stellar surface \citep{jardine17}. This `large-scale' magnetic field can be mapped using the Zeeman-Doppler imaging method \citep[ZDI; see][]{kochukhov21}. However, this method has yet to be applied to GJ~625. Nevertheless, the average \textit{total} magnetic field strength $\langle B\rangle$, which includes contributions from both the small and large-scale field, can provide a constraint on the average large-scale field strength $\langle B_V\rangle$. Slowly-rotating M~dwarfs such as GJ~625 generally show that $\langle B_V\rangle = 6$ to 65\% of $\langle B\rangle$ \citep{klein21, cristofari23, lehmann24}, and \citet{reiners22} measured a value of $\langle B\rangle = 290$~G for GJ~625. Therefore, we estimate that $\langle B_V\rangle$ ranges from 17-189 G for GJ~625.

The stellar winds of M~dwarfs such as GJ~625 are tenuous, making them difficult to measure \citep{wood21}. However, \citet{bloot25} recently presented a new method for estimating upper limits for the mass-loss rates of main-sequence star with a coronal driven wind (i.e. FGKM types) based on their detection at radio wavelengths. This method relies on an estimate for $\langle B_V\rangle$, and assumes the detected radio emission is produced by the electron cyclotron maser instability. \citet{bloot25} applied their method to the 19 radio-emitting M~dwarfs presented by \citet{Callingham2021}, one of which being GJ~625. However, \citet{bloot25} estimated values of $\langle B_V\rangle = 150$ to 1500~G for GJ~625 based on its mass. Given that our estimate of 17 to 189~G is based on the direct measurement of the magnetic field of GJ~625 by \citet{reiners22}, we opt to use these values instead. This range is also in closer agreement with the values measured for slowly-rotating M~dwarfs by \citet{lehmann24}. Following the same method as \citet{bloot25}\footnote{\url{https://github.com/SBloot/stellar-winds-ECMI}}, we estimate an upper limit for the mass-loss rate of GJ~625 of 0.68 to 25~$\dot{M}_\sun$ for our adopted range of $\langle B_V\rangle$, where $\dot{M}_\sun = 2\times10^{-14}~M_\sun$~yr$^{-1}$ \citep{cohen11}. 

With the large-scale magnetic field strength and wind mass-loss rate of GJ~625 estimated, we now follow the same methodology as \citet{fitzmaurice2024} to determine if GJ~625~b orbits sub-Alfv\'enically. For a planet to orbit inside the Alfv\'en surface, the Alfv\'enic Mach number $M_A$ of the wind at its orbit must be less than unity \citep{saur13}:
\begin{equation}
M_A = \frac{\Delta u}{u_A}.
\end{equation}
Here, $\Delta u$ is the relative velocity between the stellar wind and the planet, and $u_A$ is the Alfvén velocity of the wind:
\begin{equation}
u_A = \frac{B}{\sqrt{4\pi\rho}},
\end{equation}
where $B$ is the wind magnetic field strength, and $\rho$ is the wind density, which relates to the mass-loss rate $\dot{M}$ and velocity $u$ of the stellar wind at the planet's orbital distance $a$:
\begin{equation}
\dot{M} = 4\pi a^2 \rho u.
\end{equation}
Assuming an open magnetic field geometry for the wind at the orbit of the planet, which is likely given the long rotation period of GJ~625 \citep[see][]{fitzmaurice2024}, the Alfv\'enic Mach number can be expressed as
\begin{equation}
M_A = \frac{\Delta u}{\langle B_V\rangle} \Big(\frac{a}{R_\star}\Big)^2 \sqrt{\frac{\dot{M}}{a^2 u}}.
\label{eq:mach number}
\end{equation}
Equation~\ref{eq:mach number} illustrates how MSPI is more likely for systems with strong magnetic fields and low mass-loss rates. If $\dot{M}$ is an upper limit, then so too is the value of $M_A$.

\begin{figure}
\centering
\includegraphics[width = \columnwidth]{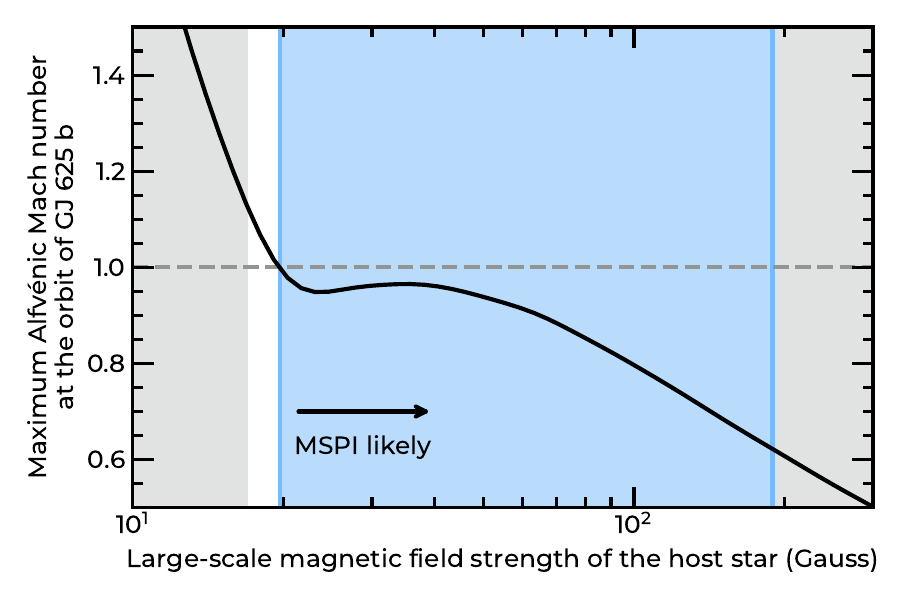}
\caption{The upper limit for the Alfvénic Mach number of the stellar wind at the orbit of GJ~625~b as a function of the average large-scale surface magnetic field strength of the host star. The grey shaded regions show where the field strength is outside the likely estimate of the magnetic field of the star from 17 to 189~G, estimated from empirical relationships. For average field strengths $\gtrsim20$~G, we estimate that the planet must orbit inside the Alfvén surface, corresponding to a Mach number $<1$ (indicated by the dashed horizontal line). In this region (highlighted in blue), magnetic star-planet interactions can occur. However, 3D MHD simulations are needed to assess these prospects further.}
\label{fig:mach number}
\end{figure}

In Figure~\ref{fig:mach number}, we show the upper limit of the Mach number computed via Equation~\ref{eq:mach number} over the range estimated for $\langle B_V\rangle$ and the corresponding upper limits for $\dot{M}$. We find that the maximum Mach number is less than unity for $\langle B_V\rangle \gtrsim 20$~G. This implies that MSPI are likely for the system, as our estimated lower limit for $\langle B_V\rangle$ is 17~G. Additionally, our mass-loss rate estimates are upper limits, so MSPI could still occur if $\langle B_V\rangle<20$~G provided the wind mass-loss rate is sufficiently low. 

While these results are promising, it is worth comparing them to a system where the sub-Alfvénic nature of a planet has been assessed in more detail using a magnetic field map derived from ZDI in combination with 3D MHD stellar wind modeling. Proxima Centauri (Prox~Cen) is a good choice for this. First, its comparable rotation period implies a similar large-scale magnetic field strength for GJ~625 \citep{Vidotto2014}. Second, \citet{kavanagh21} developed 3D~MHD models for the stellar wind of Prox~Cen based on its magnetic field map derived by \citet{klein21}.

The mass-loss rate for Prox~Cen is $<0.2~\dot{M}_\sun$ (with $\dot{M}_\sun = 2\times10^{-14}~M_\sun$~yr$^{-1}$) \citep{wood01}, and its average large-scale magnetic field strength is 200~G \citep{klein21}. Following the same methodology as described above for GJ~625, these values give a maximum Alfv\'enic Mach number of 0.16 at the orbit of Prox~Cen~b\footnote{To estimate the wind velocity of Prox~Cen, we use the code developed by \citet{kavanagh20}, adopting a mass and radius of 0.12~$M_\sun$ and 0.14~$R_\sun$ \citep{kavanagh21}, and coronal temperature of 2.70~MK \citep{johnstone15}. Note that we reduce the temperature from the corona to the wind by a factor of 1.36, as done by \citet{bloot25}.}, which implies a sub-Alfvénic nature for planet~b. However, the 3D MHD models presented by \citet{kavanagh21} indicate that planet~b orbits outside the Alfvén surface. This discrepancy is likely due to a number of factors. For instance, the 3D MHD models from \citet{kavanagh21} account for the local variations in the surface magnetic field strength encapsulated by the ZDI map. Additionally, the driving mechanism of the wind in the MHD models is Alfvén waves, as opposed to the isothermal prescription adopted in the code developed by \citet{kavanagh20}. This demonstrates that while our results show promise for establishing MSPI in GJ~625, ZDI coupled to 3D magnetohydrodynamic simulations of the stellar wind outflow are needed to assess this scenario further.

% ------------------------------------------------------
\subsection{GJ 3861: a possible RS CVn-like system?}
\label{sec:gj3861_discussion}
Our sample contains six close visual or spectroscopic multiples. In some of these systems, the coherent radio emission may be caused by particle acceleration in flares, as suggested by their relatively high chromospheric activity indicators. Alternatively, if the binaries are at a close-in orbit, the magnetospheric interactions between the two stars might be responsible instead. A notable target in this subsample is target GJ~9552 (CR~Dra), which is located in a LoTSS deep field \citep[specifically the ELAIS-N1 field; ][]{Oliver2000, Sabater2021}. The target thus obtained numerous radio observations, allowing for a detailed analysis of its radio properties \citep{Callingham2021_apr}. This analysis suggests that ECMI is the most likely emission mechanism, where the instability could be driven by a breakdown in co-rotation.

\begin{figure}
\centering
\includegraphics[width=\linewidth]{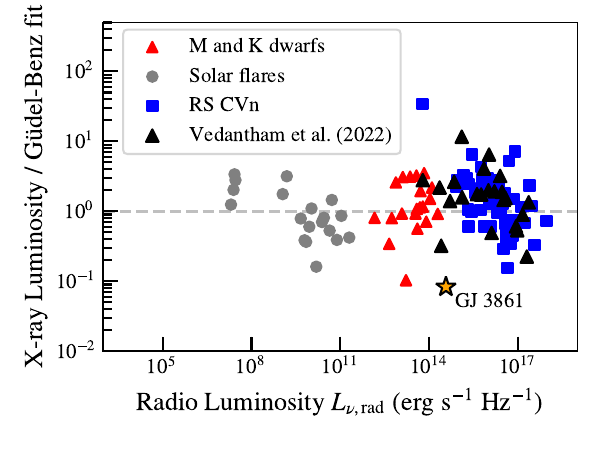}
\caption{Deviation of targets from the G\"{u}del-Benz relation \citep{Williams2014}. The targets originally used to define this relation are shown in red, grey and blue. Coherently emitting, active targets detected with LOFAR—shown as black triangles and primarily consisting of RS CVn systems—scatter around the G\"{u}del-Benz relation with a comparable spread \citep{Vedantham2022}. Our system, GJ 3861, is also indicated.}
\label{fig:GB_scatter}
\end{figure}

Close-separated binaries may fall into the category of RS Canum Venaticorum (RS~CVn) binaries. These systems are short-period binaries with orbital periods of $\sim$1–30 days, typically comprising two late-type stars (F, G or K) \citep{Hall1976}. Due to tidal locking, these systems are rapidly rotating and exhibit enhanced magnetic activity, manifesting in chromospheric emission, elevated X-ray luminosity, and significant photometric variability \citep[e.g. ][]{Walter1978, Arevalo1999, martinez2022}. Several RS CVn binaries have been observed to produce bright, coherent radio emission at LOFAR frequencies \citep{Toet2021, Callingham2023}, similar to the M\,dwarf radio sample. Interestingly, the radio emission from these systems appears to follow the G\"{u}del-Benz relation \citep[See Figure \ref{fig:GB_scatter}; adapted from][]{Vedantham2022}. This behavior is unexpected for highly circularly polarized radio emission, which is typically not associated with coronal heating processes generally understood to underpin the G\"{u}del-Benz relation. The emission mechanisms operating in these systems remain uncertain, with ECMI, plasma emission, and gyrosynchrotron emission all remaining viable interpretations \citep{Vedantham2022}.

One of the binaries in our sample, GJ~3789 (DG~CVn), composed of two M4Ve-type stars, has previously been classified as an RS CVn-like system \citep{Toet2021, Vedantham2022}. However, its estimated orbital period of approximately 7 years \citep{Beuzit2004} is significantly longer than the typical 1–30 day range for RS CVn binaries, calling this classification into question. The orbital periods of G 240-45, GJ~3729, and 2MASS J10534129+5253040 are furthermore unknown and the orbital period of GJ~9552 remains uncertain \citep{Tamazian2008, Shkolnik2010, Sperauskas2019}. Therefore it is unclear if these systems can be decisively called RS CVn-like binaries. 

In contrast, in Section \ref{sec:GJ_3861_results}, we have shown that GJ~3861 is a binary with a mass ratio of \resqfit, and an orbital period of $P=14.841181\pm0.00010$ d, making it the only system in the M\,dwarf radio sample with a confirmed short-period orbit, compatible with the RS~CVn-like classification. As further shown in Figure \ref{fig:ztf}, ground-based photometry from ZTF reveals a detection of photometric modulations at periods from 12-16 days, further confirming its activity and compatibility with an RS CVn-like binary. However, GJ 3861 deviates more significantly from the G\"{u}del-Benz relation than the other known RS CVn systems with coherent radio emission (see Figure \ref{fig:GB_scatter}), with a deviation of $2.8\sigma$, where  $\sigma$ denotes the standard deviation scatter seen in the systems discussed by \citet{Vedantham2022}. Even so, given the close-in active stars in a close-in orbital period we surmise that the most likely reason for the observed radio emission in the system is related to either the chromospheric activity and/or interaction of magnetospheres in the system.

We additionally note that GJ~3861 has not previously been recognized as a close-in binary system, which might have implications on previous work.  One such example is the work of \cite{bloot25}, which used LOFAR observations to characterize the stellar mass-loss rates of the LOFAR radio emitting sample discussed here, where they highlight GJ~3861 as a representative case in placing constraints on stellar mass-loss rates. Assuming a single star, for GJ 3861 they find a mass loss rate of $200-1120~\dot{M}_{\sun}$. However, as their method assumes a single stellar wind source, rather than a binary, the detected emission may arise from contribution from both stellar components. As such, although \cite{bloot25} place much more stringent constraints than earlier studies on the mass-loss of GJ~3861, the derived stellar mass loss rate may be overestimated due to the unaccounted contribution from the secondary.

%-------------------------------------------------------------------
%-------------------------------------------------------------------
%-------------------------------------------------------------------
\section{Summary and Conclusions}
\label{sec:conclusions}
We present the first results a spectroscopic survey to characterize recent radio detections of radio emitting M\,dwarfs identified as coherent radio sources by the LOFAR Two-Metre Sky Survey. Using high-resolution NIR/red-optical spectroscopic data from the HPF and NEID spectrographs, we homogenously derive stellar spectroscopic properties and further assess the nature of the observed radio emission. Our key findings for the sample as a whole are as follows:

\begin{enumerate}
\item \textbf{Stellar properties and MSPI targets:} We measure the sample's stellar properties, including their effective temperatures ($T_{\mathrm{eff}}$), surface gravities ($\log g$), metallicities ([Fe/H]), projected rotational velocities ($v \sin i_\star$), rotation periods ($P_\text{rot}$), stellar radii ($R_\star$), and stellar inclinations ($i_\star$). We find that the sample consists of 53\% (10/19) single, 16\% (3/19) well-separated binaries, and 31\% (6/19) close visual/spectroscopic multiples. We identify three stars in the sample---GJ~625, GJ~1151, and LHS~2395---that are chromospherically inactive and which are amenable to finding close-in planet with RVs, making them particularly favorable MSPI candidates.
\item \textbf{Distribution of stellar inclinations:} Certain models for the MSPI origin of the emission suggests that there could be an overabundance of pole-on configurations for the stellar inclinations. However, in investigating the sample of 8 stars with measured stellar inclinations, we show that by sampling from their posteriors that in 96\% of the cases, the measured distribution is indistinguishable from an isotropic distribution. However, our sample of radio-bright stars with measured inclinations are dominated by active, and in some cases rapidly-rotating, stars. 
\item \textbf{Refining the orbit of GJ~625 b:} GJ~625 hosts a previously confirmed close-in planet. By combining new RV data with archival measurements, we refine the planet’s orbital ephemeris by a factor of three in precision, providing an improved baseline for future phase-targeted radio monitoring. Furthermore, we estimate that the system conditions are suitable for MSPI, given a sufficiently strong stellar magnetic field. Future Zeeman-Doppler Imaging observations to constrain the large scale stellar magnetic field are needed to further assess the MSPI scenario.
\item \textbf{GJ~3861 as a candidate RS CVn-like system:} For the first time, we reveal GJ 3861 to be a double-lined spectroscopic binary with an orbital period of $14.841181^{+0.00011}_{-0.00010}$ days and a mass ratio of $q = 0.7663^{+0.0020}_{-0.0018}$. Its relatively short orbital period is characteristic of RS CVn systems, which have been seen to emit bright, circularly-polarized radio emission.
\end{enumerate}

% --------------------------------------------------------
% --------------------------------------------------------
% --------------------------------------------------------
\section*{Acknowledgements}
 These results are based on observations obtained with the Habitable-zone Planet Finder Spectrograph on the HET. The HPF team acknowledges support from NSF grants AST-1006676, AST-1126413, AST-1310885, AST-1517592, AST-1310875, ATI 2009889, ATI-2009982, AST-2108512, and the NASA Astrobiology Institute (NNA09DA76A) in the pursuit of precision radial velocities in the NIR. The HPF team also acknowledges support from the Heising-Simons Foundation via grant 2017-0494. This research was conducted in part under NSF grants AST-2108493, AST-2108512, AST-2108569, and AST-2108801 in support of the HPF Guaranteed Time Observations survey. Based on observations obtained with the Hobby-Eberly Telescope (HET), which is a joint project of the University of Texas at Austin, the Pennsylvania State University, Ludwig-Maximillians-Universitaet Muenchen, and Georg-August Universitaet Goettingen. The HET is named in honor of its principal benefactors, William P. Hobby and Robert E. Eberly. We acknowledge the Texas Advanced Computing Center (TACC) at the University of Texas at Austin for providing high performance computing, visualization, and storage resources that have contributed to the results reported within this paper. We thank the Resident astronomers and Telescope Operators at the HET for the skillful execution of our observations with HPF.

Data presented were obtained by the NEID spectrograph built by Penn State University and operated at the WIYN Observatory by NOIRLab, under the NN-EXPLORE partnership of the National Aeronautics and Space Administration and the National Science Foundation. The NEID archive is operated by the NASA Exoplanet Science Institute at the California Institute of Technology. We thank the NEID Queue Observers and WIYN Observing Associates for their skillful execution of our observations. Based on observations at NSF Kitt Peak National Observatory, NSF NOIRLab (NOIRLab Prop. ID 2021A-0389,
PI Paul Robertson, and NOIRLab Prop. ID 2022A-452266, PI Shubham Kanodia), which is managed by the Association of Universities for Research in Astronomy (AURA) under a cooperative agreement with the U.S. National Science Foundation. The authors are honored to be permitted to conduct astronomical research on I'oligam Du’ag (Kitt Peak), a mountain with particular significance to the Tohono O’odham.

Based on data from the CARMENES data archive at CAB (CSIC-INTA). The CARMENES archive is part of the Spanish Virtual Observatory project (http://svo.cab.inta-csic.es), funded by MCIN/AEI/10.13039/501100011033/ through grant PID2020-112949GB-I00. This research has made use of the Keck Observatory Archive (KOA), which is operated by the W. M. Keck Observatory and the NASA Exoplanet Science Institute (NExScI), under contract with the National Aeronautics and Space Administration. This research used the facilities of the Italian Center for Astronomical Archive (IA2) operated by INAF at the Astronomical Observatory of Trieste. This research made use of the NASA Exoplanet Archive, which is operated by the California Institute of Technology, under contract with the National Aeronautics and Space Administration under the Exoplanet Exploration Program.  This work has made use of data from the European Space Agency (ESA) mission Gaia processed by the Gaia Data Processing and Analysis Consortium (DPAC).

The Center for Exoplanets and Habitable Worlds and the Penn State Extraterrestrial Intelligence Center are
supported by Penn State and its Eberly College of Science. R.D.K. acknowledges funding from the Dutch Research Council (NWO) for the project ‘e-MAPS’ (project number Vi.Vidi.203.093) under the NWO talent scheme VIDI. J.I.E.-R. gratefully acknowledges support from the ANID BASAL project FB210003 and from the ANID Doctorado Nacional grant 2021-21212378. C.I.C. acknowledges support by NASA Headquarters through an appointment to the NASA Postdoctoral Program at the Goddard Space Flight Center, administered by ORAU through a contract with NASA. J.R.C. acknowledges funding from the European Union via the European Research Council (ERC) grant Epaphus (project number: 101166008). The research was carried out, in part, at the Jet Propulsion Laboratory, California Institute of Technology, under a contract with the National Aeronautics and Space Administration (80NM0018D0004).

\bibliographystyle{aa}
\bibliography{example} 

\begin{thebibliography}{131}
\expandafter\ifx\csname natexlab\endcsname\relax\def\natexlab#1{#1}\fi

\bibitem[{{Alonso-Floriano} {et~al.}(2015){Alonso-Floriano}, {Morales}, {Caballero}, {Montes}, {Klutsch}, {Mundt}, {Cort{\'e}s-Contreras}, {Ribas}, {Reiners}, {Amado}, {Quirrenbach}, \& {Jeffers}}]{Floriano2015}
{Alonso-Floriano}, F.~J., {Morales}, J.~C., {Caballero}, J.~A., {et~al.} 2015, \aap, 577, A128

\bibitem[{Anderson \& Darling(1952)}]{Anderson1952}
Anderson, T.~W. \& Darling, D.~A. 1952, The Annals of Mathematical Statistics, 23, 193–212

\bibitem[{{Ar{\'e}valo} \& {L{\'a}zaro}(1999)}]{Arevalo1999}
{Ar{\'e}valo}, M.~J. \& {L{\'a}zaro}, C. 1999, \aj, 118, 1015

\bibitem[{{Bailer-Jones} {et~al.}(2021){Bailer-Jones}, {Rybizki}, {Fouesneau}, {Demleitner}, \& {Andrae}}]{Bailer-Jones2021}
{Bailer-Jones}, C.~A.~L., {Rybizki}, J., {Fouesneau}, M., {Demleitner}, M., \& {Andrae}, R. 2021, \aj, 161, 147

\bibitem[{{Benz} \& {G\"udel}(1994)}]{Benz1994}
{Benz}, A.~O. \& {G\"udel}, M. 1994, \aap, 285, 621

\bibitem[{Beuzit {et~al.}(2004)Beuzit, Ségransan, Forveille, Udry, Delfosse, Mayor, Perrier, Hainaut, Roddier, Roddier, \& Martín}]{Beuzit2004}
Beuzit, J.-L., Ségransan, D., Forveille, T., {et~al.} 2004, Astronomy {\&} Astrophysics, 425, 997–1008

\bibitem[{{Bigg}(1964)}]{Bigg1964}
{Bigg}, E.~K. 1964, \nat, 203, 1008

\bibitem[{{Blanco-Pozo} {et~al.}(2023){Blanco-Pozo}, {Perger}, {Damasso}, {Anglada Escud{\'e}}, {Ribas}, {Baroch}, {Caballero}, {Cifuentes}, {Jeffers}, {Lafarga}, {Kaminski}, {Kaur}, {Nagel}, {Perdelwitz}, {P{\'e}rez-Torres}, {Sozzetti}, {Vigan{\`o}}, {Amado}, {Andreuzzi}, {B{\'e}jar}, {Brown}, {Del Sordo}, {Dreizler}, {Galad{\'\i}-Enr{\'\i}quez}, {Hatzes}, {K{\"u}rster}, {Lanza}, {Melis}, {Molinari}, {Montes}, {Murgia}, {Pall{\'e}}, {Pe{\~n}a-Mo{\~n}ino}, {Perrodin}, {Pilia}, {Poretti}, {Quirrenbach}, {Reiners}, {Schweitzer}, {Zapatero Osorio}, \& {Zechmeister}}]{Blanco-Pozo2023}
{Blanco-Pozo}, J., {Perger}, M., {Damasso}, M., {et~al.} 2023, \aap, 671, A50

\bibitem[{{Bloot} {et~al.}(2024){Bloot}, {Callingham}, {Vedantham}, {Kavanagh}, {Pope}, {Climent}, {Guirado}, {Pe{\~n}a-Mo{\~n}ino}, \& {P{\'e}rez-Torres}}]{Bloot2024}
{Bloot}, S., {Callingham}, J.~R., {Vedantham}, H.~K., {et~al.} 2024, \aap, 682, A170

\bibitem[{{Bloot} {et~al.}(2025){Bloot}, {Vedantham}, {Kavanagh}, {Callingham}, \& {Pope}}]{bloot25}
{Bloot}, S., {Vedantham}, H.~K., {Kavanagh}, R.~D., {Callingham}, J.~R., \& {Pope}, B.~J.~S. 2025, \aap, 695, A176

\bibitem[{Burke \& Franklin(1955)}]{Burke1955}
Burke, B.~F. \& Franklin, K.~L. 1955, Journal of Geophysical Research, 60, 213–217

\bibitem[{{Busso} {et~al.}(2022){Busso}, {Cacciari}, {Bellazzini}, {Carrasco}, {De Angeli}, {Evans}, {Fabricius}, {Montegriffo}, {Pancino}, {Rainer}, \& {Sanna}}]{GaiaDR3_documentation_ch5}
{Busso}, G., {Cacciari}, C., {Bellazzini}, M., {et~al.} 2022, {Gaia DR3 documentation Chapter 5: Photometric data}

\bibitem[{Callingham {et~al.}(2021{\natexlab{a}})Callingham, Pope, Feinstein, Vedantham, Shimwell, Zarka, Tasse, Lamy, Veken, Toet, Sabater, Best, van Weeren, R\"{o}ttgering, \& Ray}]{Callingham2021_apr}
Callingham, J.~R., Pope, B. J.~S., Feinstein, A.~D., {et~al.} 2021{\natexlab{a}}, Astronomy {\&} Astrophysics, 648, A13

\bibitem[{{Callingham} {et~al.}(2024){Callingham}, {Pope}, {Kavanagh}, {Bellotti}, {Daley-Yates}, {Damasso}, {Grie{\ss}meier}, {G{\"u}del}, {G{\"u}nther}, {Kao}, {Klein}, {Mahadevan}, {Morin}, {Nichols}, {Osten}, {P{\'e}rez-Torres}, {Pineda}, {Rigney}, {Saur}, {Stef{\'a}nsson}, {Turner}, {Vedantham}, {Vidotto}, {Villadsen}, \& {Zarka}}]{Callingham2024_nov}
{Callingham}, J.~R., {Pope}, B.~J.~S., {Kavanagh}, R.~D., {et~al.} 2024, Nature Astronomy, 8, 1359

\bibitem[{{Callingham} {et~al.}(2023){Callingham}, {Shimwell}, {Vedantham}, {Bassa}, {O'Sullivan}, {Yiu}, {Bloot}, {Best}, {Hardcastle}, {Haverkorn}, {Kavanagh}, {Lamy}, {Pope}, {R{\"o}ttgering}, {Schwarz}, {Tasse}, {van Weeren}, {White}, {Zarka}, {Bomans}, {Bonafede}, {Bonato}, {Botteon}, {Bruggen}, {Chy{\.z}y}, {Drabent}, {Emig}, {Gloudemans}, {G{\"u}rkan}, {Hajduk}, {Hoang}, {Hoeft}, {Iacobelli}, {Kadler}, {Kunert-Bajraszewska}, {Mingo}, {Morabito}, {Nair}, {P{\'e}rez-Torres}, {Ray}, {Riseley}, {Rowlinson}, {Shulevski}, {Sweijen}, {Timmerman}, {Vaccari}, \& {Zheng}}]{Callingham2023}
{Callingham}, J.~R., {Shimwell}, T.~W., {Vedantham}, H.~K., {et~al.} 2023, \aap, 670, A124

\bibitem[{Callingham {et~al.}(2021{\natexlab{b}})Callingham, Vedantham, Shimwell, Pope, Davis, Best, Hardcastle, R\"{o}ttgering, Sabater, Tasse, van Weeren, Williams, Zarka, de~Gasperin, \& Drabent}]{Callingham2021}
Callingham, J.~R., Vedantham, H.~K., Shimwell, T.~W., {et~al.} 2021{\natexlab{b}}, Nature Astronomy, 5, 1233–1239

\bibitem[{{Carleo} {et~al.}(2020){Carleo}, {Malavolta}, {Lanza}, {Damasso}, {Desidera}, {Borsa}, {Mallonn}, {Pinamonti}, {Gratton}, {Alei}, {Benatti}, {Mancini}, {Maldonado}, {Biazzo}, {Esposito}, {Frustagli}, {Gonz{\'a}lez-{\'A}lvarez}, {Micela}, {Scandariato}, {Sozzetti}, {Affer}, {Bignamini}, {Bonomo}, {Claudi}, {Cosentino}, {Covino}, {Fiorenzano}, {Giacobbe}, {Harutyunyan}, {Leto}, {Maggio}, {Molinari}, {Nascimbeni}, {Pagano}, {Pedani}, {Piotto}, {Poretti}, {Rainer}, {Redfield}, {Baffa}, {Baruffolo}, {Buchschacher}, {Billotti}, {Cecconi}, {Falcini}, {Fantinel}, {Fini}, {Galli}, {Ghedina}, {Ghinassi}, {Giani}, {Gonzalez}, {Gonzalez}, {Guerra}, {Hernandez Diaz}, {Hernandez}, {Iuzzolino}, {Lodi}, {Oliva}, {Origlia}, {Perez Ventura}, {Puglisi}, {Riverol}, {Riverol}, {San Juan}, {Sanna}, {Scuderi}, {Seemann}, {Sozzi}, \& {Tozzi}}]{Carleo2020}
{Carleo}, I., {Malavolta}, L., {Lanza}, A.~F., {et~al.} 2020, \aap, 638, A5

\bibitem[{{Cauley} {et~al.}(2019){Cauley}, {Shkolnik}, {Llama}, \& {Lanza}}]{Cauley2019}
{Cauley}, P.~W., {Shkolnik}, E.~L., {Llama}, J., \& {Lanza}, A.~F. 2019, Nature Astronomy, 3, 1128

\bibitem[{{Cohen}(2011)}]{cohen11}
{Cohen}, O. 2011, \mnras, 417, 2592

\bibitem[{Cosentino {et~al.}(2012)Cosentino, Lovis, Pepe, Collier~Cameron, Latham, Molinari, Udry, Bezawada, Black, Born, Buchschacher, Charbonneau, Figueira, Fleury, Galli, Gallie, Gao, Ghedina, Gonzalez, Gonzalez, Guerra, Henry, Horne, Hughes, Kelly, Lodi, Lunney, Maire, Mayor, Micela, Ordway, Peacock, Phillips, Piotto, Pollacco, Queloz, Rice, Riverol, Riverol, San~Juan, Sasselov, Segransan, Sozzetti, Sosnowska, Stobie, Szentgyorgyi, Vick, \& Weber}]{Cosentino2012}
Cosentino, R., Lovis, C., Pepe, F., {et~al.} 2012, in Ground-based and Airborne Instrumentation for Astronomy IV, ed. I.~S. McLean, S.~K. Ramsay, \& H.~Takami (SPIE)

\bibitem[{{Cristofari} {et~al.}(2023){Cristofari}, {Donati}, {Moutou}, {Lehmann}, {Charpentier}, {Fouqu{\'e}}, {Folsom}, {Masseron}, {Carmona}, {Delfosse}, {Petit}, {Artigau}, {Cook}, \& {SLS Consortium}}]{cristofari23}
{Cristofari}, P.~I., {Donati}, J.~F., {Moutou}, C., {et~al.} 2023, \mnras, 526, 5648

\bibitem[{{Delfosse} {et~al.}(1998){Delfosse}, {Forveille}, {Perrier}, \& {Mayor}}]{Delfosse1998}
{Delfosse}, X., {Forveille}, T., {Perrier}, C., \& {Mayor}, M. 1998, \aap, 331, 581

\bibitem[{{Dulk}(1985)}]{Dulk1985}
{Dulk}, G.~A. 1985, \araa, 23, 169

\bibitem[{El-Badry {et~al.}(2021)El-Badry, Rix, \& Heintz}]{ElBadry2021}
El-Badry, K., Rix, H.-W., \& Heintz, T.~M. 2021, Monthly Notices of the Royal Astronomical Society, 506, 2269–2295

\bibitem[{Feigelson \& Babu(2012)}]{Feigelson2012}
Feigelson, E.~D. \& Babu, G.~J. 2012, Modern Statistical Methods for Astronomy: With R Applications (Cambridge University Press)

\bibitem[{{Fernandez} {et~al.}(2017){Fernandez}, {Covey}, {De Lee}, {Chojnowski}, {Nidever}, {Ballantyne}, {Cottaar}, {Da Rio}, {Foster}, {Majewski}, {Meyer}, {Reyna}, {Roberts}, {Skinner}, {Stassun}, {Tan}, {Troup}, \& {Zasowski}}]{fernandez2017}
{Fernandez}, M.~A., {Covey}, K.~R., {De Lee}, N., {et~al.} 2017, \pasp, 129, 084201

\bibitem[{{Fitzmaurice} {et~al.}(2024){Fitzmaurice}, {Stef{\'a}nsson}, {Kavanagh}, {Mahadevan}, {Ca{\~n}as}, {Winn}, {Robertson}, {Ninan}, {Albrecht}, {Callingham}, {Cochran}, {Delamer}, {Ford}, {Kanodia}, {Lin}, {Marcussen}, {Pope}, {Ramsey}, {Roy}, {Vedantham}, \& {Wright}}]{fitzmaurice2024}
{Fitzmaurice}, E., {Stef{\'a}nsson}, G., {Kavanagh}, R.~D., {et~al.} 2024, \aj, 168, 140

\bibitem[{{Foreman-Mackey} {et~al.}(2013){Foreman-Mackey}, {Hogg}, {Lang}, \& {Goodman}}]{dfm2013}
{Foreman-Mackey}, D., {Hogg}, D.~W., {Lang}, D., \& {Goodman}, J. 2013, PASP, 125, 306

\bibitem[{{Fulton} {et~al.}(2018){Fulton}, {Petigura}, {Blunt}, \& {Sinukoff}}]{Fulton2018}
{Fulton}, B.~J., {Petigura}, E.~A., {Blunt}, S., \& {Sinukoff}, E. 2018, \pasp, 130, 044504

\bibitem[{{Gaia Collaboration} {et~al.}(2021){Gaia Collaboration}, {Smart}, {Sarro}, {Rybizki}, {Reyl{\'e}}, {Robin}, {Hambly}, {Abbas}, {Barstow}, {de Bruijne}, {Bucciarelli}, {Carrasco}, {Cooper}, {Hodgkin}, {Masana}, {Michalik}, {Sahlmann}, {Sozzetti}, {Brown}, {Vallenari}, {Prusti}, {Babusiaux}, {Biermann}, {Creevey}, {Evans}, {Eyer}, {Hutton}, {Jansen}, {Jordi}, {Klioner}, {Lammers}, {Lindegren}, {Luri}, {Mignard}, {Panem}, {Pourbaix}, {Randich}, {Sartoretti}, {Soubiran}, {Walton}, {Arenou}, {Bailer-Jones}, {Bastian}, {Cropper}, {Drimmel}, {Katz}, {Lattanzi}, {van Leeuwen}, {Bakker}, {Casta{\~n}eda}, {De Angeli}, {Ducourant}, {Fabricius}, {Fouesneau}, {Fr{\'e}mat}, {Guerra}, {Guerrier}, {Guiraud}, {Jean-Antoine Piccolo}, {Messineo}, {Mowlavi}, {Nicolas}, {Nienartowicz}, {Pailler}, {Panuzzo}, {Riclet}, {Roux}, {Seabroke}, {Sordo}, {Tanga}, {Th{\'e}venin}, {Gracia-Abril}, {Portell}, {Teyssier}, {Altmann}, {Andrae}, {Bellas-Velidis}, {Benson}, {Berthier}, {Blomme}, {Brugaletta}, {Burgess}, {Busso}, {Carry},
  {Cellino}, {Cheek}, {Clementini}, {Damerdji}, {Davidson}, {Delchambre}, {Dell'Oro}, {Fern{\'a}ndez-Hern{\'a}ndez}, {Galluccio}, {Garc{\'\i}a-Lario}, {Garcia-Reinaldos}, {Gonz{\'a}lez-N{\'u}{\~n}ez}, {Gosset}, {Haigron}, {Halbwachs}, {Harrison}, {Hatzidimitriou}, {Heiter}, {Hern{\'a}ndez}, {Hestroffer}, {Holl}, {Jan{\ss}en}, {Jevardat de Fombelle}, {Jordan}, {Krone-Martins}, {Lanzafame}, {L{\"o}ffler}, {Lorca}, {Manteiga}, {Marchal}, {Marrese}, {Moitinho}, {Mora}, {Muinonen}, {Osborne}, {Pancino}, {Pauwels}, {Recio-Blanco}, {Richards}, {Riello}, {Rimoldini}, {Roegiers}, {Siopis}, {Smith}, {Ulla}, {Utrilla}, {van Leeuwen}, {van Reeven}, {Abreu Aramburu}, {Accart}, {Aerts}, {Aguado}, {Ajaj}, {Altavilla}, {{\'A}lvarez}, {{\'A}lvarez Cid-Fuentes}, {Alves}, {Anderson}, {Anglada Varela}, {Antoja}, {Audard}, {Baines}, {Baker}, {Balaguer-N{\'u}{\~n}ez}, {Balbinot}, {Balog}, {Barache}, {Barbato}, {Barros}, {Bartolom{\'e}}, {Bassilana}, {Bauchet}, {Baudesson-Stella}, {Becciani}, {Bellazzini}, {Bernet}, {Bertone},
  {Bianchi}, {Blanco-Cuaresma}, {Boch}, {Bombrun}, {Bossini}, {Bouquillon}, {Bragaglia}, {Bramante}, {Breedt}, {Bressan}, {Brouillet}, {Burlacu}, {Busonero}, {Butkevich}, {Buzzi}, {Caffau}, {Cancelliere}, {C{\'a}novas}, {Cantat-Gaudin}, {Carballo}, {Carlucci}, {Carnerero}, {Casamiquela}, {Castellani}, {Castro-Ginard}, {Castro Sampol}, {Chaoul}, {Charlot}, {Chemin}, {Chiavassa}, {Cioni}, {Comoretto}, {Cornez}, {Cowell}, {Crifo}, {Crosta}, {Crowley}, {Dafonte}, \& {Dapergolas}}]{Gaia20201}
{Gaia Collaboration}, {Smart}, R.~L., {Sarro}, L.~M., {et~al.} 2021, \aap, 649, A6

\bibitem[{{Gaia Collaboration} {et~al.}(2023){Gaia Collaboration}, {Vallenari}, {Brown}, {Prusti}, {de Bruijne}, {Arenou}, {Babusiaux}, {Biermann}, {Creevey}, {Ducourant}, {Evans}, {Eyer}, {Guerra}, {Hutton}, {Jordi}, {Klioner}, {Lammers}, {Lindegren}, {Luri}, {Mignard}, {Panem}, {Pourbaix}, {Randich}, {Sartoretti}, {Soubiran}, {Tanga}, {Walton}, {Bailer-Jones}, {Bastian}, {Drimmel}, {Jansen}, {Katz}, {Lattanzi}, {van Leeuwen}, {Bakker}, {Cacciari}, {Casta{\~n}eda}, {De Angeli}, {Fabricius}, {Fouesneau}, {Fr{\'e}mat}, {Galluccio}, {Guerrier}, {Heiter}, {Masana}, {Messineo}, {Mowlavi}, {Nicolas}, {Nienartowicz}, {Pailler}, {Panuzzo}, {Riclet}, {Roux}, {Seabroke}, {Sordo}, {Th{\'e}venin}, {Gracia-Abril}, {Portell}, {Teyssier}, {Altmann}, {Andrae}, {Audard}, {Bellas-Velidis}, {Benson}, {Berthier}, {Blomme}, {Burgess}, {Busonero}, {Busso}, {C{\'a}novas}, {Carry}, {Cellino}, {Cheek}, {Clementini}, {Damerdji}, {Davidson}, {de Teodoro}, {Nu{\~n}ez Campos}, {Delchambre}, {Dell'Oro}, {Esquej},
  {Fern{\'a}ndez-Hern{\'a}ndez}, {Fraile}, {Garabato}, {Garc{\'\i}a-Lario}, {Gosset}, {Haigron}, {Halbwachs}, {Hambly}, {Harrison}, {Hern{\'a}ndez}, {Hestroffer}, {Hodgkin}, {Holl}, {Jan{\ss}en}, {Jevardat de Fombelle}, {Jordan}, {Krone-Martins}, {Lanzafame}, {L{\"o}ffler}, {Marchal}, {Marrese}, {Moitinho}, {Muinonen}, {Osborne}, {Pancino}, {Pauwels}, {Recio-Blanco}, {Reyl{\'e}}, {Riello}, {Rimoldini}, {Roegiers}, {Rybizki}, {Sarro}, {Siopis}, {Smith}, {Sozzetti}, {Utrilla}, {van Leeuwen}, {Abbas}, {{\'A}brah{\'a}m}, {Abreu Aramburu}, {Aerts}, {Aguado}, {Ajaj}, {Aldea-Montero}, {Altavilla}, {{\'A}lvarez}, {Alves}, {Anders}, {Anderson}, {Anglada Varela}, {Antoja}, {Baines}, {Baker}, {Balaguer-N{\'u}{\~n}ez}, {Balbinot}, {Balog}, {Barache}, {Barbato}, {Barros}, {Barstow}, {Bartolom{\'e}}, {Bassilana}, {Bauchet}, {Becciani}, {Bellazzini}, {Berihuete}, {Bernet}, {Bertone}, {Bianchi}, {Binnenfeld}, {Blanco-Cuaresma}, {Blazere}, {Boch}, {Bombrun}, {Bossini}, {Bouquillon}, {Bragaglia}, {Bramante}, {Breedt},
  {Bressan}, {Brouillet}, {Brugaletta}, {Bucciarelli}, {Burlacu}, {Butkevich}, {Buzzi}, {Caffau}, {Cancelliere}, {Cantat-Gaudin}, {Carballo}, {Carlucci}, {Carnerero}, {Carrasco}, {Casamiquela}, {Castellani}, {Castro-Ginard}, {Chaoul}, {Charlot}, {Chemin}, {Chiaramida}, {Chiavassa}, {Chornay}, {Comoretto}, {Contursi}, {Cooper}, {Cornez}, {Cowell}, {Crifo}, {Cropper}, {Crosta}, {Crowley}, {Dafonte}, {Dapergolas}, {David}, {David}, {de Laverny}, {De Luise}, \& {De March}}]{Gaia2020_DR3}
{Gaia Collaboration}, {Vallenari}, A., {Brown}, A.~G.~A., {et~al.} 2023, \aap, 674, A1

\bibitem[{{Garcia-Sage} {et~al.}(2017){Garcia-Sage}, {Glocer}, {Drake}, {Gronoff}, \& {Cohen}}]{Garcia-Sage2017}
{Garcia-Sage}, K., {Glocer}, A., {Drake}, J.~J., {Gronoff}, G., \& {Cohen}, O. 2017, \apjl, 844, L13

\bibitem[{Gould \& Chaname(2004)}]{Gould2004}
Gould, A. \& Chaname, J. 2004, The Astrophysical Journal Supplement Series, 150, 455–464

\bibitem[{{Gupta} {et~al.}(2017){Gupta}, {Ajithkumar}, {Kale}, {Nayak}, {Sabhapathy}, {Sureshkumar}, {Swami}, {Chengalur}, {Ghosh}, {Ishwara-Chandra}, {Joshi}, {Kanekar}, {Lal}, \& {Roy}}]{Gupta2017}
{Gupta}, Y., {Ajithkumar}, B., {Kale}, H.~S., {et~al.} 2017, Current Science, 113, 707

\bibitem[{{Hall}(1976)}]{Hall1976}
{Hall}, D.~S. 1976, in Astrophysics and Space Science Library, Vol.~60, IAU Colloq. 29: Multiple Periodic Variable Stars, ed. W.~S. {Fitch}, 287

\bibitem[{{Hartkopf} {et~al.}(1997){Hartkopf}, {McAlister}, {Mason}, {ten Brummelaar}, {Roberts}, {Turner}, \& {Wilson}}]{Hartkopf1997}
{Hartkopf}, W.~I., {McAlister}, H.~A., {Mason}, B.~D., {et~al.} 1997, \aj, 114, 1639

\bibitem[{{Hess} \& {Zarka}(2011)}]{Hess2011}
{Hess}, S.~L.~G. \& {Zarka}, P. 2011, \aap, 531, A29

\bibitem[{Hill {et~al.}(2021)Hill, Lee, MacQueen, Kelz, Drory, Vattiat, Good, Ramsey, Kriel, Peterson, DePoy, Gebhardt, Marshall, Tuttle, Bauer, Chonis, Fabricius, Froning, H\"{a}user, Indahl, Jahn, Landriau, Leck, Montesano, Prochaska, Snigula, Zeimann, Bryant, Damm, Fowler, Janowiecki, Martin, Mrozinski, Odewahn, Rostopchin, Shetrone, Spencer, Mentuch~Cooper, Armandroff, Bender, Dalton, Hopp, Komatsu, Nicklas, Ramsey, Roth, Schneider, Sneden, \& Steinmetz}]{Hill2021}
Hill, G.~J., Lee, H., MacQueen, P.~J., {et~al.} 2021, The Astronomical Journal, 162, 298

\bibitem[{Ilin \& Poppenhaeger(2022)}]{Ilin2022}
Ilin, E. \& Poppenhaeger, K. 2022, Monthly Notices of the Royal Astronomical Society, 513, 4579–4586

\bibitem[{{Ilin} {et~al.}(2024){Ilin}, {Poppenh{\"a}ger}, {Chebly}, {Ili{\'c}}, \& {Alvarado-G{\'o}mez}}]{Ilin2024}
{Ilin}, E., {Poppenh{\"a}ger}, K., {Chebly}, J., {Ili{\'c}}, N., \& {Alvarado-G{\'o}mez}, J.~D. 2024, \mnras, 527, 3395

\bibitem[{Ilin {et~al.}(2025)Ilin, Vedantham, Poppenh\"{a}ger, Bloot, Callingham, Brandeker, \& Chakraborty}]{Ilin2025}
Ilin, E., Vedantham, H.~K., Poppenh\"{a}ger, K., {et~al.} 2025, Nature

\bibitem[{{Jardine} {et~al.}(2017){Jardine}, {Vidotto}, \& {See}}]{jardine17}
{Jardine}, M., {Vidotto}, A.~A., \& {See}, V. 2017, \mnras, 465, L25

\bibitem[{{Johnston} {et~al.}(2008){Johnston}, {Taylor}, {Bailes}, {Bartel}, {Baugh}, {Bietenholz}, {Blake}, {Braun}, {Brown}, {Chatterjee}, {Darling}, {Deller}, {Dodson}, {Edwards}, {Ekers}, {Ellingsen}, {Feain}, {Gaensler}, {Haverkorn}, {Hobbs}, {Hopkins}, {Jackson}, {James}, {Joncas}, {Kaspi}, {Kilborn}, {Koribalski}, {Kothes}, {Landecker}, {Lenc}, {Lovell}, {Macquart}, {Manchester}, {Matthews}, {McClure-Griffiths}, {Norris}, {Pen}, {Phillips}, {Power}, {Protheroe}, {Sadler}, {Schmidt}, {Stairs}, {Staveley-Smith}, {Stil}, {Tingay}, {Tzioumis}, {Walker}, {Wall}, \& {Wolleben}}]{Johnston2008}
{Johnston}, S., {Taylor}, R., {Bailes}, M., {et~al.} 2008, Experimental Astronomy, 22, 151

\bibitem[{{Johnstone} \& {G{\"u}del}(2015)}]{johnstone15}
{Johnstone}, C.~P. \& {G{\"u}del}, M. 2015, \aap, 578, A129

\bibitem[{{Jones} {et~al.}(2024){Jones}, {Stef{\'a}nsson}, {Masuda}, {Libby-Roberts}, {Gardner}, {Holcomb}, {Beard}, {Robertson}, {Ca{\~n}as}, {Mahadevan}, {Kanodia}, {Lin}, {Kobulnicky}, {Parker}, {Bender}, {Cochran}, {Diddams}, {Fernandes}, {Gupta}, {Halverson}, {Hawley}, {Hearty}, {Hebb}, {Kowalski}, {Lubin}, {Monson}, {Ninan}, {Ramsey}, {Roy}, {Schwab}, {Terrien}, \& {Wisniewski}}]{jones2024}
{Jones}, S.~E., {Stef{\'a}nsson}, G., {Masuda}, K., {et~al.} 2024, \aj, 168, 93

\bibitem[{{Kanodia} {et~al.}(2023){Kanodia}, {Lin}, {Lubar}, {Halverson}, {Mahadevan}, {Bender}, {Logsdon}, {Ramsey}, {Ninan}, {Stef{\'a}nsson}, {Monson}, {Schwab}, {Roy}, {Paredes}, {Golub}, {Higuera}, {Klusmeyer}, {McBride}, {Blake}, {Diddams}, {Gris{\'e}}, {Gupta}, {Hearty}, {McElwain}, {Rajagopal}, {Robertson}, \& {Terrien}}]{Kanodia2023}
{Kanodia}, S., {Lin}, A. S.~J., {Lubar}, E., {et~al.} 2023, \aj, 166, 105

\bibitem[{Kanodia {et~al.}(2018)Kanodia, Mahadevan, Ramsey, Stefansson, Monson, Hearty, Blakeslee, Lubar, Bender, Ninan, Sterner, Roy, Halverson, \& Robertson}]{Kanodia018}
Kanodia, S., Mahadevan, S., Ramsey, L.~W., {et~al.} 2018, in Ground-based and Airborne Instrumentation for Astronomy VII, ed. H.~Takami, C.~J. Evans, \& L.~Simard (SPIE), 245

\bibitem[{{Kanodia} {et~al.}(2022){Kanodia}, {Ramsey}, {Maney}, {Mahadevan}, {Ca{\~n}as}, {Ninan}, {Monson}, {Kowalski}, {Goumas}, {Stefansson}, {Bender}, {Cochran}, {Diddams}, {Fredrick}, {Halverson}, {Hearty}, {Janowiecki}, {Metcalf}, {Odewahn}, {Robertson}, {Roy}, {Schwab}, \& {Terrien}}]{kanodia2022}
{Kanodia}, S., {Ramsey}, L.~W., {Maney}, M., {et~al.} 2022, \apj, 925, 155

\bibitem[{{Kavanagh} \& {Vedantham}(2023)}]{kavanagh2023}
{Kavanagh}, R.~D. \& {Vedantham}, H.~K. 2023, \mnras, 524, 6267

\bibitem[{{Kavanagh} \& {Vidotto}(2020)}]{kavanagh20}
{Kavanagh}, R.~D. \& {Vidotto}, A.~A. 2020, \mnras, 493, 1492

\bibitem[{{Kavanagh} {et~al.}(2021){Kavanagh}, {Vidotto}, {Klein}, {Jardine}, {Donati}, \& {{\'O} Fionnag{\'a}in}}]{kavanagh21}
{Kavanagh}, R.~D., {Vidotto}, A.~A., {Klein}, B., {et~al.} 2021, \mnras, 504, 1511

\bibitem[{{Kavanagh} {et~al.}(2022){Kavanagh}, {Vidotto}, {Vedantham}, {Jardine}, {Callingham}, \& {Morin}}]{Kavanagh2022}
{Kavanagh}, R.~D., {Vidotto}, A.~A., {Vedantham}, H.~K., {et~al.} 2022, \mnras, 514, 675

\bibitem[{{Kiman} {et~al.}(2019){Kiman}, {Schmidt}, {Angus}, {Cruz}, {Faherty}, \& {Rice}}]{Kiman2019}
{Kiman}, R., {Schmidt}, S.~J., {Angus}, R., {et~al.} 2019, \aj, 157, 231

\bibitem[{{Klein} {et~al.}(2021){Klein}, {Donati}, {H{\'e}brard}, {Zaire}, {Folsom}, {Morin}, {Delfosse}, \& {Bonfils}}]{klein21}
{Klein}, B., {Donati}, J.-F., {H{\'e}brard}, {\'E}.~M., {et~al.} 2021, \mnras, 500, 1844

\bibitem[{{Kochukhov}(2021)}]{kochukhov21}
{Kochukhov}, O. 2021, \aapr, 29, 1

\bibitem[{{Lamman} {et~al.}(2020){Lamman}, {Baranec}, {Berta-Thompson}, {Law}, {Schonhut-Stasik}, {Ziegler}, {Salama}, {Jensen-Clem}, {Duev}, {Riddle}, {Kulkarni}, {Winters}, \& {Irwin}}]{lamman2020}
{Lamman}, C., {Baranec}, C., {Berta-Thompson}, Z.~K., {et~al.} 2020, \aj, 159, 139

\bibitem[{{Lanzafame} {et~al.}(2023){Lanzafame}, {Brugaletta}, {Fr{\'e}mat}, {Sordo}, {Creevey}, {Andretta}, {Scandariato}, {Bus{\`a}}, {Distefano}, {Korn}, {de Laverny}, {Recio-Blanco}, {Abreu Aramburu}, {{\'A}lvarez}, {Andrae}, {Bailer-Jones}, {Bakker}, {Bellas-Velidis}, {Bijaoui}, {Brouillet}, {Burlacu}, {Carballo}, {Casamiquela}, {Chaoul}, {Chiavassa}, {Contursi}, {Cooper}, {Dafonte}, {Dapergolas}, {Delchambre}, {Demouchy}, {Dharmawardena}, {Drimmel}, {Edvardsson}, {Fouesneau}, {Garabato}, {Garc{\'\i}a-Lario}, {Garc{\'\i}a-Torres}, {Gavel}, {Gomez}, {Gonz{\'a}lez-Santamar{\'\i}a}, {Hatzidimitriou}, {Heiter}, {Jean-Antoine Piccolo}, {Kontizas}, {Kordopatis}, {Lebreton}, {Licata}, {Lindstr{\o}m}, {Livanou}, {Lobel}, {Lorca}, {Magdaleno Romeo}, {Manteiga}, {Marocco}, {Marshall}, {Mary}, {Nicolas}, {Ordenovic}, {Pailler}, {Palicio}, {Pallas-Quintela}, {Panem}, {Pichon}, {Poggio}, {Riclet}, {Robin}, {Rybizki}, {Santove{\~n}a}, {Sarro}, {Schultheis}, {Segol}, {Silvelo}, {Slezak}, {Smart}, {Soubiran},
  {S{\"u}veges}, {Th{\'e}venin}, {Torralba Elipe}, {Ulla}, {Utrilla}, {Vallenari}, {van Dillen}, {Zhao}, \& {Zorec}}]{Lanzafame2023}
{Lanzafame}, A.~C., {Brugaletta}, E., {Fr{\'e}mat}, Y., {et~al.} 2023, \aap, 674, A30

\bibitem[{Lecavelier~des Etangs {et~al.}(2013)Lecavelier~des Etangs, Sirothia, Gopal-Krishna, \& Zarka}]{LecavelierdesEtangs2013}
Lecavelier~des Etangs, A., Sirothia, S.~K., Gopal-Krishna, \& Zarka, P. 2013, Astronomy {\&} Astrophysics, 552, A65

\bibitem[{{Lehmann} {et~al.}(2024){Lehmann}, {Donati}, {Fouqu{\'e}}, {Moutou}, {Bellotti}, {Delfosse}, {Petit}, {Carmona}, {Morin}, {Vidotto}, \& {the SLS consortium}}]{lehmann24}
{Lehmann}, L.~T., {Donati}, J.~F., {Fouqu{\'e}}, P., {et~al.} 2024, \mnras, 527, 4330

\bibitem[{Mahadevan {et~al.}(2012)Mahadevan, Ramsey, Bender, Terrien, Wright, Halverson, Hearty, Nelson, Burton, Redman, Osterman, Diddams, Kasting, Endl, \& Deshpande}]{Mahadevan2012}
Mahadevan, S., Ramsey, L., Bender, C., {et~al.} 2012, in Ground-based and Airborne Instrumentation for Astronomy IV, ed. I.~S. McLean, S.~K. Ramsay, \& H.~Takami (SPIE)

\bibitem[{Mahadevan {et~al.}(2014)Mahadevan, Ramsey, Terrien, Halverson, Roy, Hearty, Levi, Stefansson, Robertson, Bender, Schwab, \& Nelson}]{Mahadevan2014}
Mahadevan, S., Ramsey, L.~W., Terrien, R., {et~al.} 2014, in Ground-based and Airborne Instrumentation for Astronomy V, ed. S.~K. Ramsay, I.~S. McLean, \& H.~Takami, Vol. 9147 (SPIE), 91471G

\bibitem[{{Mahadevan} {et~al.}(2021){Mahadevan}, {Stef{\'a}nsson}, {Robertson}, {Terrien}, {Ninan}, {Holcomb}, {Halverson}, {Cochran}, {Kanodia}, {Ramsey}, {Wolszczan}, {Endl}, {Bender}, {Diddams}, {Fredrick}, {Hearty}, {Monson}, {Metcalf}, {Roy}, \& {Schwab}}]{mahadevan2021}
{Mahadevan}, S., {Stef{\'a}nsson}, G., {Robertson}, P., {et~al.} 2021, \apjl, 919, L9

\bibitem[{Mann {et~al.}(2015)Mann, Feiden, Gaidos, Boyajian, \& Braun}]{Mann2015}
Mann, A.~W., Feiden, G.~A., Gaidos, E., Boyajian, T., \& Braun, K.~v. 2015, The Astrophysical Journal, 804, 64

\bibitem[{{Martin} {et~al.}(2017){Martin}, {Fuhrmeister}, {Mittag}, {Schmidt}, {Hempelmann}, {Gonz{\'a}lez-P{\'e}rez}, \& {Schmitt}}]{Martin2017}
{Martin}, J., {Fuhrmeister}, B., {Mittag}, M., {et~al.} 2017, \aap, 605, A113

\bibitem[{Martínez {et~al.}(2022)Martínez, Mauas, \& Buccino}]{martinez2022}
Martínez, C.~I., Mauas, P. J.~D., \& Buccino, A.~P. 2022, Monthly Notices of the Royal Astronomical Society, 512, 4835

\bibitem[{{Masci} {et~al.}(2019){Masci}, {Laher}, {Rusholme}, {Shupe}, {Groom}, {Surace}, {Jackson}, {Monkewitz}, {Beck}, {Flynn}, {Terek}, {Landry}, {Hacopians}, {Desai}, {Howell}, {Brooke}, {Imel}, {Wachter}, {Ye}, {Lin}, {Cenko}, {Cunningham}, {Rebbapragada}, {Bue}, {Miller}, {Mahabal}, {Bellm}, {Patterson}, {Juri{\'c}}, {Golkhou}, {Ofek}, {Walters}, {Graham}, {Kasliwal}, {Dekany}, {Kupfer}, {Burdge}, {Cannella}, {Barlow}, {Van Sistine}, {Giomi}, {Fremling}, {Blagorodnova}, {Levitan}, {Riddle}, {Smith}, {Helou}, {Prince}, \& {Kulkarni}}]{Masci2019}
{Masci}, F.~J., {Laher}, R.~R., {Rusholme}, B., {et~al.} 2019, \pasp, 131, 018003

\bibitem[{{Masuda} \& {Winn}(2020)}]{masuda2020}
{Masuda}, K. \& {Winn}, J.~N. 2020, \aj, 159, 81

\bibitem[{{Morin} {et~al.}(2008){Morin}, {Donati}, {Petit}, {Delfosse}, {Forveille}, {Albert}, {Auri{\`e}re}, {Cabanac}, {Dintrans}, {Fares}, {Gastine}, {Jardine}, {Ligni{\`e}res}, {Paletou}, {Ramirez Velez}, \& {Th{\'e}ado}}]{Morin2008}
{Morin}, J., {Donati}, J.~F., {Petit}, P., {et~al.} 2008, \mnras, 390, 567

\bibitem[{{Newton} {et~al.}(2017){Newton}, {Irwin}, {Charbonneau}, {Berlind}, {Calkins}, \& {Mink}}]{Newton2017}
{Newton}, E.~R., {Irwin}, J., {Charbonneau}, D., {et~al.} 2017, \apj, 834, 85

\bibitem[{Ninan {et~al.}(2019)Ninan, Mahadevan, Stefansson, Bender, Roy, Kaplan, Fredrick, Metcalf, Monson, Terrien, Ramsey, \& Diddams}]{Ninan2019}
Ninan, J.~P., Mahadevan, S., Stefansson, G., {et~al.} 2019, Journal of Astronomical Telescopes, Instruments, and Systems, 5, 1

\bibitem[{{Oliver} {et~al.}(2000){Oliver}, {Rowan-Robinson}, {Alexander}, {Almaini}, {Balcells}, {Baker}, {Barcons}, {Barden}, {Bellas-Velidis}, {Cabrera-Guerra}, {Carballo}, {Cesarsky}, {Ciliegi}, {Clements}, {Crockett}, {Danese}, {Dapergolas}, {Drolias}, {Eaton}, {Efstathiou}, {Egami}, {Elbaz}, {Fadda}, {Fox}, {Franceschini}, {Genzel}, {Goldschmidt}, {Graham}, {Gonzalez-Serrano}, {Gonzalez-Solares}, {Granato}, {Gruppioni}, {Herbstmeier}, {H{\'e}raudeau}, {Joshi}, {Kontizas}, {Kontizas}, {Kotilainen}, {Kunze}, {La Franca}, {Lari}, {Lawrence}, {Lemke}, {Linden-V{\o}rnle}, {Mann}, {M{\'a}rquez}, {Masegosa}, {Mattila}, {McMahon}, {Miley}, {Missoulis}, {Mobasher}, {Morel}, {N{\o}rgaard-Nielsen}, {Omont}, {Papadopoulos}, {Perez-Fournon}, {Puget}, {Rigopoulou}, {Rocca-Volmerange}, {Serjeant}, {Silva}, {Sumner}, {Surace}, {Vaisanen}, {van der Werf}, {Verma}, {Vigroux}, {Villar-Martin}, \& {Willott}}]{Oliver2000}
{Oliver}, S., {Rowan-Robinson}, M., {Alexander}, D.~M., {et~al.} 2000, \mnras, 316, 749

\bibitem[{Owen \& Adams(2014)}]{Owen2014}
Owen, J.~E. \& Adams, F.~C. 2014, Monthly Notices of the Royal Astronomical Society, 444, 3761–3779

\bibitem[{Parviainen(2016)}]{pyde}
Parviainen, H. 2016, PyDE: v1.5

\bibitem[{{Penoyre} {et~al.}(2022){Penoyre}, {Belokurov}, \& {Evans}}]{Penoyre2022}
{Penoyre}, Z., {Belokurov}, V., \& {Evans}, N.~W. 2022, \mnras, 513, 2437

\bibitem[{{P{\'e}rez-Torres} {et~al.}(2021){P{\'e}rez-Torres}, {G{\'o}mez}, {Ortiz}, {Leto}, {Anglada}, {G{\'o}mez}, {Rodr{\'\i}guez}, {Trigilio}, {Amado}, {Alberdi}, {Anglada-Escud{\'e}}, {Osorio}, {Umana}, {Berdi{\~n}as}, {L{\'o}pez-Gonz{\'a}lez}, {Morales}, {Rodr{\'\i}guez-L{\'o}pez}, \& {Chibueze}}]{Perez-Torres2021}
{P{\'e}rez-Torres}, M., {G{\'o}mez}, J.~F., {Ortiz}, J.~L., {et~al.} 2021, \aap, 645, A77

\bibitem[{{Perger} {et~al.}(2021){Perger}, {Ribas}, {Anglada-Escud{\'e}}, {Morales}, {Amado}, {Caballero}, {Quirrenbach}, {Reiners}, {B{\'e}jar}, {Dreizler}, {Galad{\'\i}-Enr{\'\i}quez}, {Hatzes}, {Henning}, {Jeffers}, {Kaminski}, {K{\"u}rster}, {Lafarga}, {Montes}, {Pall{\'e}}, {Rodr{\'\i}guez-L{\'o}pez}, {Schweitzer}, {Zapatero Osorio}, \& {Zechmeister}}]{perger2021}
{Perger}, M., {Ribas}, I., {Anglada-Escud{\'e}}, G., {et~al.} 2021, \aap, 649, L12

\bibitem[{Pineda \& Villadsen(2023)}]{Pineda2023}
Pineda, J.~S. \& Villadsen, J. 2023, Nature Astronomy, 7, 569–578

\bibitem[{Pope {et~al.}(2021)Pope, Callingham, Feinstein, G\"{u}nther, Vedantham, Ansdell, \& Shimwell}]{Pope2021}
Pope, B. J.~S., Callingham, J.~R., Feinstein, A.~D., {et~al.} 2021, The Astrophysical Journal Letters, 919, L10

\bibitem[{{Presa} {et~al.}(2024){Presa}, {Driessen}, \& {Vidotto}}]{Presa2024}
{Presa}, A., {Driessen}, F.~A., \& {Vidotto}, A.~A. 2024, \mnras, 534, 3622

\bibitem[{{Quirrenbach} {et~al.}(2020){Quirrenbach}, {CARMENES Consortium}, {Amado}, {Ribas}, {Reiners}, {Caballero}, {Aceituno}, {Alacid}, {Alonso-Floriano}, {Anglada-Escud{\'e}}, {Azzaro}, {Baroch}, {Bauer}, {Becerril}, {B{\'e}jar}, {Bluhm}, {Calvo Ortega}, {Cardona Guill{\'e}n}, {Casasayas-Barris}, {Chaturvedi}, {Cifuentes}, {Colom{\'e}}, {Conte}, {Cort{\'e}s-Contreras}, {Czesla}, {D{\'\i}ez-Alonso}, {Dom{\'\i}nguez Fern{\'a}ndez}, {Dreizler}, {Duque-Arribas}, {Espinoza}, {Fuhrmeister}, {Galad{\'\i}-Enr{\'\i}quez}, {Gar{\textasciiacute}a Quintana}, {Gonz{\'a}lez-Alvare}, {Gonz{\'a}lez Cuesta}, {Gonz{\'a}lez Hern{\'a}ndez}, {Guenther}, {de Guindos}, {Hatzes}, {Henning}, {Herbort}, {Herrero}, {Hintz}, {Iglesias-P{\'a}ra}, {Jeffers}, {Johnson}, {de Juan}, {Kaminski}, {Kemmer}, {Khaimova}, {Khalafinejad}, {Klahr}, {Kossakowski}, {Kreidberg}, {K{\"u}rster}, {Labarga}, {Lafarga}, {Lamp{\'o}n}, {Lara}, {Lillo-Box}, {Lodieu}, {L{\'o}pez Gallifa}, {L{\'o}pez Gonz{\'a}lez}, {L{\'o}pez-Puertas}, {Luque}, {Marfil},
  {Mart{\'\i}n-Ruiz}, {Matth{\'e}}, {Molaverdikhani}, {Montes}, {Morales}, {Morales-Calder{\'o}on}, {Nagel}, {Nortmann}, {Nowak}, {Ofir}, {Oshaghi}, {Pall{\'e}}, {Passegger}, {Pavlov}, {Pedraz}, {Perdelwitz}, {Perger}, {Reffert}, {Revilla}, {Rodr{\'\i}guez}, {Rodr{\'\i}guez L{\'o}pez}, {Sabotta}, {Sadegi}, {Sairam}, {Salz}, {S{\'a}nchez-L{\'o}pez}, {Sanz-Forcada}, {Sarkis}, {Sch{\"a}fer}, {Schiller}, {Schlecker}, {Schmitt}, {Sch{\"o}fer}, {Schweitzer}, {Seiferta}, {Shan}, {Shulyak}, {Skrzypinski}, {Solano}, {Soto}, {Stahl}, {Stangret}, {Stock}, {Strachan}, {Stuber}, {St{\"u}rmer}, {Tabernero}, {Tal-Or}, {Tala-Pinto}, {Trifonov}, {Vanaverbeke}, {Yan}, {Zapatero Osorio}, \& {Zechmeister}}]{Quirrenbach2020}
{Quirrenbach}, A., {CARMENES Consortium}, {Amado}, P.~J., {et~al.} 2020, in Society of Photo-Optical Instrumentation Engineers (SPIE) Conference Series, Vol. 11447, Ground-based and Airborne Instrumentation for Astronomy VIII, ed. C.~J. {Evans}, J.~J. {Bryant}, \& K.~{Motohara}, 114473C

\bibitem[{Ramsey {et~al.}(1998)Ramsey, Adams, Barnes~III, Booth, Cornell, Fowler, Gaffney, Glaspey, Good, Hill, Kelton, Krabbendam, Long, MacQueen, Ray, Ricklefs, Sage, Sebring, Spiesman, \& Steiner}]{Ramsey1998}
Ramsey, L.~W., Adams, M.~T., Barnes~III, T.~G., {et~al.} 1998, in Advanced Technology Optical/IR Telescopes VI, ed. L.~M. Stepp (SPIE)

\bibitem[{{Reiners}(2007)}]{Reiners2007}
{Reiners}, A. 2007, \aap, 467, 259

\bibitem[{{Reiners} {et~al.}(2022){Reiners}, {Shulyak}, {K{\"a}pyl{\"a}}, {Ribas}, {Nagel}, {Zechmeister}, {Caballero}, {Shan}, {Fuhrmeister}, {Quirrenbach}, {Amado}, {Montes}, {Jeffers}, {Azzaro}, {B{\'e}jar}, {Chaturvedi}, {Henning}, {K{\"u}rster}, \& {Pall{\'e}}}]{reiners22}
{Reiners}, A., {Shulyak}, D., {K{\"a}pyl{\"a}}, P.~J., {et~al.} 2022, \aap, 662, A41

\bibitem[{Ribas {et~al.}(2023)Ribas, Reiners, Zechmeister, Caballero, Morales, Sabotta, Baroch, Amado, Quirrenbach, Abril, Aceituno, Anglada-Escudé, Azzaro, Barrado, Béjar, Benítez~de Haro, Bergond, Bluhm, Calvo~Ortega, Cardona~Guillén, Chaturvedi, Cifuentes, Colomé, Cont, Cortés-Contreras, Czesla, Díez-Alonso, Dreizler, Duque-Arribas, Espinoza, Fernández, Fuhrmeister, Galadí-Enríquez, García-López, González-Álvarez, González~Hernández, Guenther, de~Guindos, Hatzes, Henning, Herrero, Hintz, Huelmo, Jeffers, Johnson, de~Juan, Kaminski, Kemmer, Khaimova, Khalafinejad, Kossakowski, K\"{u}rster, Labarga, Lafarga, Lalitha, Lampón, Lillo-Box, Lodieu, López~González, López-Puertas, Luque, Magán, Mancini, Marfil, Martín, Martín-Ruiz, Molaverdikhani, Montes, Nagel, Nortmann, Nowak, Pallé, Passegger, Pavlov, Pedraz, Perdelwitz, Perger, Ramón-Ballesta, Reffert, Revilla, Rodríguez, Rodríguez-López, Sadegi, Sánchez~Carrasco, Sánchez-López, Sanz-Forcada, Sch\"{a}fer, Schlecker, Schmitt,
  Sch\"{o}fer, Schweitzer, Seifert, Shan, Skrzypinski, Solano, Stahl, Stangret, Stock, St\"{u}rmer, Tabernero, Tal-Or, Trifonov, Vanaverbeke, Yan, \& Zapatero~Osorio}]{Ribas2023}
Ribas, I., Reiners, A., Zechmeister, M., {et~al.} 2023, Astronomy {\&} Astrophysics, 670, A139

\bibitem[{{Robertson} {et~al.}(2019){Robertson}, {Anderson}, {Stefansson}, {Hearty}, {Monson}, {Mahadevan}, {Blakeslee}, {Bender}, {Ninan}, {Conran}, {Levi}, {Lubar}, {Cole}, {Dykhouse}, {Kanodia}, {Nitroy}, {Smolsky}, {Tuggle}, {Blank}, {Nelson}, {Blake}, {Halverson}, {Henderson}, {Kaplan}, {Li}, {Logsdon}, {McElwain}, {Rajagopal}, {Ramsey}, {Roy}, {Schwab}, {Terrien}, \& {Wright}}]{Robertson2019}
{Robertson}, P., {Anderson}, T., {Stefansson}, G., {et~al.} 2019, Journal of Astronomical Telescopes, Instruments, and Systems, 5, 015003

\bibitem[{{Robertson} {et~al.}(2016){Robertson}, {Bender}, {Mahadevan}, {Roy}, \& {Ramsey}}]{Robertson2016}
{Robertson}, P., {Bender}, C., {Mahadevan}, S., {Roy}, A., \& {Ramsey}, L.~W. 2016, \apj, 832, 112

\bibitem[{{Robertson} {et~al.}(2020){Robertson}, {Stefansson}, {Mahadevan}, {Endl}, {Cochran}, {Beard}, {Bender}, {Diddams}, {Duong}, {Ford}, {Fredrick}, {Halverson}, {Hearty}, {Holcomb}, {Juan}, {Kanodia}, {Lubin}, {Metcalf}, {Monson}, {Ninan}, {Palafoutas}, {Ramsey}, {Roy}, {Schwab}, {Terrien}, \& {Wright}}]{Robertson2020}
{Robertson}, P., {Stefansson}, G., {Mahadevan}, S., {et~al.} 2020, \apj, 897, 125

\bibitem[{{Rosenthal} {et~al.}(2021){Rosenthal}, {Fulton}, {Hirsch}, {Isaacson}, {Howard}, {Dedrick}, {Sherstyuk}, {Blunt}, {Petigura}, {Knutson}, {Behmard}, {Chontos}, {Crepp}, {Crossfield}, {Dalba}, {Fischer}, {Henry}, {Kane}, {Kosiarek}, {Marcy}, {Rubenzahl}, {Weiss}, \& {Wright}}]{rosenthal2021}
{Rosenthal}, L.~J., {Fulton}, B.~J., {Hirsch}, L.~A., {et~al.} 2021, \apjs, 255, 8

\bibitem[{{Rucinski}(1999)}]{Rucinski1999}
{Rucinski}, S. 1999, Turkish Journal of Physics, 23, 271

\bibitem[{{Rucinski}(1992)}]{Rucinski1992}
{Rucinski}, S.~M. 1992, \aj, 104, 1968

\bibitem[{{Sabater} {et~al.}(2021){Sabater}, {Best}, {Tasse}, {Hardcastle}, {Shimwell}, {Nisbet}, {Jelic}, {Callingham}, {R{\"o}ttgering}, {Bonato}, {Bondi}, {Ciardi}, {Cochrane}, {Jarvis}, {Kondapally}, {Koopmans}, {O'Sullivan}, {Prandoni}, {Schwarz}, {Smith}, {Wang}, {Williams}, \& {Zaroubi}}]{Sabater2021}
{Sabater}, J., {Best}, P.~N., {Tasse}, C., {et~al.} 2021, \aap, 648, A2

\bibitem[{{Saur} {et~al.}(2013){Saur}, {Grambusch}, {Duling}, {Neubauer}, \& {Simon}}]{saur13}
{Saur}, J., {Grambusch}, T., {Duling}, S., {Neubauer}, F.~M., \& {Simon}, S. 2013, \aap, 552, A119

\bibitem[{{Sch{\"o}fer} {et~al.}(2019){Sch{\"o}fer}, {Jeffers}, {Reiners}, {Shulyak}, {Fuhrmeister}, {Johnson}, {Zechmeister}, {Ribas}, {Quirrenbach}, {Amado}, {Caballero}, {Anglada-Escud{\'e}}, {Bauer}, {B{\'e}jar}, {Cort{\'e}s-Contreras}, {Dreizler}, {Guenther}, {Kaminski}, {K{\"u}rster}, {Lafarga}, {Montes}, {Morales}, {Pedraz}, \& {Tal-Or}}]{Schofer2019}
{Sch{\"o}fer}, P., {Jeffers}, S.~V., {Reiners}, A., {et~al.} 2019, \aap, 623, A44

\bibitem[{Schwab {et~al.}(2016)Schwab, Rakich, Gong, Mahadevan, Halverson, Roy, Terrien, Robertson, Hearty, Levi, Monson, Wright, McElwain, Bender, Blake, St\"{u}rmer, Gurevich, Chakraborty, \& Ramsey}]{Schwab2016}
Schwab, C., Rakich, A., Gong, Q., {et~al.} 2016, in Ground-based and Airborne Instrumentation for Astronomy VI, ed. C.~J. Evans, L.~Simard, \& H.~Takami, Vol. 9908 (SPIE), 99087H

\bibitem[{Shetrone {et~al.}(2007)Shetrone, Cornell, Fowler, Gaffney, Laws, Mader, Mason, Odewahn, Roman, Rostopchin, Schneider, Umbarger, \& Westfall}]{Shetrone2007}
Shetrone, M., Cornell, M., Fowler, J., {et~al.} 2007, Publications of the Astronomical Society of the Pacific, 119, 556–566

\bibitem[{{Shields} {et~al.}(2016){Shields}, {Ballard}, \& {Johnson}}]{Shields2016}
{Shields}, A.~L., {Ballard}, S., \& {Johnson}, J.~A. 2016, \physrep, 663, 1

\bibitem[{Shimwell {et~al.}(2019)Shimwell, Tasse, Hardcastle, Mechev, Williams, Best, R\"{o}ttgering, Callingham, Dijkema, de~Gasperin, Hoang, Hugo, Mirmont, Oonk, Prandoni, Rafferty, Sabater, Smirnov, van Weeren, White, Atemkeng, Bester, Bonnassieux, Br\"{u}ggen, Brunetti, Chyży, Cochrane, Conway, Croston, Danezi, Duncan, Haverkorn, Heald, Iacobelli, Intema, Jackson, Jamrozy, Jarvis, Lakhoo, Mevius, Miley, Morabito, Morganti, Nisbet, Orrú, Perkins, Pizzo, Schrijvers, Smith, Vermeulen, Wise, Alegre, Bacon, van Bemmel, Beswick, Bonafede, Botteon, Bourke, Brienza, Calistro~Rivera, Cassano, Clarke, Conselice, Dettmar, Drabent, Dumba, Emig, Enßlin, Ferrari, Garrett, Génova-Santos, Goyal, G\"{u}rkan, Hale, Harwood, Heesen, Hoeft, Horellou, Jackson, Kokotanekov, Kondapally, Kunert-Bajraszewska, Mahatma, Mahony, Mandal, McKean, Merloni, Mingo, Miskolczi, Mooney, Nikiel-Wroczyński, O’Sullivan, Quinn, Reich, Roskowiński, Rowlinson, Savini, Saxena, Schwarz, Shulevski, Sridhar, Stacey, Urquhart, van~der Wiel,
  Varenius, Webster, \& Wilber}]{Shimwell2019}
Shimwell, T.~W., Tasse, C., Hardcastle, M.~J., {et~al.} 2019, Astronomy {\&} Astrophysics, 622, A1

\bibitem[{{Shkolnik}(2004)}]{Shkolnik2004}
{Shkolnik}, E. 2004, PhD thesis, University of British Columbia, Canada

\bibitem[{{Shkolnik} {et~al.}(2010){Shkolnik}, {Hebb}, {Liu}, {Reid}, \& {Collier Cameron}}]{Shkolnik2010}
{Shkolnik}, E.~L., {Hebb}, L., {Liu}, M.~C., {Reid}, I.~N., \& {Collier Cameron}, A. 2010, \apj, 716, 1522

\bibitem[{{Shulyak} {et~al.}(2010){Shulyak}, {Reiners}, {Wende}, {Kochukhov}, {Piskunov}, \& {Seifahrt}}]{Shulyak2010}
{Shulyak}, D., {Reiners}, A., {Wende}, S., {et~al.} 2010, \aap, 523, A37

\bibitem[{{Skinner} {et~al.}(2018){Skinner}, {Covey}, {Bender}, {Rivera}, {De Lee}, {Souto}, {Chojnowski}, {Troup}, {Badenes}, {Bizyaev}, {Blake}, {Burgasser}, {Ca{\~n}as}, {Carlberg}, {G{\'o}mez Maqueo Chew}, {Deshpande}, {Fleming}, {Fern{\'a}ndez-Trincado}, {Garc{\'\i}a-Hern{\'a}ndez}, {Hearty}, {Kounkel}, {Longa-Pe{\~n}e}, {Mahadevan}, {Majewski}, {Minniti}, {Nidever}, {Oravetz}, {Pan}, {Stassun}, {Terrien}, \& {Zamora}}]{skinner2018}
{Skinner}, J., {Covey}, K.~R., {Bender}, C.~F., {et~al.} 2018, \aj, 156, 45

\bibitem[{{Sperauskas} {et~al.}(2019){Sperauskas}, {Deveikis}, \& {Tokovinin}}]{Sperauskas2019}
{Sperauskas}, J., {Deveikis}, V., \& {Tokovinin}, A. 2019, \aap, 626, A31

\bibitem[{{Stassun} {et~al.}(2018){Stassun}, {Oelkers}, {Pepper}, {Paegert}, {De Lee}, {Torres}, {Latham}, {Charpinet}, {Dressing}, {Huber}, {Kane}, {L{\'e}pine}, {Mann}, {Muirhead}, {Rojas-Ayala}, {Silvotti}, {Fleming}, {Levine}, \& {Plavchan}}]{Stassun2018}
{Stassun}, K.~G., {Oelkers}, R.~J., {Pepper}, J., {et~al.} 2018, \aj, 156, 102

\bibitem[{Stefansson {et~al.}(2020)Stefansson, Cañas, Wisniewski, Robertson, Mahadevan, Maney, Kanodia, Beard, Bender, Brunt, Clemens, Cochran, Diddams, Endl, Ford, Fredrick, Halverson, Hearty, Hebb, Huehnerhoff, Jennings, Kaplan, Levi, Lubar, Metcalf, Monson, Morris, Ninan, Nitroy, Ramsey, Roy, Schwab, Sigurdsson, Terrien, \& Wright}]{Stefansson2020_feb}
Stefansson, G., Cañas, C., Wisniewski, J., {et~al.} 2020, The Astronomical Journal, 159, 100

\bibitem[{Stefansson {et~al.}(2016)Stefansson, Hearty, Robertson, Mahadevan, Anderson, Levi, Bender, Nelson, Monson, Blank, Halverson, Henderson, Ramsey, Roy, Schwab, \& Terrien}]{Stefansson2016}
Stefansson, G., Hearty, F., Robertson, P., {et~al.} 2016, The Astrophysical Journal, 833, 175

\bibitem[{{Stefansson} {et~al.}(2020){Stefansson}, {Mahadevan}, {Maney}, {Ninan}, {Robertson}, {Rajagopal}, {Haase}, {Allen}, {Ford}, {Winn}, {Wolfgang}, {Dawson}, {Wisniewski}, {Bender}, {Ca{\~n}as}, {Cochran}, {Diddams}, {Fredrick}, {Halverson}, {Hearty}, {Hebb}, {Kanodia}, {Levi}, {Metcalf}, {Monson}, {Ramsey}, {Roy}, {Schwab}, {Terrien}, \& {Wright}}]{stefansson2020k225}
{Stefansson}, G., {Mahadevan}, S., {Maney}, M., {et~al.} 2020, \aj, 160, 192

\bibitem[{{Stef{\'a}nsson} {et~al.}(2023){Stef{\'a}nsson}, {Mahadevan}, {Miguel}, {Robertson}, {Delamer}, {Kanodia}, {Ca{\~n}as}, {Winn}, {Ninan}, {Terrien}, {Holcomb}, {Ford}, {Zawadzki}, {Bowler}, {Bender}, {Cochran}, {Diddams}, {Endl}, {Fredrick}, {Halverson}, {Hearty}, {Hill}, {Lin}, {Metcalf}, {Monson}, {Ramsey}, {Roy}, {Schwab}, {Wright}, \& {Zeimann}}]{Stefansson2023}
{Stef{\'a}nsson}, G., {Mahadevan}, S., {Miguel}, Y., {et~al.} 2023, Science, 382, 1031

\bibitem[{{Stef{\`a}nsson} {et~al.}(2022){Stef{\`a}nsson}, {Mahadevan}, {Petrovich}, {Winn}, {Kanodia}, {Millholland}, {Maney}, {Ca{\~n}as}, {Wisniewski}, {Robertson}, {Ninan}, {Ford}, {Bender}, {Blake}, {Cegla}, {Cochran}, {Diddams}, {Dong}, {Endl}, {Fredrick}, {Halverson}, {Hearty}, {Hebb}, {Hirano}, {Lin}, {Logsdon}, {Lubar}, {McElwain}, {Metcalf}, {Monson}, {Rajagopal}, {Ramsey}, {Roy}, {Schwab}, {Schweiker}, {Terrien}, \& {Wright}}]{Stefansson2022_jun}
{Stef{\`a}nsson}, G., {Mahadevan}, S., {Petrovich}, C., {et~al.} 2022, \apjl, 931, L15

\bibitem[{{Stef{\'a}nsson} {et~al.}(2025){Stef{\'a}nsson}, {Mahadevan}, {Winn}, {Marcussen}, {Kanodia}, {Albrecht}, {Fitzmaurice}, {Mikulskyt{\.{e}}}, {Ca{\~n}as}, {Espinoza-Retamal}, {Zwart}, {Krolikowski}, {Hotnisky}, {Robertson}, {Alvarado-Montes}, {Bender}, {Blake}, {Callingham}, {Cochran}, {Delamer}, {Diddams}, {Dong}, {Fernandes}, {Giovinazzi}, {Halverson}, {Libby-Roberts}, {Logsdon}, {McElwain}, {Ninan}, {Rajagopal}, {Reji}, {Roy}, {Schwab}, \& {Wright}}]{stefansson2025}
{Stef{\'a}nsson}, G., {Mahadevan}, S., {Winn}, J.~N., {et~al.} 2025, \aj, 169, 107

\bibitem[{{Stefansson} {et~al.}(2024){Stefansson}, {Patapis}, {Callingham}, {Cugno}, {Delamer}, {Mahadevan}, {Pope}, \& {Vedantham}}]{Stefansson2024_jwst}
{Stefansson}, G., {Patapis}, P.~A., {Callingham}, J., {et~al.} 2024, {Resolving the Radio Riddle: Unveiling the Origins of Radio Emission in a Red Dwarf and its Wide-Orbit Companion}, JWST Proposal. Cycle 3, ID. \#5497

\bibitem[{Suárez~Mascareño {et~al.}(2017)Suárez~Mascareño, González~Hernández, Rebolo, Velasco, Toledo-Padrón, Affer, Perger, Micela, Ribas, Maldonado, Leto, Zanmar~Sanchez, Scandariato, Damasso, Sozzetti, Esposito, Covino, Maggio, Lanza, Desidera, Rosich, Bignamini, Claudi, Benatti, Borsa, Pedani, Molinari, Morales, Herrero, \& Lafarga}]{SurezMascareo2017}
Suárez~Mascareño, A., González~Hernández, J.~I., Rebolo, R., {et~al.} 2017, Astronomy {\&} Astrophysics, 605, A92

\bibitem[{{Tamazian} {et~al.}(2008){Tamazian}, {Docobo}, {Balega}, {Melikian}, {Maximov}, \& {Malogolovets}}]{Tamazian2008}
{Tamazian}, V.~S., {Docobo}, J.~A., {Balega}, Y.~Y., {et~al.} 2008, \aj, 136, 974

\bibitem[{{Toet} {et~al.}(2021){Toet}, {Vedantham}, {Callingham}, {Veken}, {Shimwell}, {Zarka}, {R{\"o}ttgering}, \& {Drabent}}]{Toet2021}
{Toet}, S.~E.~B., {Vedantham}, H.~K., {Callingham}, J.~R., {et~al.} 2021, \aap, 654, A21

\bibitem[{Tofflemire(2019)}]{Tofflemire2019}
Tofflemire, B.~M. 2019, tofflemire/saphires: Zenodo archive

\bibitem[{{Treumann}(2006)}]{Treumann2006}
{Treumann}, R.~A. 2006, \aapr, 13, 229

\bibitem[{{Trigilio} {et~al.}(2023){Trigilio}, {Biswas}, {Leto}, {Umana}, {Busa}, {Cavallaro}, {Das}, {Chandra}, {Perez-Torres}, {Wade}, {Bordiu}, {Buemi}, {Bufano}, {Ingallinera}, {Loru}, \& {Riggi}}]{Trigilio2023}
{Trigilio}, C., {Biswas}, A., {Leto}, P., {et~al.} 2023, arXiv e-prints, arXiv:2305.00809

\bibitem[{{Tuomi} {et~al.}(2018){Tuomi}, {Jones}, {Barnes}, {Anglada-Escud{\'e}}, {Butler}, {Kiraga}, \& {Vogt}}]{Tuomi2018}
{Tuomi}, M., {Jones}, H. R.~A., {Barnes}, J.~R., {et~al.} 2018, \aj, 155, 192

\bibitem[{Turner {et~al.}(2021)Turner, Zarka, Grießmeier, Lazio, Cecconi, Emilio~Enriquez, Girard, Jayawardhana, Lamy, Nichols, \& de~Pater}]{Turner2021}
Turner, J.~D., Zarka, P., Grießmeier, J.-M., {et~al.} 2021, Astronomy {\&} Astrophysics, 645, A59

\bibitem[{{van Haarlem} {et~al.}(2013){van Haarlem}, {Wise}, {Gunst}, {Heald}, {McKean}, {Hessels}, {de Bruyn}, {Nijboer}, {Swinbank}, {Fallows}, {Brentjens}, {Nelles}, {Beck}, {Falcke}, {Fender}, {H{\"o}randel}, {Koopmans}, {Mann}, {Miley}, {R{\"o}ttgering}, {Stappers}, {Wijers}, {Zaroubi}, {van den Akker}, {Alexov}, {Anderson}, {Anderson}, {van Ardenne}, {Arts}, {Asgekar}, {Avruch}, {Batejat}, {B{\"a}hren}, {Bell}, {Bell}, {van Bemmel}, {Bennema}, {Bentum}, {Bernardi}, {Best}, {B{\^\i}rzan}, {Bonafede}, {Boonstra}, {Braun}, {Bregman}, {Breitling}, {van de Brink}, {Broderick}, {Broekema}, {Brouw}, {Br{\"u}ggen}, {Butcher}, {van Cappellen}, {Ciardi}, {Coenen}, {Conway}, {Coolen}, {Corstanje}, {Damstra}, {Davies}, {Deller}, {Dettmar}, {van Diepen}, {Dijkstra}, {Donker}, {Doorduin}, {Dromer}, {Drost}, {van Duin}, {Eisl{\"o}ffel}, {van Enst}, {Ferrari}, {Frieswijk}, {Gankema}, {Garrett}, {de Gasperin}, {Gerbers}, {de Geus}, {Grie{\ss}meier}, {Grit}, {Gruppen}, {Hamaker}, {Hassall}, {Hoeft}, {Holties},
  {Horneffer}, {van der Horst}, {van Houwelingen}, {Huijgen}, {Iacobelli}, {Intema}, {Jackson}, {Jelic}, {de Jong}, {Juette}, {Kant}, {Karastergiou}, {Koers}, {Kollen}, {Kondratiev}, {Kooistra}, {Koopman}, {Koster}, {Kuniyoshi}, {Kramer}, {Kuper}, {Lambropoulos}, {Law}, {van Leeuwen}, {Lemaitre}, {Loose}, {Maat}, {Macario}, {Markoff}, {Masters}, {McFadden}, {McKay-Bukowski}, {Meijering}, {Meulman}, {Mevius}, {Middelberg}, {Millenaar}, {Miller-Jones}, {Mohan}, {Mol}, {Morawietz}, {Morganti}, {Mulcahy}, {Mulder}, {Munk}, {Nieuwenhuis}, {van Nieuwpoort}, {Noordam}, {Norden}, {Noutsos}, {Offringa}, {Olofsson}, {Omar}, {Orr{\'u}}, {Overeem}, {Paas}, {Pandey-Pommier}, {Pandey}, {Pizzo}, {Polatidis}, {Rafferty}, {Rawlings}, {Reich}, {de Reijer}, {Reitsma}, {Renting}, {Riemers}, {Rol}, {Romein}, {Roosjen}, {Ruiter}, {Scaife}, {van der Schaaf}, {Scheers}, {Schellart}, {Schoenmakers}, {Schoonderbeek}, {Serylak}, {Shulevski}, {Sluman}, {Smirnov}, {Sobey}, {Spreeuw}, {Steinmetz}, {Sterks}, {Stiepel}, {Stuurwold},
  {Tagger}, {Tang}, {Tasse}, {Thomas}, {Thoudam}, {Toribio}, {van der Tol}, {Usov}, {van Veelen}, {van der Veen}, {ter Veen}, {Verbiest}, {Vermeulen}, {Vermaas}, {Vocks}, {Vogt}, {de Vos}, {van der Wal}, {van Weeren}, {Weggemans}, {Weltevrede}, {White}, {Wijnholds}, {Wilhelmsson}, {Wucknitz}, {Yatawatta}, {Zarka}, \& {Zensus}}]{Haarlem2013}
{van Haarlem}, M.~P., {Wise}, M.~W., {Gunst}, A.~W., {et~al.} 2013, \aap, 556, A2

\bibitem[{Vedantham {et~al.}(2022)Vedantham, Callingham, Shimwell, Benz, Hajduk, Ray, Tasse, \& Drabent}]{Vedantham2022}
Vedantham, H.~K., Callingham, J.~R., Shimwell, T.~W., {et~al.} 2022, The Astrophysical Journal Letters, 926, L30

\bibitem[{{Vedantham} {et~al.}(2020{\natexlab{a}}){Vedantham}, {Callingham}, {Shimwell}, {Dupuy}, {Best}, {Liu}, {Zhang}, {De}, {Lamy}, {Zarka}, {R{\"o}ttgering}, \& {Shulevski}}]{Vedantham2020_nov}
{Vedantham}, H.~K., {Callingham}, J.~R., {Shimwell}, T.~W., {et~al.} 2020{\natexlab{a}}, \apjl, 903, L33

\bibitem[{{Vedantham} {et~al.}(2020{\natexlab{b}}){Vedantham}, {Callingham}, {Shimwell}, {Tasse}, {Pope}, {Bedell}, {Snellen}, {Best}, {Hardcastle}, {Haverkorn}, {Mechev}, {O'Sullivan}, {R{\"o}ttgering}, \& {White}}]{Vedantham2020}
{Vedantham}, H.~K., {Callingham}, J.~R., {Shimwell}, T.~W., {et~al.} 2020{\natexlab{b}}, Nature Astronomy, 4, 577

\bibitem[{{Vidotto} {et~al.}(2014){Vidotto}, {Gregory}, {Jardine}, {Donati}, {Petit}, {Morin}, {Folsom}, {Bouvier}, {Cameron}, {Hussain}, {Marsden}, {Waite}, {Fares}, {Jeffers}, \& {do Nascimento}}]{Vidotto2014}
{Vidotto}, A.~A., {Gregory}, S.~G., {Jardine}, M., {et~al.} 2014, \mnras, 441, 2361

\bibitem[{Vogt {et~al.}(1994)Vogt, Allen, Bigelow, Bresee, Brown, Cantrall, Conrad, Couture, Delaney, Epps, Hilyard, Hilyard, Horn, Jern, Kanto, Keane, Kibrick, Lewis, Osborne, Pardeilhan, Pfister, Ricketts, Robinson, Stover, Tucker, Ward, \& Wei}]{Vogt1994}
Vogt, S.~S., Allen, S.~L., Bigelow, B.~C., {et~al.} 1994, in Instrumentation in Astronomy VIII, ed. D.~L. Crawford \& E.~R. Craine (SPIE)

\bibitem[{{Walter} {et~al.}(1978){Walter}, {Charles}, \& {Bowyer}}]{Walter1978}
{Walter}, F., {Charles}, P., \& {Bowyer}, S. 1978, \apjl, 225, L119

\bibitem[{Williams {et~al.}(2014)Williams, Cook, \& Berger}]{Williams2014}
Williams, P. K.~G., Cook, B.~A., \& Berger, E. 2014, The Astrophysical Journal, 785, 9

\bibitem[{{Wilson}(1941)}]{wilson1941}
{Wilson}, O.~C. 1941, \apj, 93, 29

\bibitem[{{Wood} {et~al.}(2001){Wood}, {Linsky}, {M{\"u}ller}, \& {Zank}}]{wood01}
{Wood}, B.~E., {Linsky}, J.~L., {M{\"u}ller}, H.-R., \& {Zank}, G.~P. 2001, \apjl, 547, L49

\bibitem[{{Wood} {et~al.}(2021){Wood}, {M{\"u}ller}, {Redfield}, {Konow}, {Vannier}, {Linsky}, {Youngblood}, {Vidotto}, {Jardine}, {Alvarado-G{\'o}mez}, \& {Drake}}]{wood21}
{Wood}, B.~E., {M{\"u}ller}, H.-R., {Redfield}, S., {et~al.} 2021, \apj, 915, 37

\bibitem[{{Zarka}(2018)}]{Zarka2018}
{Zarka}, P. 2018, in Handbook of Exoplanets, ed. H.~J. {Deeg} \& J.~A. {Belmonte}, 22

\bibitem[{{Zechmeister} {et~al.}(2018){Zechmeister}, {Reiners}, {Amado}, {Azzaro}, {Bauer}, {B{\'e}jar}, {Caballero}, {Guenther}, {Hagen}, {Jeffers}, {Kaminski}, {K{\"u}rster}, {Launhardt}, {Montes}, {Morales}, {Quirrenbach}, {Reffert}, {Ribas}, {Seifert}, {Tal-Or}, \& {Wolthoff}}]{Zechmeister2018}
{Zechmeister}, M., {Reiners}, A., {Amado}, P.~J., {et~al.} 2018, \aap, 609, A12

\end{thebibliography}

\appendix

%-------------------------------------------------------------------
%-------------------------------------------------------------------
%-------------------------------------------------------------------
\section{RVs}
\label{sec:rvs}
Table \ref{tab:gj625rvs}, and Table \ref{tab:gj3861rvs} lists the RVs of GJ~625 and GJ 3861, respectively.

%-------------------------------------------------------------------
\begin{table}[H]
\caption{The first and last 20 GJ~625 RVs used in its orbit analysis. The full table of RVs are available at \href{https://zenodo.org/}{Zenodo} under DOI 10.5281/zenodo.15491027.} \label{tab:gj625rvs}
\begin{tabular}{lccc}
\hline \hline
Time [$\mathrm{BJD}_{\mathrm{TDB}}$]  &  $v$  [m/s] & $\sigma$ [m/s] & Instrument \\
\hline
2450862.148819 & 0.8 & 2.3 & HIRES-pre \\
2450956.003773 & 0.9 & 2.3 & HIRES-pre \\
2451009.896343 & 0.6 & 1.8 & HIRES-pre \\
2451227.151979 & 3.5 & 2.2 & HIRES-pre \\
2451312.984491 & 5.7 & 2.3 & HIRES-pre \\
2451342.912187 & -1.9 & 2.6 & HIRES-pre \\
2451373.911308 & 10.3 & 1.8 & HIRES-pre \\
2451703.899375 & 0.7 & 2.5 & HIRES-pre \\
2452004.016910 & -6.6 & 2.6 & HIRES-pre \\
2452098.893947 & -2.7 & 1.8 & HIRES-pre \\
2452390.080266 & -4.6 & 2.8 & HIRES-pre \\
2452537.721424 & -4.2 & 2.5 & HIRES-pre \\
2452713.085660 & -5.8 & 2.7 & HIRES-pre \\
2452806.003912 & -1.2 & 2.4 & HIRES-pre \\
2452850.890451 & 3.8 & 2.4 & HIRES-pre \\
2453179.928067 & -1.2 & 2.3 & HIRES-pre \\
2453430.073368 & -5.8 & 1.6 & HIRES-post \\
2453842.056840 & -2.2 & 1.7 & HIRES-post \\
2454248.013345 & 1.1 & 1.7 & HIRES-post \\
2454248.981030 & 2.9 & 1.9 & HIRES-post \\
... & ... & ... & ... \\
... & ... & ... & ... \\
... & ... & ... & ... \\
... & ... & ... & ... \\
... & ... & ... & ... \\
2459429.599710 & -1.9 & 1.4 & HARPSN \\
2459431.575983 & 2.8 & 1.2 & HARPSN \\
2459433.542890 & 7.18 & 0.78 & HARPSN \\
2459441.397804 & -2.0 & 1.1 & HARPSN \\
2459442.361047 & 0.4 & 1.1 & HARPSN \\
2459444.419671 & 1.8 & 1.1 & HARPSN \\
2459446.444683 & 3.37 & 0.94 & HARPSN \\
2459449.514670 & 2.48 & 0.89 & HARPSN \\
2459451.373740 & 0.20 & 0.80 & HARPSN \\
2459453.354846 & -0.39 & 0.81 & HARPSN \\
2459454.367793 & -3.1 & 1.3 & HARPSN \\
2459459.391279 & -3.1 & 1.1 & HARPSN \\
2459461.391605 & -4.4 & 1.0 & HARPSN \\
2459464.400146 & -0.7 & 1.3 & HARPSN \\
2459468.362289 & -3.20 & 0.74 & HARPSN \\
2459472.408774 & -2.85 & 0.82 & HARPSN \\
2459476.373683 & 0.8 & 1.0 & HARPSN \\
2459478.374010 & 0.1 & 1.4 & HARPSN \\
2459481.338990 & 2.2 & 1.3 & HARPSN \\
2459713.899768 & 2.65 & 0.49 & NEID \\
\hline
\end{tabular}
\end{table}

\begin{table*}
\caption{GJ 3861 RVs. $v_1$ and $\sigma_1$ lists the RVs and corresponding uncertainties of the primary star, and $v_2$ and $\sigma_2$ list the rvs RVs and uncertainties of the secondary star.}
\begin{tabular}{lcccc}
 \hline \hline
Time [$\mathrm{BJD}_{\mathrm{TDB}}$]  &  $v_1$ [km/s] & $\sigma_1$ [km/s] & $v_2$ [km/s]    & $\sigma_2$ [km/s] \\ \hline
2459302.962248                        &    0.381      &    0.044          &  -35.155 &    0.042 \\
2459649.019935                        &   -5.507      &    0.054          &  -27.650 &    0.147 \\
2459662.982436                        &   -0.090      &    0.050          &  -34.588 &    0.010 \\
2460310.038762                        &  -21.996      &    0.034          &   -6.197 &    0.116 \\
2460334.959251                        &  -29.021      &    0.068          &    3.112 &    0.092 \\
2460339.951484                        &  -19.374      &    0.023          &   -9.548 &    0.035 \\
2460342.936435                        &    5.506      &    0.031          &  -42.109 &    0.020 \\
2460359.893783                        &    2.849      &    0.017          &  -38.604 &    0.044 \\
2460391.810179                        &  -10.455      &    0.036          &  -20.967 &    0.066 \\
2460444.681975                        &   -9.374      &    0.049          &  -22.377 &    0.070 \\
2460736.862509                        &  -35.722      &    0.030          &   11.730 &    0.066 \\
\hline
\end{tabular}
\label{tab:gj3861rvs}
\end{table*}

% --------------------------------------------------------
% --------------------------------------------------------
% --------------------------------------------------------

\section{Ca IRT Emission}
\begin{figure*}[h!]
\centering
\includegraphics[width=\linewidth]{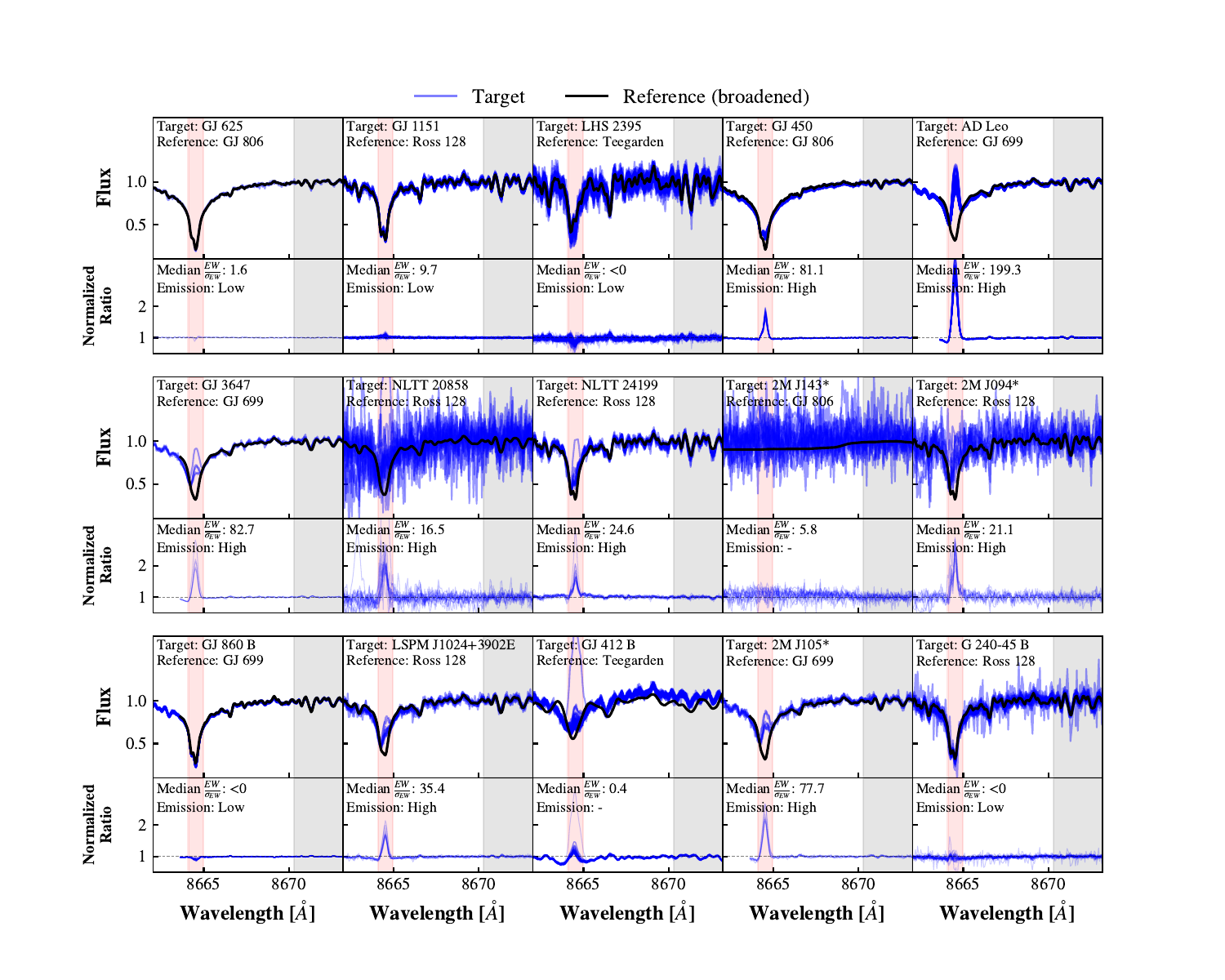}
\caption{HPF observations of the Ca II IRT line at 8664$\AA$ for all non-SB2 targets in the M\,dwarf radio sample (blue curves), compared to a single high S/N as-observed HPF spectrum of slowly rotating quiescent reference stars (black). The reference spectrum is rotationally broadened to match the observed broadening of the lines, by using the estimates of the $v\sin i$ obtained by \texttt{HPF-SpecMatch}. The red shaded region marks the line center, which is the emission window used to compute the Equivalent Width. The lower panels show the normalized ratio residuals (target/reference), highlighting any excess emission as a peak in the residuals. A continuum region (8670.300–8673.190$\AA$, grey shaded) is used to normalize the residuals to unity. The lower panels also denote our overall Ca II IRT emission designation, as either `Low' ($\frac{EW}{\sigma_{EW}} < 10$), or `High'  ($\frac{EW}{\sigma_{EW}} > 10$). We do not give an emission designation to targets 2MASS J143 and GJ~412, as the observations do not match well to their respective reference spectra, making it unadvisable to interpret the obtained $\frac{EW}{\sigma_{EW}}$ value.}
\label{fig:cairt}
\end{figure*}

Figure \ref{fig:cairt} shows the Ca IRT line of all our observations (blue), with a slowly-rotating, quiescent reference star shown in black.

% --------------------------------------------------------
% --------------------------------------------------------
% --------------------------------------------------------
\section{Broadening Functions of Single-Lined Systems}
\label{sec:bfssingle}
\begin{figure*}[h!]
\centering
\includegraphics[width=0.95\linewidth]{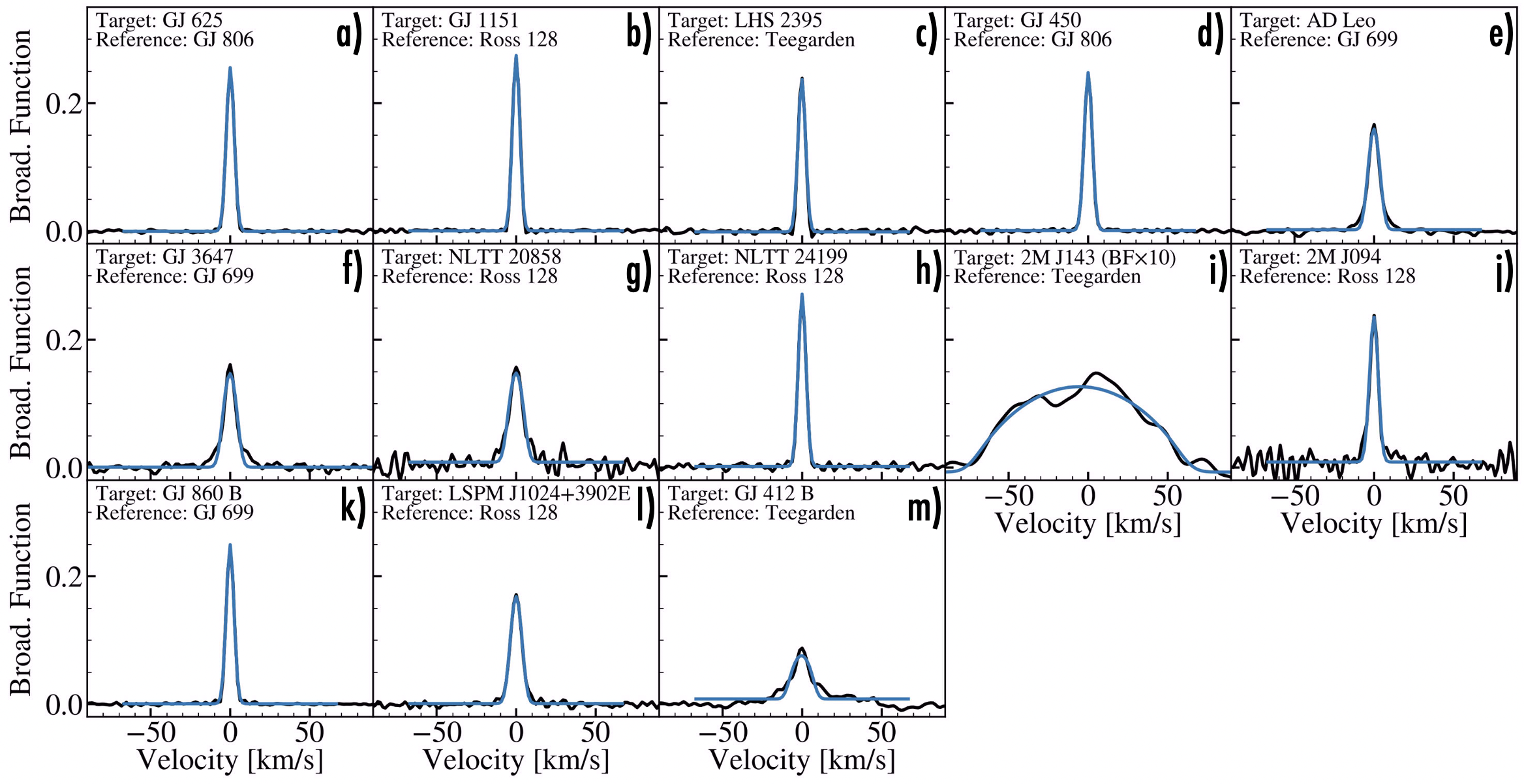}
\caption{Broadening functions for the single-lined systems. The black curves show the broadening function from the highest SNR HPF spectrum. The blue curve shows a rotationally broadened line profile accounting for the resolution of HPF of 55,000, using the best-fit $v \sin i_\star$ listed in Table \ref{tab:general_properties}.}
\label{fig:bfs}
\end{figure*}

Figure \ref{fig:bfs} shows the broadening functions for the single-lined systems.

% --------------------------------------------------------
% --------------------------------------------------------
% --------------------------------------------------------
\section{ZTF Light Curves}
\begin{figure*}
\centering
\includegraphics[width=0.95\linewidth]{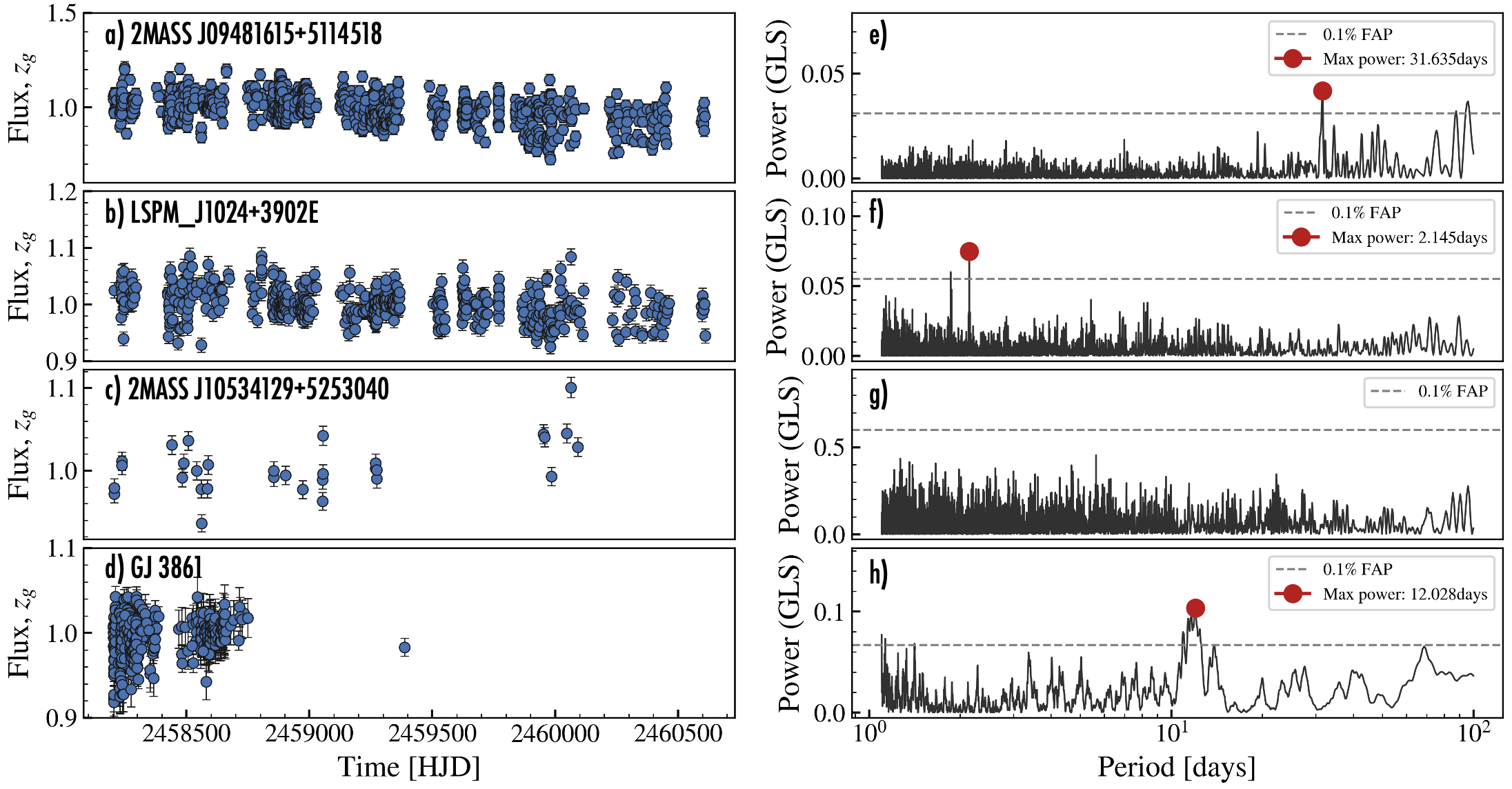}
\caption{ZTF $z_g$-band photometry of a) 2MASS J09481615+5114518, b) LSPM J1024+3902E, c) 2MASS 10534129+5253040, and d) GJ 3861. Panels e), f), g) and h) show the corresponding Lomb-Scargle periodograms. The grey vertical dashed lines show the 0.1\% false alarm probability, and the red dots highlight the highest peaks with false alarms lower than 0.1\% probability. }
\label{fig:ztf}
\end{figure*}

Figure \ref{fig:ztf}, panel A), shows the ZTF $z_g$-band light-curves of several targets with unknown $P_{\text{rot}}$ values. Panel B) shows the corresponding Lomb-Scargle periodograms, which reveal the $P_{\text{rot}}$ for targets 2MASS J09481615+5114518, LSPM J1024+3902E, and GJ 3861. The data for 2MASS 10534129+5253040 was not precise enough to constrain its rotation period.

\end{document}